\documentclass{aa}  

\usepackage{graphicx}
\usepackage{txfonts}
\usepackage[colorlinks = true,
            linkcolor = blue,
            urlcolor  = blue,
            citecolor = blue,
            anchorcolor = blue]{hyperref}

\usepackage{booktabs}
\usepackage[flushleft]{threeparttable}
\usepackage{lipsum}  
\usepackage{balance}

\makeatletter
\renewcommand*\aa@pageof{
, page \thepage{} of \pageref*{LastPage}}
\makeatother

\begin{document}

\title{
AGN feedback in the Local Universe: \\
multiphase outflow 
of the Seyfert galaxy NGC~5506}

\author{
Federico Esposito \inst{1,2} \and
Almudena Alonso-Herrero \inst{3} \and
Santiago García-Burillo \inst{4} \and
Viviana Casasola \inst{5} \and
Françoise Combes \inst{6} \and
Daniele Dallacasa \inst{1,5} \and
Richard Davies \inst{7} \and
Ismael García-Bernete \inst{8} \and
Begoña García-Lorenzo \inst{9,10} \and
Laura Hermosa Muñoz \inst{3} \and
Luis Peralta de Arriba \inst{3} \and
Miguel Pereira-Santaella \inst{11} \and
Francesca Pozzi \inst{1,2} \and
Cristina Ramos Almeida \inst{9,10} \and
Thomas Taro Shimizu \inst{7} \and
Livia Vallini \inst{2} \and
Enrica Bellocchi \inst{12,13} \and
Omaira González-Martín \inst{14} \and
Erin K. S. Hicks \inst{15} \and
Sebastian Hönig \inst{16} \and
Alvaro Labiano \inst{17} \and
Nancy A. Levenson \inst{18} \and
Claudio Ricci \inst{19,20} \and
David J. Rosario \inst{21}
}

\institute{
Dipartimento di Fisica e Astronomia, 
Università degli Studi di Bologna, 
Via P. Gobetti 93/2, I-40129 Bologna, Italy\\
\email{federico.esposito7@unibo.it}
\and
Osservatorio di Astrofisica e Scienza dello Spazio 
(INAF–OAS), Via P. Gobetti 93/3, I-40129 Bologna, Italy
\and
Centro de Astrobiología (CAB), CSIC-INTA, 
Camino Bajo del Castillo s/n, 
E-28692 Villanueva de la Cañada, Madrid, Spain
\and
Observatorio de Madrid, OAN-IGN, 
Alfonso XII, 3, E-28014 Madrid, Spain
\and
INAF – Istituto di Radioastronomia, Via P. Gobetti 101, 40129, Bologna, Italy
\and
Observatoire de Paris, LERMA, Collège de France, CNRS, 
PSL University, Sorbonne University, 75014, Paris
\and
Max-Planck-Institut für Extraterrestrische Physik (MPE),
Giessenbachstraße 1, D-85748 Garching, Germany
\and
Department of Physics, University of Oxford, 
Keble Road, Oxford OX1 3RH, UK
\and
Instituto de Astrofísica de Canarias (IAC), 
Calle Vía Láctea, s/n, 38205 La Laguna, Tenerife, Spain
\and
Departamento de Astrofísica, Universidad de La Laguna, 
38206 La Laguna, Tenerife, Spain
\and
Instituto de Física Fundamental, CSIC, 
Calle Serrano 123, 28006 Madrid, Spain
\and
Departamento de Física de la Tierra y Astrofísica, 
Fac. de CC Físicas, Universidad Complutense de Madrid,
28040 Madrid, Spain
\and
Instituto de Física de Partículas y del Cosmos IPARCOS,
Fac. de CC Físicas, Universidad Complutense de Madrid,
28040 Madrid, Spain
\and
Instituto de Radioastronomía and Astrofísica (IRyA-UNAM),
3-72 (Xangari), 8701, Morelia, Mexico
\and
Department of Physics \& Astronomy, 
University of Alaska Anchorage, AK 99508-4664, USA
\and
Department of Physics \& Astronomy, University of 
Southampton, Hampshire, SO17 1BJ, Southampton, UK
\and
Telespazio UK for the European Space Agency (ESA), 
ESAC, Camino Bajo del Castillo s/n, 
28692 Villanueva de la Cañada, Spain
\and
Space Telescope Science Institute, 3700 San Martin Drive,
Baltimore, Maryland 21218, USA
\and
Instituto de Estudios Astrofísicos, Facultad de Ingeniería 
y Ciencias, Universidad Diego Portales, Avenida Ejercito
Libertador 441, Santiago, Chile
\and
Kavli Institute for Astronomy and Astrophysics, 
Peking University, Beijing 100871, China
\and
School of Mathematics, Statistics and Physics, 
Newcastle University, Newcastle upon Tyne, NE1 7RU, UK
}

\date{Received XXX; accepted YYY}

 
  \abstract{
   We present new optical
   GTC/MEGARA seeing-limited ($0.9\arcsec$)
   integral-field 
   observations of NGC~5506,
   together with
   ALMA observations 
   of the CO($3-2$) transition at a
   $0.2 \arcsec$ ($\sim 25$ pc)
   resolution.
   NGC~5506 is a luminous (bolometric luminosity
   of $\sim 10^{44}$ erg s$^{-1}$) nearby
   (26 Mpc) Seyfert galaxy, part of the
   Galaxy Activity, Torus, and Outflow Survey (GATOS).
   We modelled the CO($3-2$) kinematics with
   \textsuperscript{3D}\textsc{Barolo},
   revealing a rotating and outflowing
   cold gas ring within the central 1.2 kpc.
   We derived an integrated cold molecular gas mass
   outflow rate for the ring of 
   $\sim 8$ M$_{\odot}$ yr$^{-1}$.
   We fitted the optical emission lines 
   with a maximum of two Gaussian components
   to separate rotation from non-circular motions.
   We detected high [OIII]$\lambda 5007$ 
   projected velocities
   (up to $\sim 1000$ km s$^{-1}$)
   at the active galactic nucleus (AGN) 
   position, decreasing with radius
   to an average $\sim 330$ km s$^{-1}$
   around $\sim 350$ pc.
   We also modelled the [OIII] gas kinematics
   with a non-parametric method,
   estimating the ionisation parameter and
   electron density in every spaxel,
   from which we derived an ionised mass
   outflow rate of $0.076$ M$_{\odot}$ yr$^{-1}$
   within the central 1.2 kpc.
   Regions of high CO($3-2$) velocity dispersion,
   extending to projected distances 
   of $\sim 350$ pc from the AGN,
   appear to be the result from the interaction 
   of the AGN wind with molecular gas 
   in the galaxy's disc.
   Additionally, we find the ionised outflow to
   spatially correlate with radio and 
   soft X-ray emission in the central kiloparsec.
   We conclude that the effects of AGN feedback 
   in NGC~5506 manifest as a large-scale
   ionised wind interacting with the molecular
   disc, resulting in outflows extending to 
   radial distances of 610 pc.
   }


   \maketitle


\section{Introduction}

The feedback from active galactic nuclei (AGN)
has been proposed as a key mechanism
influencing the course of galaxy evolution
\citep[see e.g. the reviews by][]{alexander12,
fabian12, kormendy13, heckman14, harrison18}.
AGN are powered by supermassive black holes
(SMBHs) and their accretion discs located at
galactic centres, emitting intense
radiation and influencing the excitation
and kinematics of the surrounding gas.
This feedback process initiates from a very small region 
\citep[$10^{-3}$ pc is a typical
accretion disc radius, see][]{cornachione20, guo22, jha22},
yet its effects can extend to influence the entire
galaxy structure
\citep{okamoto05, croton06, gaibler12, 
mcnamara12, feruglio15, dubois16, anglesalcazar17}.

Analysing AGN feedback is a complex task due to
its impact on different phases of the interstellar
medium (ISM) across different physical scales,
from the sub-parsec highly ionised 
ultra-fast outflows
\citep[UFOs,][]{tombesi10, fukumura15, nomura16},
warm absorbers \citep{blustin05, laha14},
and broad absorption lines
\citep[BALs,][]{weymann91, proga00},
to the kiloparsec-scale ionised 
\citep{mccarthy96, baum00, liu13, perna20, fluetsch21}
and molecular 
\citep{feruglio10, cicone14, bischetti19, ramosalmeida22}
outflows,
up to the megaparsec-scale emission
of giant radio galaxies
\citep[GRGs,][]{ishwarachandra99, kuzmicz18, dabhade20}
and X-ray groups and clusters
\citep{mccarthy10, fabian11, mcnamara12, pasini20}.

Furthermore, AGN activity is considered to be
intermittent over the course of a galaxy's lifetime
\citep{king04, hopkins06, schawinski15, king15b},
or even on timescales of days or less
\citep{dultzinhacyan92, wagner95}.
Consequently, comprehending the overall impact 
of AGN feedback is highly challenging, requiring 
the use of multiwavelength, multi-scale, and multi-time
observations as essential tools.

The molecular phase of the ISM
is of paramount importance, since it is
the fuel for star formation and the phase
in which the bulk of the gaseous mass 
in star-forming galaxies resides
\citep[e.g.][]{casasola20}.
The AGN radiation heats the molecular gas
by creating X-ray-dominated regions within the 
ISM \citep{maloney96, esposito22, esposito24, wolfire22},
and it perturbs its kinematics,
driving outflows
\citep{cicone14, fiore17, veilleux20, lamperti22}.

Molecular gas typically forms a rotating disc
associated with the galaxy gravitational potential.
In AGN-host galaxies, a common form of perturbation
involves the interaction between the molecular disc 
and the AGN hot wind, which manifests as outflowing 
ionised gas observable in X-rays 
\citep{cappi06, tombesi13, giustini23}, 
UV \citep{hewett03, rankine20}
and optical \citep{fabian12, mullaney13}
wavelengths
\citep[see also the review by][
and references therein, 
for the hot-cold gas coupling]{veilleux20}.
In this regard, a multiwavelength approach is essential
to effectively trace the multiphase outflow
\citep{davies14, cicone18, garciabernete21, speranza24}.
Nearby AGN serve as a perfect laboratory
for studying these feedback signatures in detail,
particularly with the increasingly improved
spatial resolution and spectral coverage 
of today's instruments.

The Galactic Activity, Torus, and Outflow Survey (GATOS) 
aims to understand the obscuring material 
(torus) and the nuclear gas cycle (inflows and outflows)
in the immediate surroundings of
the nuclear region of local AGN
\citep{garciaburillo21, alonsoherrero21, garciabernete24}.
The GATOS sample includes Seyfert galaxies
with distances $10-40$ Mpc,
selected from the 70-month Burst Alert Telescope
(BAT) catalogue of AGN
\citep{baumgartner13}, some of which have been
observed at different wavelengths,
including optical and near-infrared 
integral field unit (IFU) spectroscopy,
JWST, and ALMA observations.

One of the key findings of the GATOS survey is the
existence of an anti-correlation
between the nuclear molecular gas
concentration and the AGN power:
for a sample of 18
galaxies \citep{garciaburillo21} 
and an extended sample 
(García-Burillo et al., in preparation).
Molecular gas, traced by low-$J$ CO lines
and observed by ALMA 
at a spatial resolution $\sim 10$ pc,
is also detected in outflows in those sources
showing the most extreme nuclear-scale 
gas deficiencies, hence
suggesting that the AGN power plays a role
in clearing the nuclear region.
These outflows have been observed
and analysed in detail for some 
GATOS selected sources, which have been
analysed in detail for the molecular 
and ionised phases 
\citep{alonsoherrero18, alonsoherrero19, 
alonsoherrero23, garciabernete21,
peraltadearriba23}.

In this study, we investigate the molecular 
and ionised gas phases of NGC~5506, 
an Sa spiral galaxy in the GATOS sample at a 
redshift-independent distance
of 26 Mpc \citep{karachentsev06}.
At this distance the spatial scale is 
$125$ pc/$\arcsec$.
NGC~5506 has an AGN bolometric luminosity of 
$\sim 1.3 \times 10^{44}$ erg s$^{-1}$
\citep{davies14} and is classified as an
optically obscured narrow line Seyfert 1
\citep[NLSy1,][]{nagar02}.
The black hole mass is
$M_{\rm BH} = 2.0^{+8.0}_{-1.6} \times 10^7$
M$_{\odot}$ \citep{gofford15}, 
yielding an Eddington ratio of
$\lambda_{\rm Edd} \equiv L_{\rm bol}/L_{\rm Edd}
= 0.05^{+0.21}_{-0.04}$.
NGC~5506 is notable in the GATOS sample
for having one of the highest
molecular gas nuclear deficiencies
\citep[see Fig. 18 of][]{garciaburillo21},
suggesting a potential imprint of AGN 
feedback on the molecular gas.
Furthermore, NGC~5506 hosts a sub-parsec bent radio jet
\citep{roy00, roy00b, kinney00}
and a UFO \citep{gofford13, gofford15},
making it an intriguing target for
investigating multiphase (and multiscale) outflows.
As a NLSy1, NGC~5506 is expected to be in a
young AGN phase, characterised by a
small black hole mass and a high accretion rate
\citep[see e.g.][]{crenshaw03, tarchi11, salome23}.

Evidence of complex kinematics from
the long-slit optical spectrum 
was found by \cite{wilson85}, 
who suggested radial motion for the ionised gas.
\cite{maiolino94} refined this model, 
identifying outflowing velocities of up to 400 km s$^{-1}$
for [OIII], [NII], and H$\alpha$, 
with the outflow cone inclined at $-15^{\circ}$ 
from the north. 
Additionally, \cite{fischer13} estimated 
an ionised outflow velocity of 500 km s$^{-1}$
using slitless \textit{Hubble} Space Telescope (HST) 
observations \citep[see also][]{ruiz05}, 
modelling a biconical outflow.
\cite{davies20} carried out a detailed
analysis of optical data from observations
made with X-shooter at VLT,
finding a [OIII] outflow with a maximum 
velocity of 792 km s$^{-1}$ and
$\dot{M}_{\rm out} = 0.21$ M$_{\odot}$ yr$^{-1}$.
\cite{riffel17, riffel21} and \cite{bianchin22} 
studied the outflow
of the ionised gas in the near-IR
(with GEMINI NIFS),
finding a mass outflow rate ranging
from 0.11 to 12.49  M$_{\odot}$ yr$^{-1}$
(by adopting two fixed $n_e$ values
- 500 cm$^{-3}$ and $10^4$ cm$^{-3}$ - 
and exploring different geometries).
The highest outflow values
would result in a kinetic efficiency
$\dot{E}_{\rm out} / L_{\rm bol} = 0.71$.
They also calculate, from $L_{\rm bol}$,
a mass accretion rate to the SMBH
of 0.067 M$_{\odot}$ yr$^{-1}$.
%
\begin{figure*}
\centering
\includegraphics[width=0.85\textwidth]{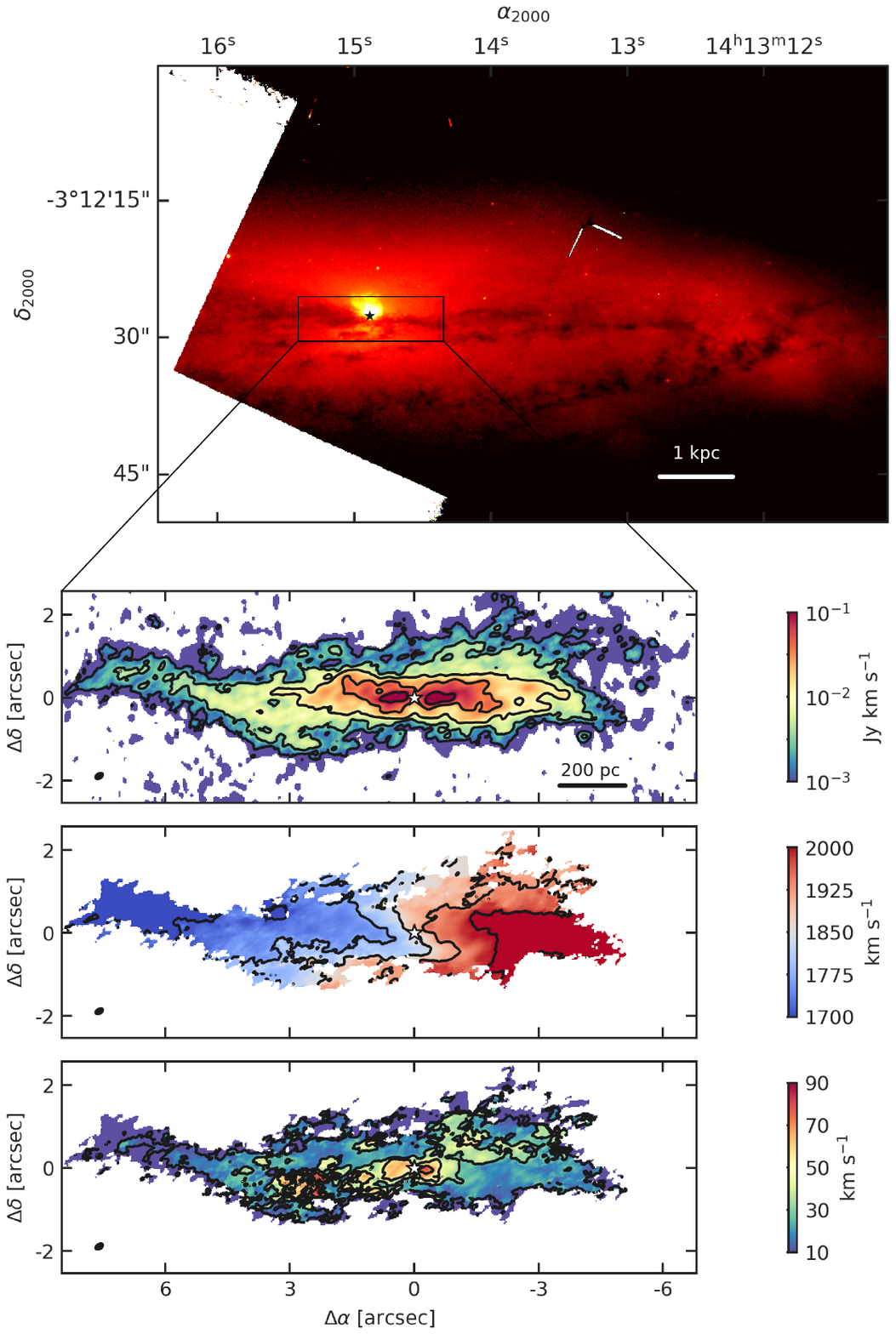}
\caption{Optical and molecular views of NGC 5506.
\textit{Top}. 
HST/F606W image of NGC~5506
from \cite{malkan98}. 
The black rectangle identifies a region
of $15.3 \arcsec \times 5.1 \arcsec$
(corresponding to $1.9 \times 0.6$ kpc$^2$).
\textit{Bottom}. ALMA CO($3-2$)
intensity, velocity and velocity dispersion maps,
clipped at a signal-to-noise ratio of 3.
The contours are between
$10^{-3}$ and $10^{-1}$ Jy km s$^{-1}$
(with 0.5 dex steps) for the intensity map,
between 1700 and 2000 km s$^{-1}$
(with 75 km s$^{-1}$ steps)
for the velocity map,
and between 10 and 90 km s$^{-1}$
(with 20 km s$^{-1}$ steps)
for the velocity dispersion map.
North is up and east is left,
and offsets in the ALMA maps
are measured relative to
the 870 $\mu$m continuum peak
\citep[as in][]{garciaburillo21}, marked
with a star symbol in every panel.
The ALMA beam ($0.21 \arcsec \times 0.13 \arcsec$)
appears in every bottom panel
as a black ellipse in the lower left.}
\label{fig:ngc5506}
\end{figure*}
\begin{figure*}
\centering
\includegraphics[width=0.98\textwidth]{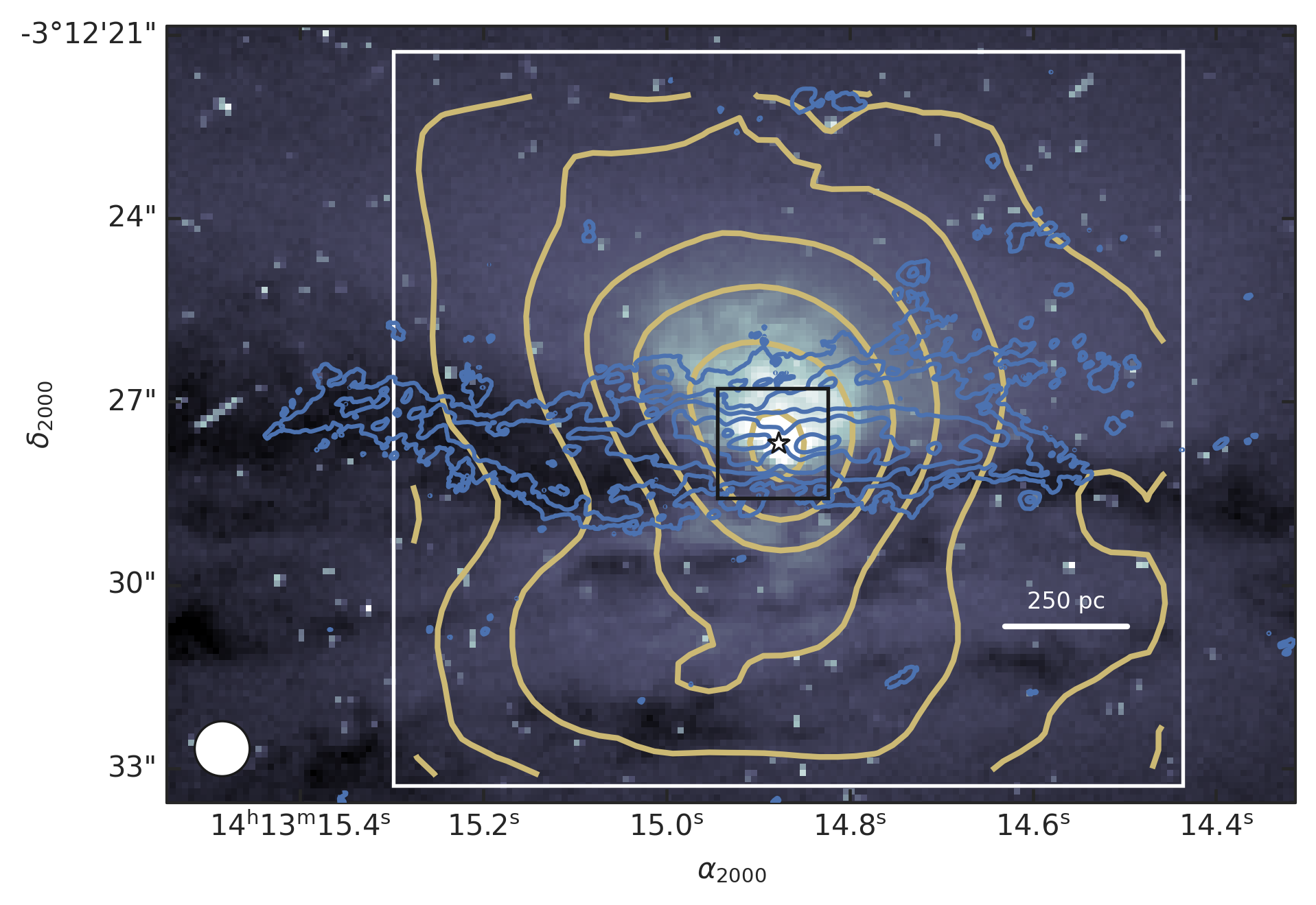}
\includegraphics[width=\textwidth]{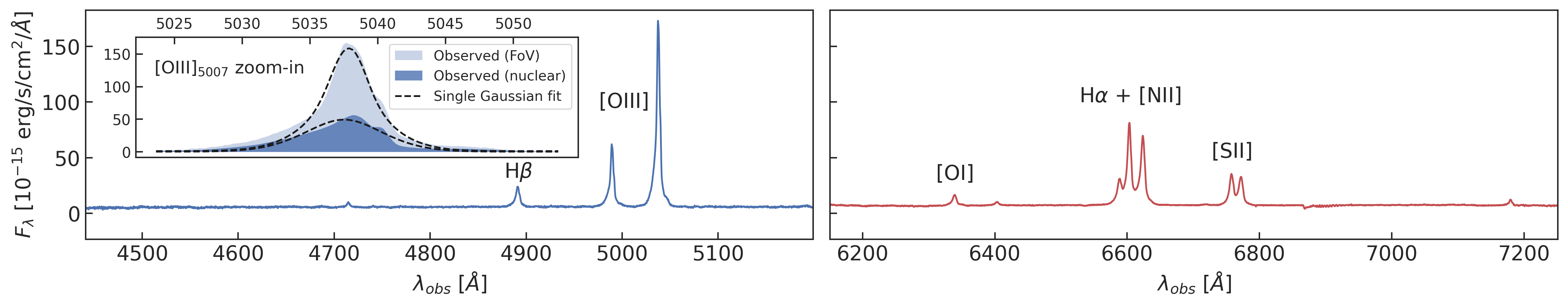}
\caption{Optical image and spectra of the
central region of NGC 5506.
\textit{Top}. GTC/MEGARA
[OIII] ($\lambda_e = 5007$ \AA, in orange)
and CO ($\lambda_e = 870 \mu m$, in blue)
contours over the HST/F606W image of NGC~5506
\citep{malkan98}.
The [OIII] contour levels, from the
single-component Gaussian fit, have a 
logarithmic spacing from $3 \sigma$
to $80\%$ of the peak intensity in steps
of $0.5$ dex, while the CO($3-2$) 
contours are the same of
Fig.~\ref{fig:ngc5506}.
The white star symbol is the
AGN position, as determined in
Section~\ref{sec:GTC_obs}.
The black and white squares are the nuclear region,
with size 1.8 arcsec $\sim 225$ pc,
observed by X-shooter
\citep[see][]{davies20}, and the
MEGARA FoV ($12.5 \arcsec \times 11.3 \arcsec$
$\sim$ 1.5 kpc $\times$ 1.4 kpc),
respectively.
The white circle in the bottom left
is the MEGARA seeing conditions
(diameter $0.9 \arcsec$).
\textit{Bottom}.
Left and right panels contain
the spectra (integrated within the 
MEGARA FoV)
revealed with the MEGARA LR-B and LR-R gratings, 
respectively,
with names of identified emission lines and doublets.
The inset is the zoom-in of a [OIII] line
(after continuum subtraction):
the blue shadings are
the observed spectra of the MEGARA FoV
and of the nuclear region,
the black dashed lines are the fits
with a single Gaussian.
The inset axes have the same units
of the outer panel.
}
\label{fig:oiii_co_hst}
\end{figure*}

In this work, we present new
IFU observations made
with the Multi-Espectrógrafo en GTC 
de Alta Resolución para Astronomía
(MEGARA) at the Gran Telescopio Canarias (GTC),
which cover several optical emission lines,
together with ALMA Band 7 observations 
of the CO($3-2$) transition.
The paper is structured as follows.
In Section~\ref{sec:observations} we present
the ALMA Band 7 and GTC/MEGARA observations
In Section~\ref{sec:morphology} we describe the
morphology of the molecular and ionised
gas emission lines, 
while we model the kinematics of the two phases in
Sections~\ref{sec:mol_model} and 
\ref{sec:ion_model}, respectively.
In Section~\ref{sec:discussion} we discuss the results
of this work and we compare our data with the
available literature, and we draw our conclusions
in Section~\ref{sec:conclusions}.

\section{Observations}
\label{sec:observations}

\subsection{ALMA Band 7}

We observed NGC~5506 with the 
Band 7 ALMA receiver and a single pointing
(project-ID: \#2017.1.00082.S; 
PI: S. García-Burillo).
We analysed the moderate resolution
datacube from \cite{garciaburillo21}.
The datacube has a 
$0.21 \arcsec \times 0.13 \arcsec$
(26 pc $\times$ 16 pc)
beam (with PA $=-60^{\circ}$, measured
anticlockwise from the northern direction),
a $17 \arcsec$ (2.1 kpc)
field of view (FoV) and
a largest angular scale of 
$4 \arcsec$ (0.5 kpc).
To check the astrometry, we first
aligned the HST/F606W image
(top panel of Fig.~\ref{fig:ngc5506})
with the position of stars 
(from the Gaia mission), and then we aligned
the ALMA continuum peak
and the HST peak, resulting in
$\alpha_{2000} = 
14^{\text{h}}13^{\text{m}}14.877^{\text{s}}$,
$\delta_{2000} = -03^{\circ}12\arcmin 27.67\arcsec$
\citep[as in][]{garciaburillo21}.

\begin{table}
\centering
\caption{Fundamental parameters for NGC~5506.}
\label{tab:tab_data}
\begin{threeparttable}
\begin{tabular}{lll}
\toprule
Parameter & Value & 
Reference\textsuperscript{a} \\
\midrule
$\alpha_{2000}$ & 
$14^{\text{h}}13^{\text{m}}14.877^{\text{s}}$ & (1) \\
$\delta_{2000}$ & 
$-03^{\circ}12\arcmin 27.67\arcsec$ & (1) \\
$V_{\text{hel}}$\textsuperscript{b} & 
$1882 \pm 11$ km s$^{-1}$ & (1) \\
RC3 Type & Sa pec edge-on & (2) \\
Nuclear activity & Optically obscured NLSy1 & (3) \\
Distance & 26 Mpc ($1 \arcsec = 125$ pc) & (4) \\
$D_{25}$ & 2.82\arcmin & (5) \\
Inclination & $80^{\circ}$ & (1) \\
Position Angle & $265^{\circ}$ & (1) \\
$M_{\text{BH}}$ & $2.0^{+8.0}_{-1.6} \times 10^7$
M$_{\odot}$ & (6) \\
$L_{\text{bol}}$ & 
$1.3 \times 10^{44}$ erg s$^{-1}$ & (7) \\
$L_{\text{IR}}$ & 
$3.1 \times 10^{10}$ L$_{\odot}$ & (8) \\
$\lambda_{Edd}$ & $0.05^{+0.21}_{-0.04}$ & (1) \\
\bottomrule
\end{tabular}
\begin{tablenotes}
\item \textbf{Notes.} \textsuperscript{a}
(1) This paper; (2) \cite{devaucouleurs91};
(3) \cite{nagar02}; (4) \cite{karachentsev06};
(5) \cite{baillard11}; (6) \cite{gofford15}; 
(7) \cite{davies20}; (8) \cite{sanders03}.
\textsuperscript{b} Heliocentric velocity
is the mean between the systemic velocities
derived for the molecular and the ionised gas
\end{tablenotes}
\end{threeparttable}
\end{table}
\subsection{GTC/MEGARA Bands B and R}
\label{sec:GTC_obs}

We observed the central region of NGC~5506 on 20/03/2021
(Program GTC27-19B; PI: A. Alonso-Herrero), 
with MEGARA in IFU mode
\citep{gildepaz16, carrasco18}.
We used two low resolution (LR) volume phase
holographic gratings:
the LR-B (spectral range $\sim 4300 - 5200$ \AA, 
resolution $R \sim 5000$), to observe
H$\beta$ and the [OIII]$\lambda \lambda 4959, 5007$
doublet (exposure time 480 s), and the LR-R
($\sim 6100 - 7300$ \AA, $R \sim 5900$), 
to observe H$\alpha$, [OI]$\lambda6300$, and the
[NII]$\lambda\lambda6548, 6583$ and 
[SII]$\lambda\lambda6716, 6731$ doublets
(exposure time 400 s).
The observed FoV is 
$12.5 \arcsec \times 11.3 \arcsec$, 
corresponding to $1.6 \times 1.4$ kpc$^2$.

The data reduction was performed by
following \cite{peraltadearriba23}
and using the official MEGARA pipeline
\citep{pascual21}.
The resolution of the GTC/MEGARA observations 
was limited by the seeing conditions.
We plotted it with a circle of diameter
$0.9 \arcsec$ in all the relevant figures.
The final datacubes were produced with
a spaxel size of $0.3 \arcsec$, as recommended
by the pipeline developers 
\citep{pascual21, peraltadearriba23}:
this corresponds to a physical spaxel size of 37.5 pc.
We corrected the maps astrometry 
from the two configurations
by aligning their continuum peaks
with the ALMA Band 7 (870 $\mu$m)
and HST (F606W filter) ones;
to this point as the AGN position.
We note that optical extinction may have an impact 
on the observed optical nucleus and
actual AGN location on scales below the
MEGARA seeing of $0.9 \arcsec$.

\section{Morphology and kinematics}
\label{sec:morphology}

\subsection{ALMA CO(3-2)}
Fig.~\ref{fig:ngc5506} shows
the Hubble Space Telescope (HST) image
and the ALMA CO($3-2$) first three moments maps.
The CO intensity map reveals an
edge-on disc with a nuclear deficit
(with respect to the circumnuclear region)
of diameter $\sim 100$ pc.
This molecular gas depletion 
in the very centre
has already been observed and analysed
in \cite{garciaburillo21}.
The circumnuclear disc 
is symmetric up to a diameter of 
$\sim 7 \arcsec = 875$ pc.
At radii larger than $3-4 \arcsec$
there is an extended gas tail
in the eastern direction,
which traces the dust lane visible
in the HST image
(see also Fig.~\ref{fig:oiii_co_hst}).

The bottom panels of Fig.~\ref{fig:ngc5506}
show the velocity and velocity dispersion of CO($3-2$). 
The velocity field is centred at 1850 km s$^{-1}$.
It appears to be dominated by rotation,
redshifted on the western side and
blueshifted on the eastern side.
However, it also exhibits perturbations
due to non-circular motion.
The velocity dispersion has a median value
of 17 km s$^{-1}$, and displays
higher values along the NW-SE axis,
with a maximum value of 86 km s$^{-1}$
at $\delta \alpha \sim 3\arcsec$.

\begin{figure}
\centering
\includegraphics[width=0.49\textwidth]{
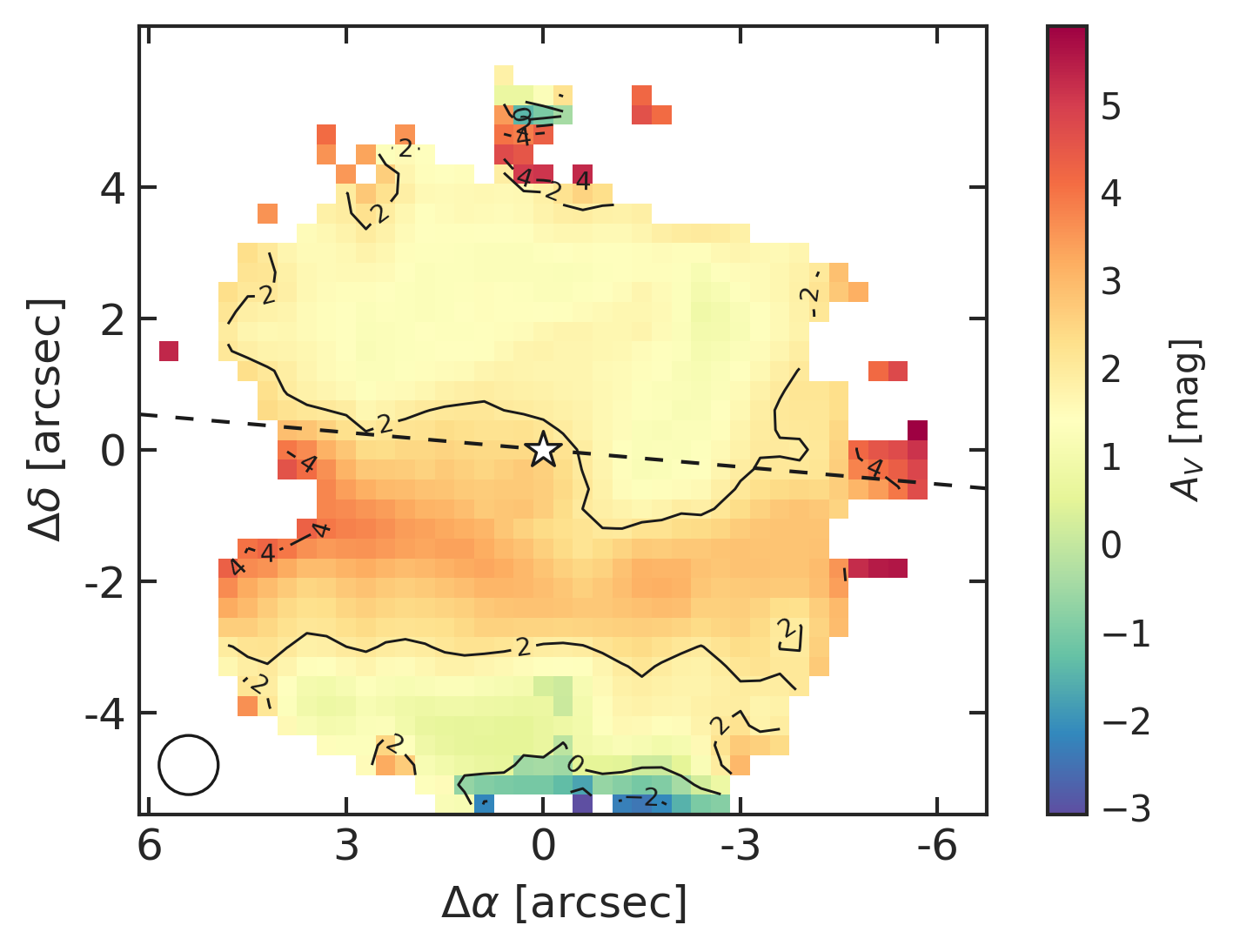}
\caption{Visual extinction map
of the MEGARA FoV, calculated from
the H$\alpha$/H$\beta$ ratio and 
with $R_V = 3.1$ \citep{cardelli89}.
The white star symbol is the
AGN position, and distances are
relative to it.
The dashed black line is our fiducial
major kinematic axis with PA $= 265^{\circ}$
(see Section~\ref{sec:mol_model}).
The white circle in the bottom left
is the MEGARA seeing conditions.
}
\label{fig:extinction}
\end{figure}

\subsection{GTC/MEGARA emission lines}
\label{sec:gtc_morpho}

To derive the line intensity and kinematics, 
we extracted and fitted every spaxel of the
MEGARA FoV with the ALUCINE\footnote{
Available at 
\url{https://gitlab.com/lperalta_ast/alucine}}
\citep[Ajuste de Líneas para Unidades de
Campo Integral de Nebulosas en Emisión,][]{peraltadearriba23}, 
initially with a single Gaussian 
and an amplitude-over-noise (AoN) of 3 or higher.
Fig.~\ref{fig:oiii_co_hst} shows the contours of
the [OIII] doublet
intensity, which
are the brightest lines in the MEGARA spectrum
(also in Fig.~\ref{fig:oiii_co_hst}).
The other identified emission lines, namely
H$\beta$, [OI], H$\alpha$ + [NII] doublet,
and [SII] doublet, are labelled in
Fig.~\ref{fig:oiii_co_hst}.

The [OIII] emission of Fig.~\ref{fig:oiii_co_hst}
nicely follows the HST image,
and it is shaped as a bicone, typical
of narrow line regions
\citep[NLRs][]{pogge88, wilson93, schmitt03}.
The bicone emerges almost vertically in projection
from the dusty molecular disc
\citep[$\sim 20^{\circ}$ anticlockwise from the north,
as reported by][]{fischer13, garciaburillo21}.
We note here that \cite{fischer13} detected
a one-sided ionisation cone
(the northern side)
using slitless spectroscopy of the [OIII] line.
This is also
evident in the HST map
(Figs.~\ref{fig:ngc5506} and \ref{fig:oiii_co_hst}).
With MEGARA we detect the southern
side as well, although it appears more
extinguished.
To check this we derived, following
\cite{cardelli89}, the visual extinction map
from the H$\alpha$/H$\beta$ line ratios
(whereas the line fluxes come from the
single Gaussian fit spaxel-by-spaxel).
The resulting map (Fig.~\ref{fig:extinction})
shows a clear dust band crossing
the southern side of NGC~5506 nuclear region.
This piece of information also suggests that
the southern side is the near side of the galaxy 
\citep[in accordance with][analysis]{garciaburillo21}.

It is tempting to interpret the
comparison of [OIII] and CO
contours in Fig.~\ref{fig:oiii_co_hst}
as an ionised outflow that escapes the 
galaxy disc following the path of less
resistance \citep{fauchergiguere12}.
We explore this possibility
in Section~\ref{sec:discussion}.


\begin{figure}
\centering
\includegraphics[width=0.49\textwidth]{
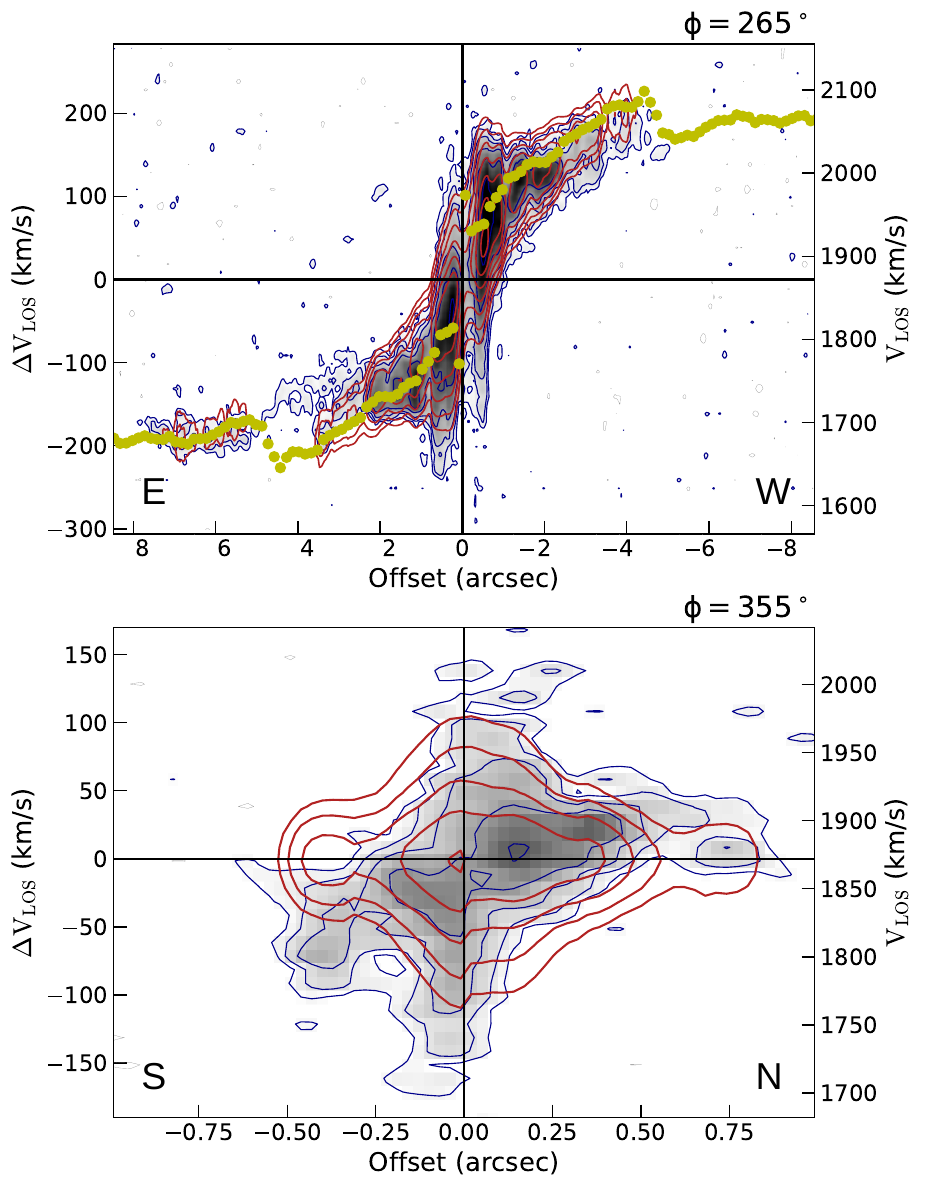}
\caption{ALMA CO($3-2$) PV diagrams
generated with 3DB
along the kinematic major (\textit{top panel})
and minor (\textit{bottom panel}) axes.
The grey scale and blue contours are 
the ALMA CO($3-2$) observations $> 3\sigma$,
while the red contours are the 
3DB rotating disc model (without a radial
velocity component).
The yellow dots are the fitted rotation curve.
The approximate 
eastern, western, southern, and northern
directions are marked in the panels.}
\label{fig:barolo_PV_VRAD0}
\end{figure}

\section{Modelling the molecular gas kinematics}
\label{sec:mol_model}

The CO($3-2$) velocity field map 
(Fig.~\ref{fig:ngc5506}) shows the
typical signatures of a rotating disc with
some deviations from non-circular motions.
We modelled the CO($3-2$) datacube with 
\textsuperscript{3D}\textsc{Barolo}\footnote{
Available at \url{https://bbarolo.readthedocs.io}}
\citep[hereafter 3DB]{diteodoro15}, 
which creates a disc model 
for the rotating gas by dividing the
emission into concentric rings, 
and fits the following parameters for every ring: 
the kinematic centre coordinates, 
the scale-height of the disc ($z_0$),
the inclination ($i$) of the disc with respect to the
line of sight, the position angle (PA, measured
anticlockwise from the northern direction
for the receding side of the rotating disc) of the 
major kinematic axis, the systemic velocity $v_{\rm sys}$
with which the whole galaxy is receding from us, 
the rotational velocity ($v_{\rm rot}$) of the gas, 
the velocity dispersion ($\sigma_{\rm gas}$), 
and the radial velocity ($v_{\rm rad}$).

We note here that 3DB is designed to model
the gas kinematics within a rotating disc
(plus a radial velocity component).
Our strategy is to use 3DB to identify 
and quantify any radial motion within the disc,
which may point to an inflow or outflow of gas.
If the radial flow forms an angle $\theta_{\rm out}$
with the galaxy disc, only the velocity 
projected on the disc, that is,
$v_{\rm out} \cos \theta_{\rm out}$,
will be detected by 3DB
\citep[see also][]{diteodoro21, bacchini23}.

We fixed the kinematic centre at the position
of the continuum peak.
We set a ring radial size
of $0.15 \arcsec$ ($\simeq 19$ pc), 
similar to the ALMA beam
($0.21 \arcsec \times 0.13 \arcsec$),
and a total of 60 rings,
thus reaching out to a distance
of $9 \arcsec$ ($\simeq 1.1$ kpc) from the centre.

\begin{figure}
\centering
\includegraphics[width=0.49\textwidth]{
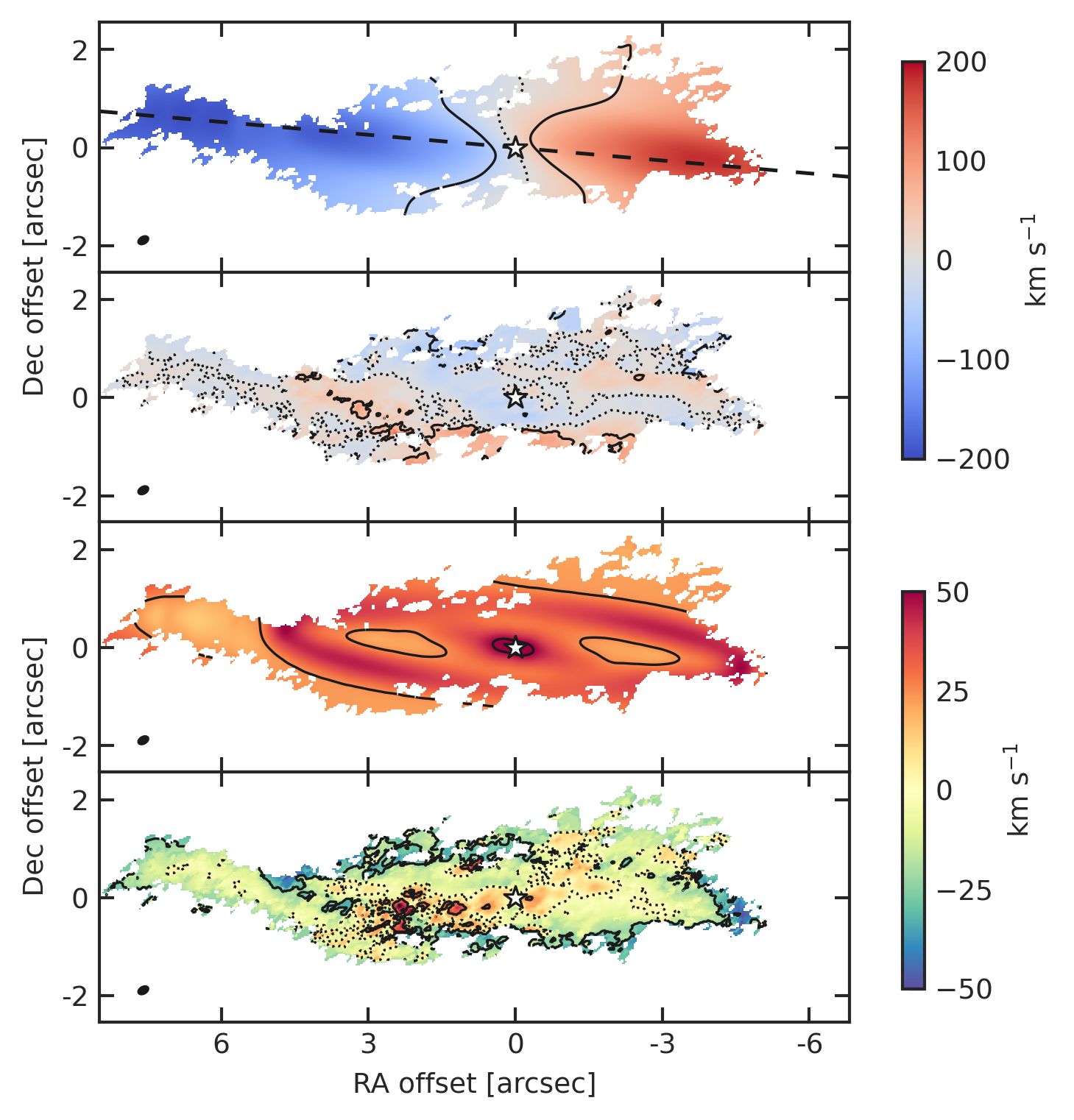}
\caption{Best-fit model and residuals
(i.e. observation minus the model) obtained 
with 3DB for the rotating disc with a radial velocity
component. Top and bottom panels show
the mean velocity field and the velocity dispersion 
field of CO($3-2$), respectively.
In the first panel, $v_{\rm sys}=1872$ km s$^{-1}$
has been subtracted from the model velocities.
Velocity contours (top panels)
are at $-50$ and 50 km s$^{-1}$
(solid) and at 0 km s$^{-1}$ (dotted),
while dispersion contours (bottom panels)
are at $-50$, $-25$, 25, and 50 km s$^{-1}$
(solid) and at 0 km s$^{-1}$ (dotted).
The dashed black line in the first panel is 
the kinematic major axis with PA $=265^{\circ}$.
The ALMA beam appears in every panel
as a black ellipse in the bottom left.}
\label{fig:barolo_map_combined}
\end{figure}

\subsection{Rotating disc}
\label{sec:rotation_only}

We performed a first 3DB run with
$v_{\rm rad} = 0$ km s$^{-1}$,
and $z_0$, $i$, PA, $v_{\rm sys}$,
$v_{\rm rot}$, and $\sigma_{\rm gas}$ as
free parameters.
In this way we derived 
$z_0 = 0.2 \arcsec \simeq 25$ pc,
$i=80^{\circ}$, PA $=265^{\circ}$, 
and $v_{\rm sys}=1872 \pm 10$ km s$^{-1}$
(where the error is given by the ALMA
datacube spectral step).
The inclination is the same as that found
by \cite{garciaburillo21} with the
software \texttt{kinemetry}
\citep{krajnovic06}, while the PA
is slightly different 
(they found PA $=275^{\circ}$).
The $v_{\rm sys}$ value is 
in agreement with several works:
\cite{fischer13} reported 1823 km s$^{-1}$,
\cite{riffel17} 1878 km s$^{-1}$,
\cite{davies20} 1962 km s$^{-1}$,
\cite{garciaburillo21} 1840 km s$^{-1}$,
and the average of these values
is 1876 km s$^{-1}$, only 4 km s$^{-1}$
over our estimate.

We then performed a 3DB run with
$v_{\rm rot}$ and $\sigma_{\rm gas}$
as the only free parameters, while
the others were fixed to 
the values determined in the first run.
This approach assumes the absence of
radial motions associated 
with molecular inflows or outflows.
Fig.~\ref{fig:barolo_PV_VRAD0} displays
the position-velocity (PV) diagrams resulting
from this run. Overall, the 3DB model contours 
(red lines in Fig.~\ref{fig:barolo_PV_VRAD0})
reasonably reproduce the observed
PV values (plotted with blue colours).
From the major-axis PV diagram
(Fig.~\ref{fig:barolo_PV_VRAD0}, top panel)
we can appreciate the goodness of the 
$v_{\rm sys}$ estimate, as the CO($3-2$)
emission is symmetric with respect to $v_{\rm sys}$.
Along the kinematic minor axis 
(Fig.~\ref{fig:barolo_PV_VRAD0}, bottom panel),
there are indications of non-circular motions 
in the central $1 \arcsec$, which we explore
further in the next section.

\begin{figure}
\centering
\includegraphics[width=0.49\textwidth]{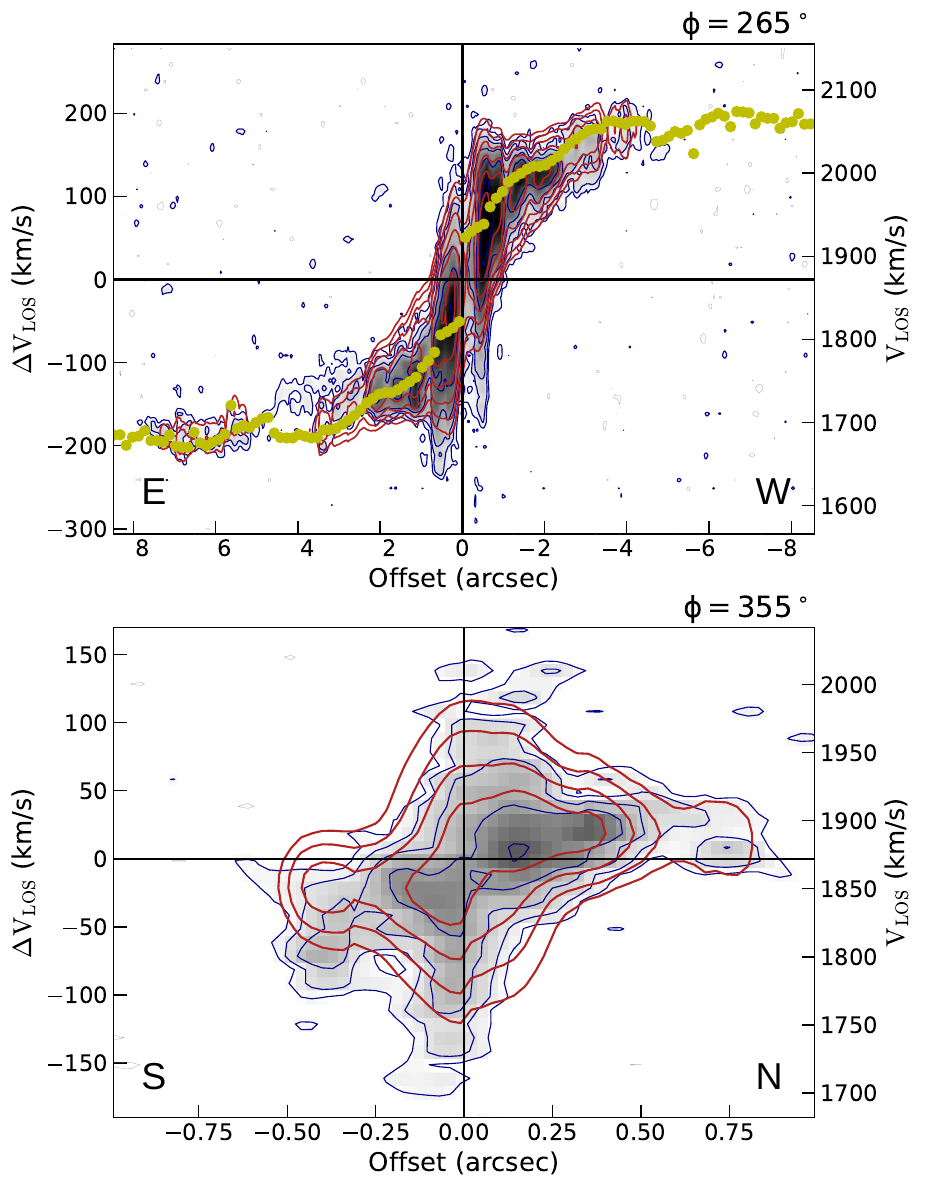}
\caption{Same as Fig.~\ref{fig:barolo_PV_VRAD0},
but for the 3DB run with a radial velocity component.}
\label{fig:barolo_PV}
\end{figure}

\subsection{Rotating disc with a
radial velocity component}

Within the approximate inner (projected)
$1 \arcsec$,
the minor axis PV diagram shows 
redshifted motions to the north of the AGN 
and blueshifted to the south 
(see top-left and bottom-right quadrants, respectively, 
of Fig.~\ref{fig:barolo_PV_VRAD0}, bottom panel).
Since the south is the near side of the galaxy 
(see Fig.~\ref{fig:extinction} and
related discussion in 
Section~\ref{sec:gtc_morpho}),
this suggests the presence of a CO outflow
component in the plane of the disc.
We thus run another 3DB model including a 
radial velocity ($v_{\rm rad}$) component.
The other free parameters are $v_{\rm rot}$
and $\sigma_{\rm gas}$, while the others
have been set as the previous run.

Fig.~\ref{fig:barolo_map_combined}
shows the 3DB models and residuals for this run,
for the first and second moments 
(mean velocity and mean velocity dispersion).
The velocity and velocity dispersion
absolute residuals have median values
of 16 km s$^{-1}$ and 14 km s$^{-1}$,
respectively.
The highest velocity residuals
(Fig.~\ref{fig:barolo_map_combined},
second panel) are in the SE direction,
where also the highest values of dispersion
(Fig.~\ref{fig:ngc5506}, bottom panel)
and dispersion residuals
(Fig.~\ref{fig:barolo_map_combined},
fourth panel) reside.

\begin{figure}
\centering
\includegraphics[width=0.49\textwidth]{
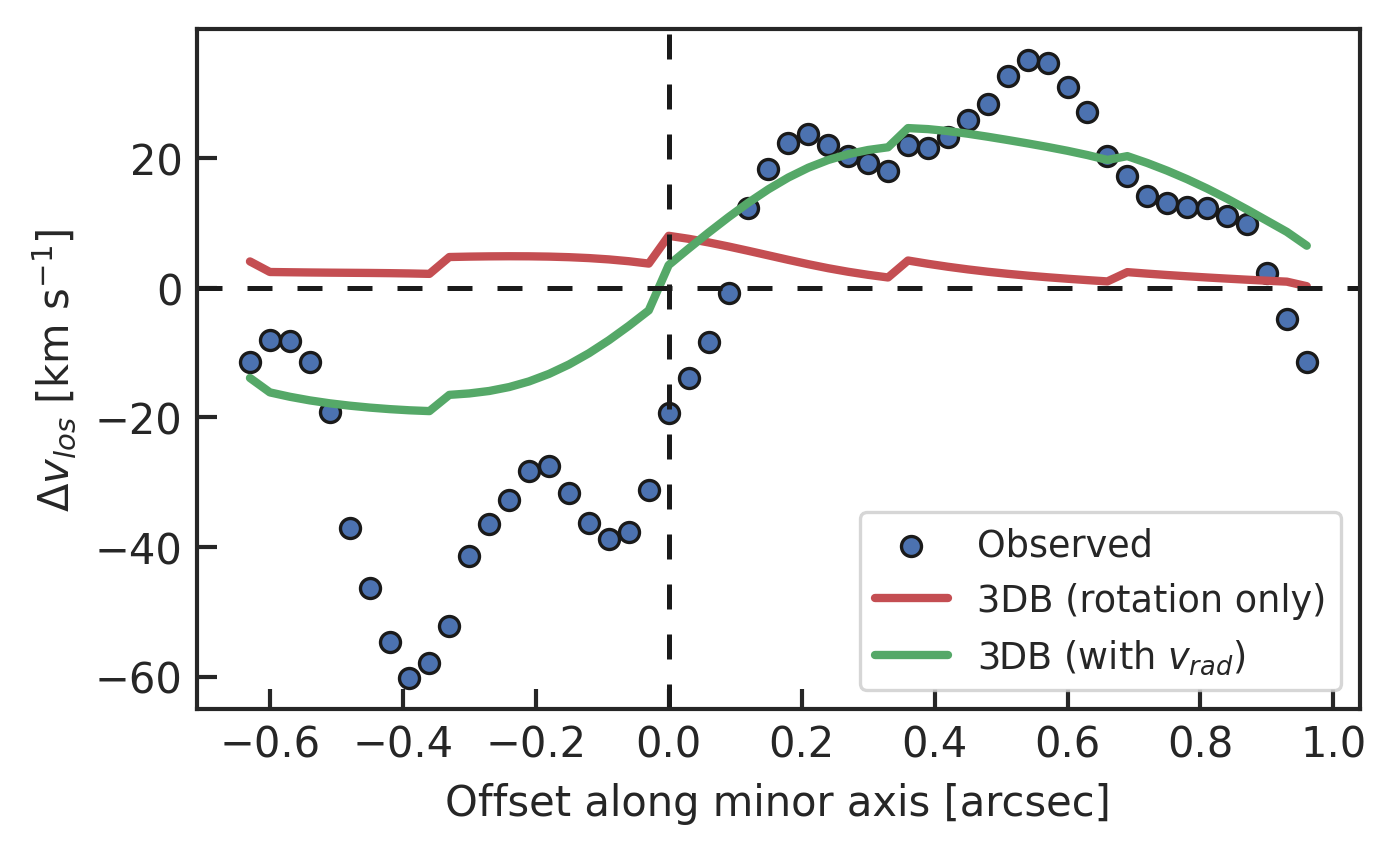}
\caption{Relative velocities of CO($3-2$)
along the minor axis, extracted and averaged from a 
slit width of 3 pixels (corresponding to a projected
width of $0.09\arcsec$). The blue circles are the
observed values, not weighted for the emitted flux.
The green and red lines are the 3DB models with and
without a radial velocity component.}
\label{fig:co_minor_radial}
\end{figure}
\begin{figure}
\centering
\includegraphics[width=0.49\textwidth]{
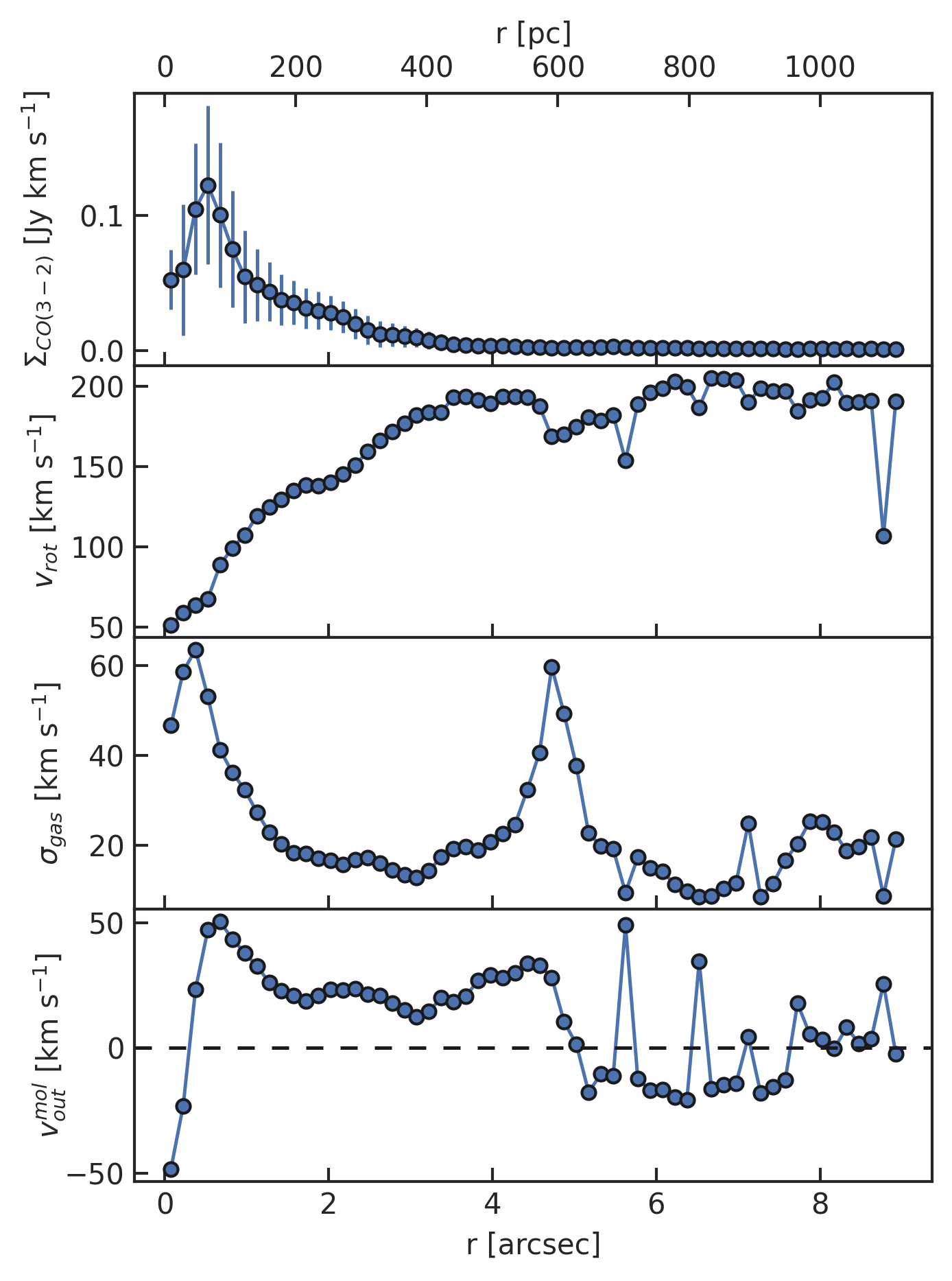}
\caption{Radial profiles of the molecular gas
derived with 3DB. From top to bottom:
the CO($3-2$) surface density,
the rotational velocity
(same as the yellow dots in the top panel
of Fig.~\ref{fig:barolo_PV}),
the velocity dispersion and
the outflow velocity, all as a function
of deprojected distance from the AGN 
(on the plane of the galaxy).
The dashed black line in the bottom panel
is the zero line, dividing between ouflow
($v_{\rm out}^{\rm mol} > 0$) and inflow
($v_{\rm out}^{\rm mol} < 0$).}
\label{fig:barolo_radial}
\end{figure}

Fig.~\ref{fig:barolo_PV} shows
the PV diagrams of this run, where we 
can appreciate a better fit along
the minor axis
(Fig.~\ref{fig:barolo_PV}, bottom panel).
Especially, the 3DB model now
follows the northern red - southern blue
asymmetry along the CO($3-2$) minor axis.
We also plot the mean velocities
along the minor axis in Fig.~\ref{fig:co_minor_radial},
where we compare the 3DB results for the two models
(with and without the radial velocity component).
While not perfect, the model incorporating $v_{\rm rad}$
more closely aligns with the observed data, 
particularly at positive offsets from the centre
(i.e. in the northern direction).
The yellow dots in the top panel
represent the mean $v_{\rm rot}$
(also plotted in the second panel of 
Fig.~\ref{fig:barolo_radial}).
We find $v_{\rm rot}$ reaching $193$ km s$^{-1}$
at $r = 3.5 \arcsec$ 
(440 pc), in reasonable agreement with 
the rotational velocity of $181 \pm 5$ km s$^{-1}$
measured from HI absorption \citep{gallimore99}.

The four panels in
Fig.~\ref{fig:barolo_radial} show,
from top to bottom,
the CO($3-2$) surface density $\Sigma_{\rm CO(3-2)}$,
and the modelled rotational velocity $v_{\rm rot}$,
velocity dispersion $\sigma_{\rm gas}$,
and radial velocity $v_{\rm out}^{\rm mol}$
(which is the same as $v_{\rm rad}$, with the
positive sign meaning outflowing and negative
meaning inflowing gas).
We distinguish significative changes in the curve
profiles at two particular radii: 
$0.4 \arcsec$ and $5 \arcsec$.

At $r \sim 0.4 \arcsec$ (50 pc) 
we find the maximum value of
$\Sigma_{\rm CO(3-2)}$, which corresponds
to the inner radius of the ring.
Within this radius $v_{\rm out}^{\rm mol}<0$, which
is indicative of
inflowing gas: it could be an indication
of AGN feeding from the molecular disc
\citep[see e.g.][]{combes21},
but since we only have two radial points
we could not confirm this finding.
At $r \sim 0.4 \arcsec$ we also have a peak in the 
$\sigma_{\rm gas}$ and $v_{\rm out}^{\rm mol}$ profiles:
this means that the molecular ring is not only
rotating, but also outflowing
\citep[as in NGC 1068, see][]{garciaburillo19}.

At $r \sim 5 \arcsec$ (610 pc) there is another peak
of $\sigma_{\rm gas}$, and $v_{\rm out}^{\rm mol}$ goes
from positive to negative, suggesting
a transition from outflow to inflow
\citep[][found a similar result 
for NGC 1068]{garciaburillo14}.
However, at $r > 5 \arcsec$ there is a lot of
oscillation between inflow and outflow,
probably due to the small number of datapoints
(see the asymmetry of the CO emission
in Fig.~\ref{fig:ngc5506}),
so we did not take into consideration these
outer radii.

The CO($3-2$) radial motion on the 
molecular plane could be explained 
with \textit{(i)} inflowing/outflowing gas
\citep{garciabernete21, ramosalmeida22}
or \textit{(ii)} elliptical orbits associated
with a bar
\citep{buta96, casasola11, audibert19}.
Since the presence of a bar is not evident
on the CO($3-2$) PV diagrams
\citep[Figs.~\ref{fig:barolo_PV_VRAD0} 
and~\ref{fig:barolo_PV}, cf.][]{
alonsoherrero23}, it probably does not
dominate the motion of the molecular gas.
Nevertheless, due to this possibility,
we conservatively assume that 
the outflow velocities we derive
between $R_{out,min}^{\rm mol} = 0.4 \arcsec$ (50 pc) 
and $R_{out,max}^{\rm mol} = 5 \arcsec$ (610 pc)
are upper limits.
We discuss the presence of a bar in NGC~5506
in Section~\ref{sec:bar}.

\begin{figure}
\centering
\includegraphics[width=0.49\textwidth]{
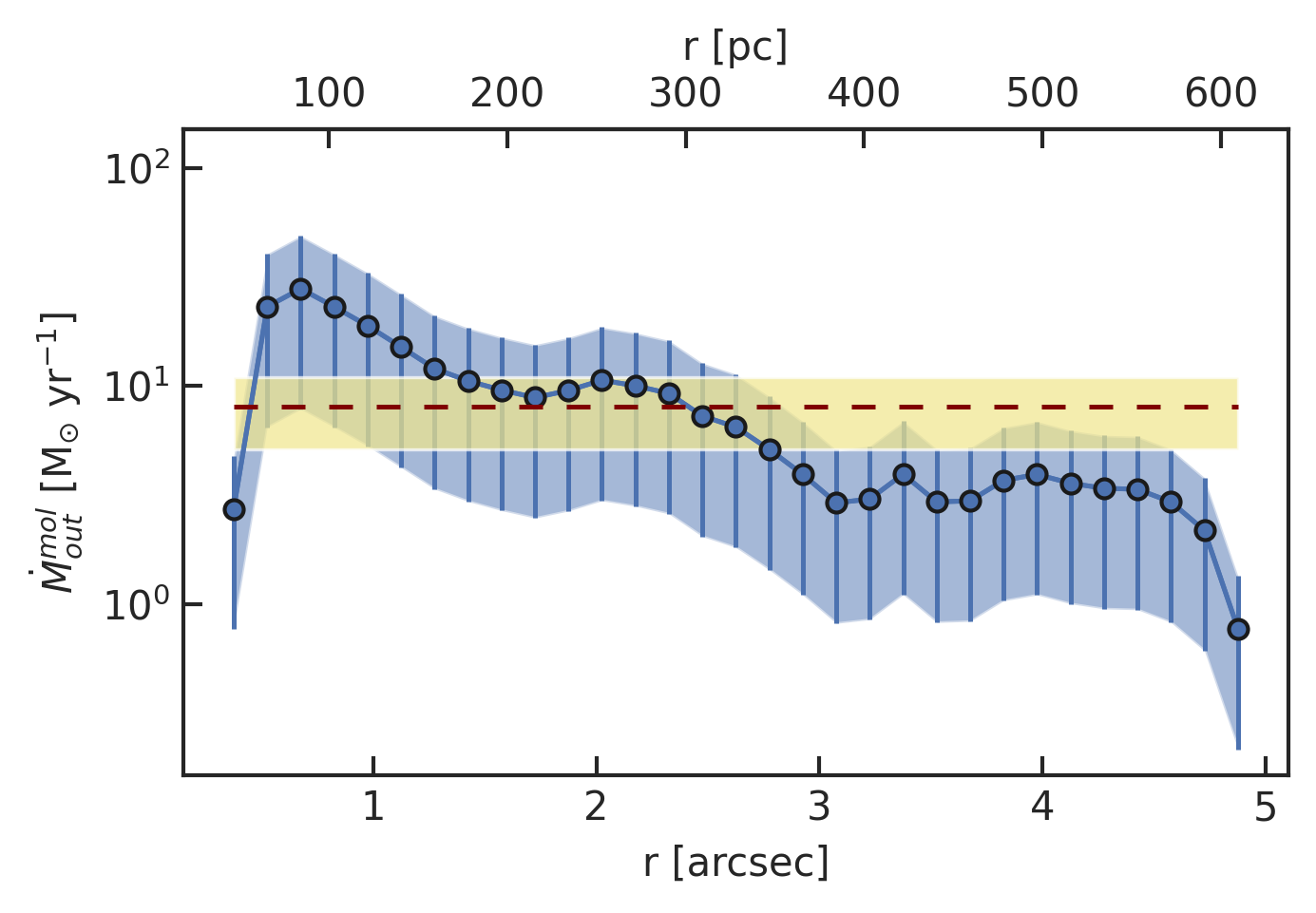}
\caption{Molecular gas mass outflow rate
as a function of the deprojected distance
from the AGN (on the plane of the galaxy). 
The molecular gas mass includes 
the helium contribution.
The blue dots correspond to the values 
computed with the 3DB model radial velocities 
and CO($3-2$) intensities of
Fig.~\ref{fig:barolo_radial}.
The blue shading and errorbars represent the
variation in $r_{31}$ and $X_{\text{CO}}$
(see text for details).
The red dashed line is the integrated mass outflow rate,
$8 \pm 3$ M$_{\odot}$ yr$^{-1}$,
with the shaded yellow region
representing its uncertainty.}
\label{fig:mol_MOR}
\end{figure}

\subsection{The molecular mass outflow rate}
\label{sec:mol_outflowrate}

We used the CO($3-2$) emission
between $R_{out,min}^{\rm mol} = 0.4 \arcsec$ (50 pc) 
and $R_{out,max}^{\rm mol} = 5 \arcsec$ (610 pc)
to calculate the main properties of the outflow, 
such as the amount of molecular gas 
it is driving outwards
($M_{\rm out}^{\rm mol}$).
To do so, we first converted it
to CO($1-0$), using a typical
brightness temperatures ratio for galaxy discs of
$r_{31} \equiv 
T_{B, \rm CO(3-2)} / T_{B, \rm CO(1-0)} = 0.7$:
this is the average value found by \cite{israel20}
in 126 nearby galaxy centres, and we adopt it 
for consistency with \cite{garciaburillo21}.
We then used a Galactic CO-to-H$_2$ conversion factor
of $X_{\rm CO} = 2 \times 10^{20}$ mol cm$^{-2}$
(K km s$^{-1}$)$^{-1}$
\citep{bolatto13}.
We chose the Galactic value to better compare
our results with most of the literature,
and also because NGC~5506 does not show any indication
of merger and is not particularly luminous
in the infrared
\citep[$L_{IR} = 10^{10.49}$ L$_{\odot}$,][]{sanders03}.
We calculate 
$M_{\rm out}^{\rm mol} = 1.75 \times 10^8$ M$_{\odot}$
between $R_{out,min}^{\rm mol}$ and $R_{out,max}^{\rm mol}$.
This result depends on our choice of 
$r_{31}$ and $X_{\rm CO}$. Specifically, a higher
brightness temperatures ratio 
$r_{31} \sim 1$,
as found in the central $\sim 1 \arcsec$
of NGC~1068 \citep{garciaburillo14, viti14},
would decrease $M_{\rm out}^{\rm mol}$.
Conversely, a lower $r_{31} \sim 0.4$ \citep[i.e. within
$1\sigma$ of the values collected by][]{israel20},
would increase $M_{\rm out}^{\rm mol}$.
Also a lower $X_{\rm CO} \sim 0.8 \times 10^{20}$
mol cm$^{-2}$ (K km s$^{-1}$)$^{-1}$, 
usually associated to starburst galaxies 
\citep{bolatto13, pereztorres21}, would
decrease $M_{\rm out}^{\rm mol}$.
The combined uncertainty of these two
conversions could decrease the calculated
mass by a multiplicative factor 
of $\approx 0.28$ or increase
it by a factor of $\approx 1.75$.

Assuming a simple shell geometry
\citep[as in][]{alonsoherrero23},
we can write the mass outflow rate as
\begin{equation}
\label{eq:Mdot_out}
\dot{M}_{\rm out}^{\rm mol} = 
\frac{M_{\rm out}^{\rm mol} \, v_{\rm out}^{\rm mol}}{R_{\rm out}^{\rm mol}} 
\; \; \; ,
\end{equation}
where $v_{\rm out}^{\rm mol}$ is defined 
as the average velocity measured
between $R_{out,min}^{\rm mol}$ and $R_{out,max}^{\rm mol}$.
By taking the standard deviation as its uncertainty,
we find $v_{\rm out}^{\rm mol} = 25.6 \pm 9.4$ km s$^{-1}$,
from which we infer a molecular mass outflow rate of
$\dot{M}_{\rm out}^{\rm mol} = 8 \pm 3$ 
M$_{\odot}$ yr$^{-1}$
(which includes the helium contribution).

Fig.~\ref{fig:mol_MOR} shows
the radial profile of the mass outflow rate, 
i.e. the same calculation of 
Equation~\ref{eq:Mdot_out}
for every radial ring.
To account for different $r_{31}$ 
and $X_{\text{CO}}$,
we plotted errorbars in Fig.~\ref{fig:mol_MOR}
corresponding to the typical
ranges $r_{31} = 0.4 - 1$ and
$X_{\text{CO}} = (0.8 - 2) \times 10^{20}$
mol cm$^{-2}$ (K km s$^{-1}$)$^{-1}$.
We find a strong peak of $\dot{M}_{\rm out}^{\rm mol}$
at the inner radius of the molecular ring
($R \sim 85$ pc),
which is outflowing (while rotating) up to
$\dot{M}_{out,max}^{\rm mol} = 28$ M$_{\odot}$ yr$^{-1}$.
A second (minor) peak is visible around
250 pc ($\sim 2 \arcsec$), within which resides
half of the molecular mass, and which corresponds
to a small $v_{\rm out}$ peak
(bottom panel of Fig.~\ref{fig:barolo_radial}).

\begin{figure*}
\centering
\includegraphics[width=0.98\textwidth]{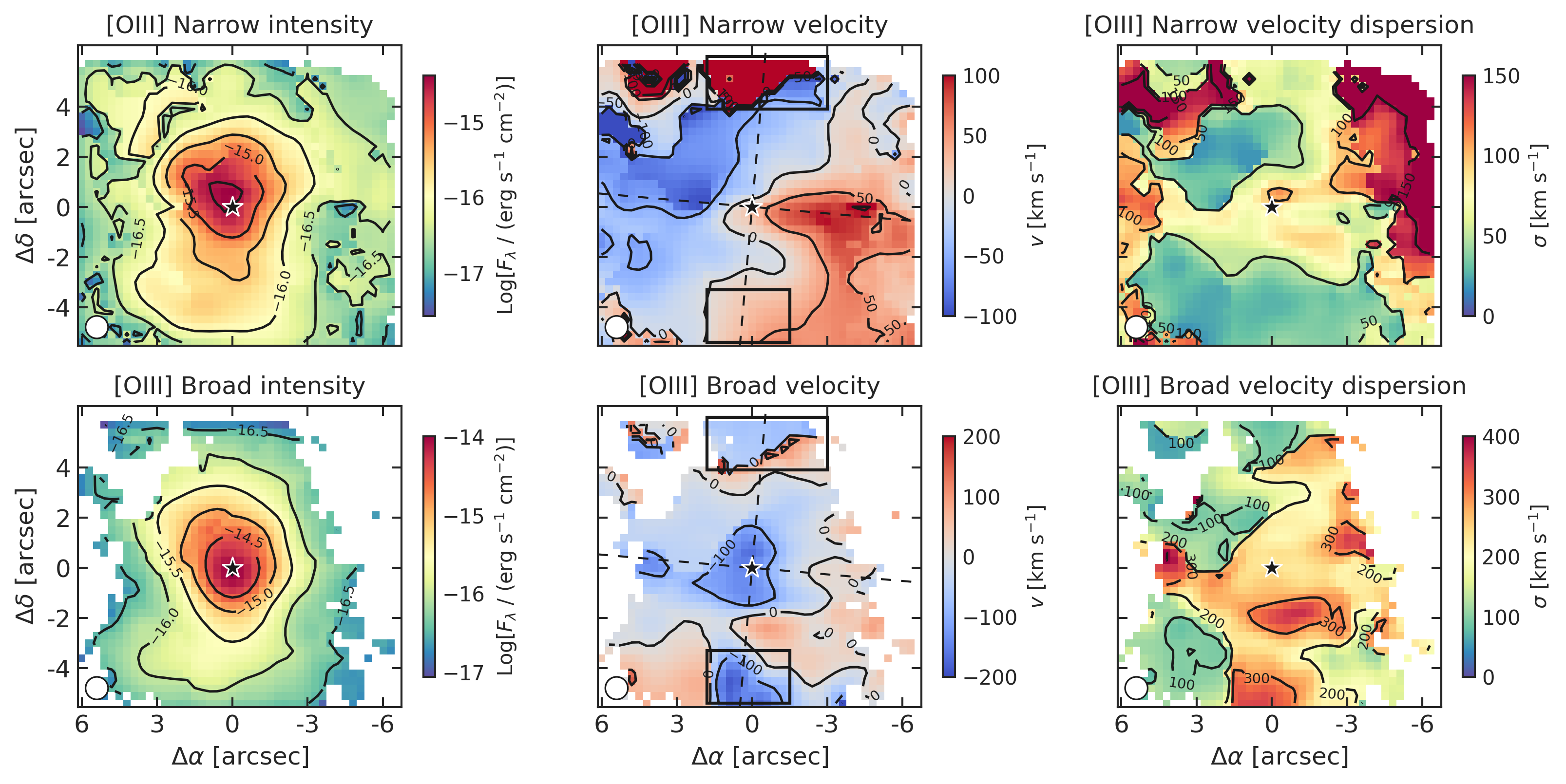}
\caption{[OIII] double Gaussian fit 
made with ALUCINE. Top and bottom rows
are for the narrow and broad component,
respectively. From left to right,
the three columns show the intensity,
velocity, and velocity dispersion
of both components.
The AGN position is marked with a 
black star symbol, and distances are measured
from it.
The white circle in the bottom left
of each panel is the MEGARA seeing conditions.
The velocity panels (central column)
show the PA $= 265^{\circ}$ and $355^{\circ}$
dashed black lines, and two 
black rectangles that
highlight the northern and southern edges
of the velocity field.}
\label{fig:OIII_maps}
\end{figure*}
The average value of
$\dot{M}_{\rm out}^{\rm mol} = 8 \pm 3$ M$_{\odot}$ yr$^{-1}$
is similar to those of other local Seyferts,
which range from  $\sim 1$ M$_{\odot}$ yr$^{-1}$
to a few tens of M$_{\odot}$ yr$^{-1}$ 
\citep{combes13, garciaburillo14, morganti15,
alonsoherrero19, dominguezfernandez20,
garciabernete21, alonsoherrero23}.



\section{Modelling the ionised gas kinematics}
\label{sec:ion_model}

\subsection{Gaussian decomposition}
\label{sec:alucine}

In the inset of
Fig.~\ref{fig:oiii_co_hst}
we presented the single Gaussian
fit of the [OIII] line.
It is evident that a single Gaussian
cannot accurately reproduce the complex shape
of the line profile. Consequently,
we decided to fit the observed lines
(listed in Fig.~\ref{fig:oiii_co_hst})
with two Gaussians, spaxel by spaxel.
The ALUCINE code determines,
based on the AoN $>3$ cut, whether
one or two Gaussians are necessary
for each spaxel.
As input parameters, 
ALUCINE needs also the wavelength 
range for subtracting the 
continuum, and a systemic velocity.
At first we set 
$v_{sys, \text{CO}}=1872$ km s$^{-1}$
as the CO($3-2$), but we achieved 
better results by setting 
$v_{sys, \text{[OIII]}}=1893$ km s$^{-1}$.
It is worth noting that this 21 km s$^{-1}$ 
difference is only 1.5 times the MEGARA
spectral step ($\sim 14$ km s$^{-1}$).
A detailed comparison of $v_{\rm sys}$
values from the literature 
is available in \cite{davies20}
and in Section~\ref{sec:rotation_only}.

We name the two Gaussians 'narrow' and 'broad'
component, where their width is the
discriminant factor.
We focus mainly on the [OIII] line
in the analysis, since it shows the
highest signal (Fig.~\ref{fig:oiii_co_hst})
and is the one usually studied for
AGN ionised winds
\citep{weedman70, heckman81, 
veilleux91, crenshaw00, harrison14}.
The results for the [OIII] line are
in Fig.~\ref{fig:OIII_maps}.
The top panels show the narrow component:
from the velocity map we can identify
a rotation pattern, with velocities
up to $-100$ and $100$ km s$^{-1}$,
oriented roughly in the same way
of the CO disc (Fig.~\ref{fig:ngc5506}).
The external parts of the narrow component
can be hardly associated to rotation however:
at the northern end of the FoV the gas
reaches $340$ km s$^{-1}$ (with relatively low
dispersions around 60 km s$^{-1}$),
while at NE and NW there are areas with
very high dispersion (up to $\sim 200$ km s$^{-1}$).
It could be that these extreme northern regions
trace the external part of the ionised outflow.

The broad component of [OIII] 
(bottom panels of Fig.~\ref{fig:OIII_maps})
contains fewer pixels than the narrow one,
since for some spaxels
a single Gaussian component was sufficient 
to obtain a proper modelling
(or the broad Gaussian had AoN$<3$).
The velocity map of this component displays
a central blueshifted region
(up to $-170$ km s$^{-1}$), and some
positive and negative velocities all over the FoV.
The velocity dispersion map reaches higher
values than the narrow one
(up to $400$ km s$^{-1}$).

\begin{figure}
\centering
\includegraphics[width=0.49\textwidth]{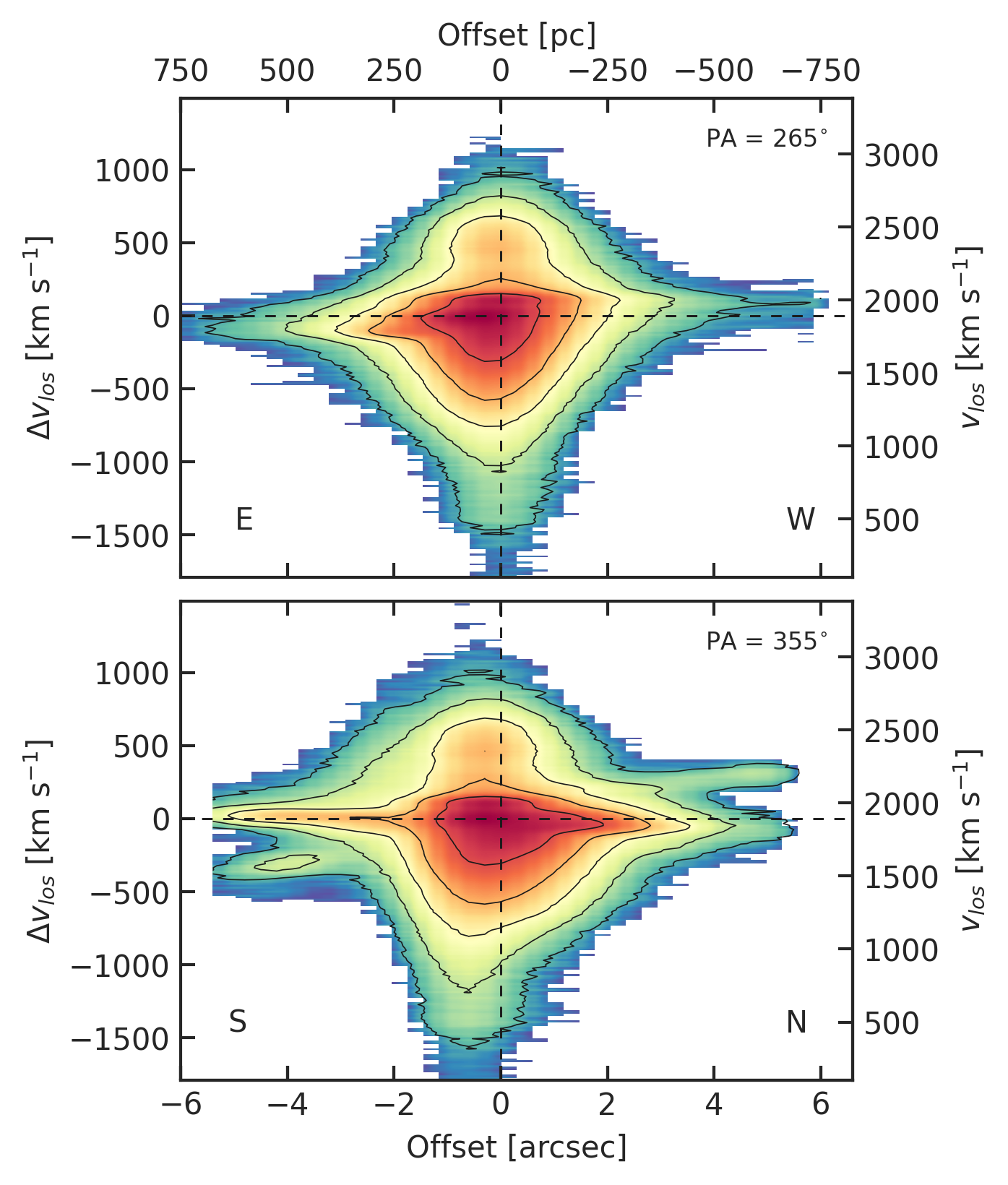}
\caption{PV diagrams of the 
observed [OIII] ($\lambda_e = 5007$\AA) line, 
clipped at $3 \sigma$, 
along major (\textit{top panel}) 
and minor (\textit{bottom panel}) kinematic axes,
with PA $= 265^{\circ}$ and $355^{\circ}$, 
respectively.
Contours are at $[10, 30, 100, 300, 1000] \sigma$.
The vertical dashed line is the AGN position,
and the horizontal dashed line is the systemic
velocity $v_{\text{sys}}^{\text{ion}} = 1893$ km s$^{-1}$.
The approximate eastern, western, southern, and northern
directions are marked in the panels.
At $\Delta v_{los} < -1500$ km s$^{-1}$
contamination with the secondary [OIII]
line ($\lambda_e = 4959$\AA) is probable.
}
\label{fig:oiii_pv}
\end{figure}

\begin{figure}
\centering
\includegraphics[width=0.49\textwidth]{
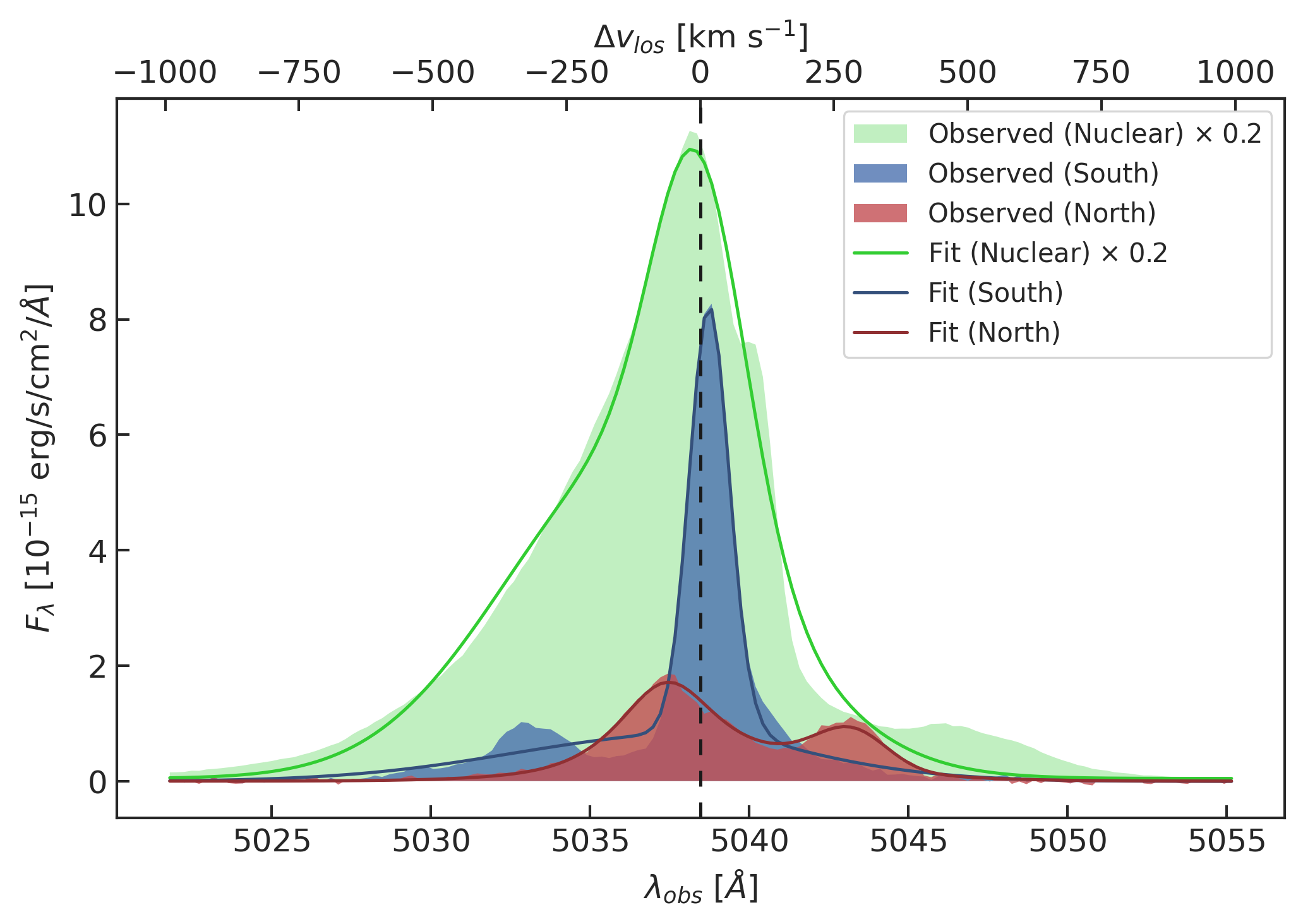}
\caption{Observed spectra 
(continuum-subtracted) of the [OIII]$\lambda 5007$
line at three locations along the kinematical minor
axis (PA $= 355^{\circ}$).
In green, blue, and red shadings, the spectra
extracted within the nuclear region
(black square in Fig.~\ref{fig:oiii_co_hst}),
the southern region, and the northern region
(black rectangles in Fig.~\ref{fig:OIII_maps}),
respectively.
With the same colours, the lines are the
ALUCINE two-component fits for the same regions.
Note that for the nuclear and southern regions 
a third component would be needed to fit 
the residual blueshifted and redshifted components, 
respectively.
The nuclear spectrum is multiplied by a 
0.2 factor for a better comparison.
The vertical dashed line is the redshifted
(with $v_{\rm sys}^{\rm ion} = 1893$ km s$^{-1}$)
[OIII] line.
}
\label{fig:edgespectra}
\end{figure}

\begin{figure*}
\centering
\includegraphics[width=0.98\textwidth]{
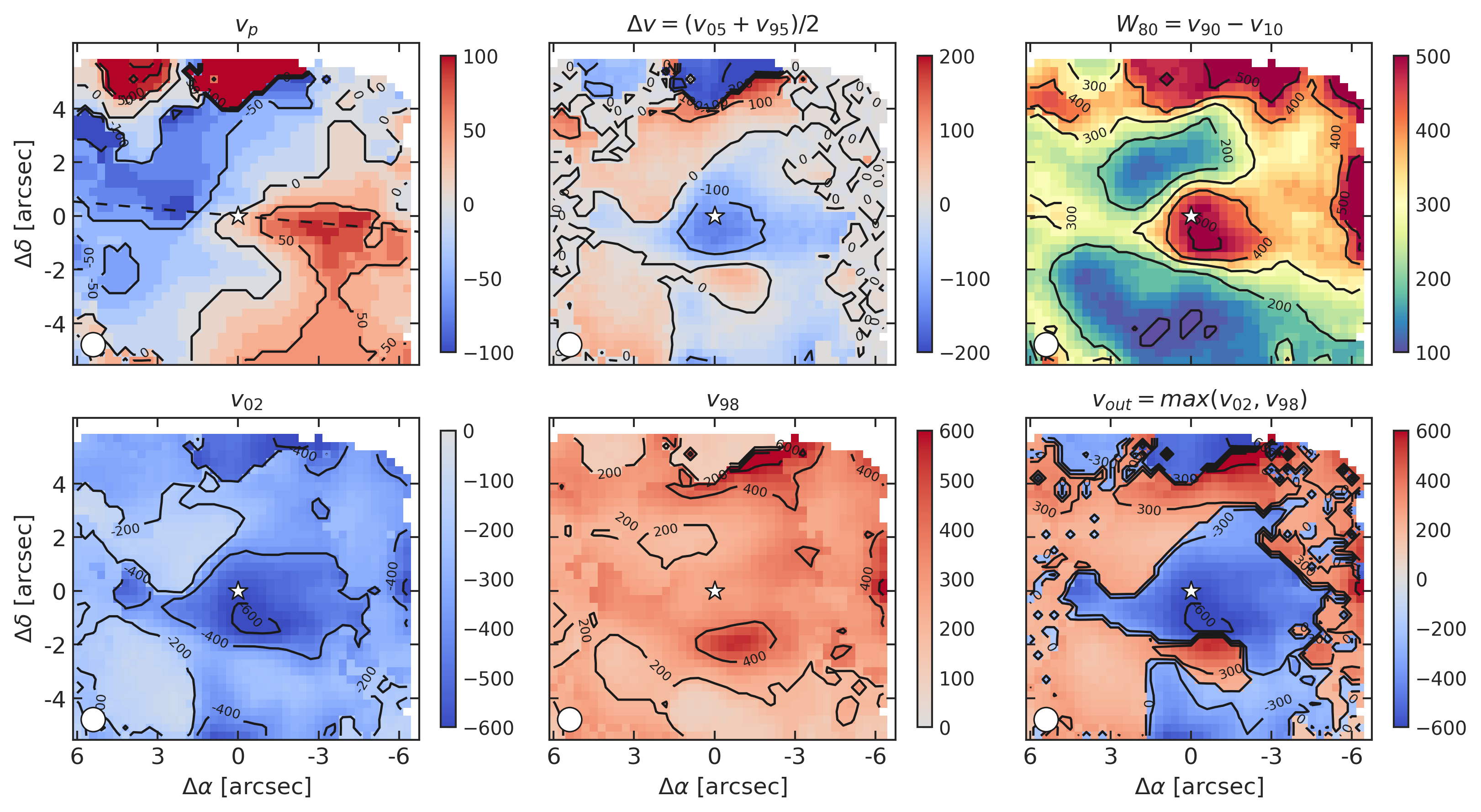}
\caption{Non-parametric velocity components for the 
[OIII] line. The top panels show, from left to right,
the peak velocity $v_p$, 
the broad velocity $\Delta v$,
and the 80\% width $W_{80}$.
The bottom panels show, from left to right,
the velocity at the 2\textsuperscript{nd} 
flux percentile ($v_{02}$), at the 
98\textsuperscript{th} ($v_{98}$),
and the positive or negative velocities
that have the maximum absolute value
between these two (for every spaxel),
which is our estimate for the outflow
velocity $v_{\rm out}$.
The white star symbol marks the AGN position,
and the dashed black line in the top-left panel
is the kinematic major axis (PA $= 265^{\circ}$).
The white circle in the bottom left
of each panel is the MEGARA seeing conditions.
Velocities in all the panels are
in km s$^{-1}$.
}
\label{fig:OIII_velocities}
\end{figure*}

The Gaussian decomposition made with
ALUCINE is able to separate the [OIII]
rotation from the outflow component
(except for the extreme northern regions
at high velocity or velocity dispersion).
The same applies for the other emission lines
in the MEGARA spectrum, for which the 
results are shown in Appendix~\ref{sec:more_megara}.
Compared to [OIII], 
the narrow component intensity maps of 
[NII], [SII], and [OI]
are more consistent with an ionised rotating disc
(aligned with the HST and ALMA discs,
i.e. with PA $\sim 265^{\circ}$),
while the broad component is more
elongated on the north-south direction, 
(as [OIII], H$\alpha$ and H$\beta$).
This north-south elongation is especially evident
in the velocity dispersion of the broad 
components of [NII] and [SII]
(Figs.~\ref{fig:nii} and~\ref{fig:sii}).

We also show, in Appendix~\ref{sec:bpt},
the Baldwin, Phillips, Telervich (BPT) diagrams
\citep{baldwin81, veilleux87, kewley01, kauffmann03}
made with the same fitted lines.
From such diagrams 
(Figs.~\ref{fig:bpt1}~-~\ref{fig:bpt3})
we can conclude that most of the observed
central emission is due to the AGN activity
rather than star formation, both for
the narrow and broad components.

We explain how we used this decomposition to
calculate the [OIII] mass of the broad component
in Section~\ref{sec:ion_mass}.
However, the velocities obtained with ALUCINE
represent mean velocities within each spaxel.
To explore the full range of velocities
of the ionised gas, we produced 
[OIII] PV diagrams (Fig.~\ref{fig:oiii_pv}),
along the same PAs as the CO emission
(Fig.~\ref{fig:barolo_PV}).
In both major- and minor-axis PV diagrams, 
the ionised gas exhibits velocities 
exceeding 1000 km s$^{-1}$,
both redshifted and blueshifted.
The most extreme blueshifted velocities
($\Delta v_{los} < -1500$ km s$^{-1}$)
are possibly contaminated with emission 
from the secondary [OIII] doublet line
($\lambda_e=4959$ \AA).
The major-axis PV diagram
(Fig.~\ref{fig:oiii_pv}, top panel)
displays a rotation curve between
$-100$ and $100$ km s$^{-1}$. However,
most of the emission, along both PAs, 
appears to be dominated by the outflowing gas.
Along the minor axis
(bottom panel of Fig.~\ref{fig:oiii_pv}),
there is also an observable X shape, elongated
at large radii, likely due to the
northern red and southern blue regions
in the central panels of Fig.~\ref{fig:OIII_maps}
(within the black rectangles).
These regions may represent the locations 
where the outflow is emerging from the nuclear zone.

To have a better understanding of the observed
PV diagrams, we show, in Fig.~\ref{fig:edgespectra},
the [OIII] line profile
at three locations along the minor axis.
The northern (in red) and
southern (in blue) regions exhibit
two distinct components: one travelling 
at approximately the systemic velocity,
and the other outflowing at $\sim \pm 300$
km s$^{-1}$.
The southern blue emission also displays
a stronger centred emission,
which is evident in the PV diagram
(Fig.~\ref{fig:oiii_pv}, bottom panel)
at $\gtrsim 4 \arcsec$ south
(while its northern counterpart is fainter).
In the Gaussian decomposition 
(Fig.~\ref{fig:OIII_maps}), the northern
region with high redshifted velocities
likely belongs to the outflowing and broad
component rather than to
the rotational/narrow one.
However, the associated flux (and consequently,
the ionised mass within it) 
is negligible in our analysis
(see Section~\ref{sec:ion_mass}).

We note here that,
even if several works allow three or more
Gaussians to fit the emission of [OIII]
in AGN with possible outflows
\citep[e.g][]{harrison14, 
dallagnoldeoliveira21, speranza22, hermosamunoz23},
we limited our analysis to two components
to have a simpler interpretation of them:
we associated the narrow component to the ionised
gas rotation, and the broad one to the outflow.
Adding more Gaussians to the ALUCINE
fit would result in higher velocities and velocity
dispersions for the broader components
(but we refer to the next section for a better
characterisation of the outflow velocities).
However, such broader components would add
a small contribution to the modelled flux
(see Fig.~\ref{fig:residuals}), 
hence to the outflow mass.

\subsection{Non-parametric [OIII] velocities}
\label{sec:nonparam}

In the previous section we saw that 
the ionised gas (traced by [OIII]) shows 
very high velocities probably due to an AGN wind.
However, due to the complex line profiles, 
it is hard to see the different velocities
from the Gaussian decomposition
of Fig.~\ref{fig:OIII_maps}.
In this section, we make use of a
non-parametric method to 
measure the outflow velocities.

We followed the method described by \cite{harrison14}
to spatially resolve the velocities of
the [OIII] emission line.
This method uses the [OIII] line produced 
as the sum of the two fitted Gaussians
(Section~\ref{sec:alucine}).
For every spaxel we calculate 
the velocities corresponding
to different percentiles
to the flux contained in the modelled line profile,
namely the velocities at the 
2\textsuperscript{nd}, 5\textsuperscript{th}, 
10\textsuperscript{th}, 90\textsuperscript{th}, 
95\textsuperscript{th}, and 98\textsuperscript{th}
percentiles, respectively called 
$v_{02}$, $v_{05}$, $v_{10}$, $v_{90}$, $v_{95}$, 
and $v_{98}$. We also calculate, for every spaxel,
the velocity of the emission line peak $v_p$.

The results are illustrated in
Fig.~\ref{fig:OIII_velocities}.
The top-left panel shows $v_p$, which is
similar to the velocity of the narrow
component of the Gaussian decomposition
(cf. Fig.~\ref{fig:OIII_maps}).
It has been shown that $v_p$ traces
the ionised gas rotation
\citep{rupke13, harrison14}.
In the case of NGC~5506, $v_p$ is 
similar to the mean-velocity field 
of the molecular gas, whose 
kinematic PA $=265^{\circ}$
is plotted with a dashed black line.
The region $\sim 5 \arcsec$ 
N from the AGN, redshifted at 
$\sim 300$ km s$^{-1}$, is not following
the rotation pattern.

The top-central panel of 
Fig.~\ref{fig:OIII_velocities}
shows the $\Delta v = (v_{05} + v_{95})/2$ map.
This is very similar to the velocity
map of the broad component
modelled by ALUCINE (cf. Fig.~\ref{fig:OIII_maps}),
and so represents its velocity offset.
There are differences between the two maps though,
especially around $\sim 3 \arcsec$ 
NE from the nucleus,
where the $\Delta v$ plot shows redshifted
velocities around $50$ km s$^{-1}$.
This may be an outflow feature lost
in the ALUCINE decomposition map.

The top-right panel of 
Fig.~\ref{fig:OIII_velocities}
is the $W_{80} = v_{90} - v_{10}$ width,
which represents the width containing 80 percent 
of the [OIII] emitted flux.
In the case of a single modelled Gaussian,
this would correspond approximately to the FWHM.
In our decomposition, 
$W_{80}$ is, in a way, a combination
of the velocity dispersions of the two components
of Fig.~\ref{fig:OIII_maps}.
However, $W_{80}$ exhibits larger values across
the entire FoV, particularly at the
AGN position (reaching up to 500 km s$^{-1}$)
and in the $\sim 5 \arcsec$ N region
(with an average $<W_{80, N}> \sim 500$ km s$^{-1}$).
The maximum value observed is 
$W_{80, max} = 826$ km s$^{-1}$
located $\sim 6 \arcsec$ W.

The bottom three panels of 
Fig.~\ref{fig:OIII_velocities}
show the velocities found in the 
2\textsuperscript{nd} and 98\textsuperscript{th}
percentiles of the flux
(the third panel is showing the 
positive or negative velocities
that have the maximum absolute value
among the two).
These correspond to the projected maximum values
for the outflow velocities
\citep[as in][]{rupke13, harrison14, davies20}.
We find the highest blueshifted
velocities around the AGN
($-565$ km s$^{-1}$ at the AGN position,
$-620$ km s$^{-1}$ at $\sim 1 \arcsec$ S-SW)
and in the $\sim 5 \arcsec$ N region
(up to $-702$ km s$^{-1}$).
The highest redshifted values are found
at $\sim 1.6 \arcsec$ (200 pc) S-SW from the AGN
(up to $551$ km s$^{-1}$), and
very close to the $\sim 5 \arcsec$ N region
(up to $689$ km s$^{-1}$).

The prevalence of blueshifted velocities
in the nuclear region was previously identified
in the X-shooter spectrum,
extracted with a FoV of $1.8 \times 1.8$ arcsec$^2$
\citep{davies20}, which is also visible
in Fig.~\ref{fig:edgespectra}
(green profile).
With the MEGARA data,
we observe high-velocity components,
not associated with rotation,
both blueshifted and redshifted,
in all panels of Fig.~\ref{fig:OIII_velocities}
and in nearly every direction,
particularly in the central 
$4 \times 4$ arcsec$^2$,
as evident in Fig.~\ref{fig:oiii_pv}.
This may be due to a wide bicone aperture,
where any given line of sight intersects both
approaching and receding clouds of gas 
simultaneously.

\begin{figure}
\centering
\includegraphics[width=0.49\textwidth]{
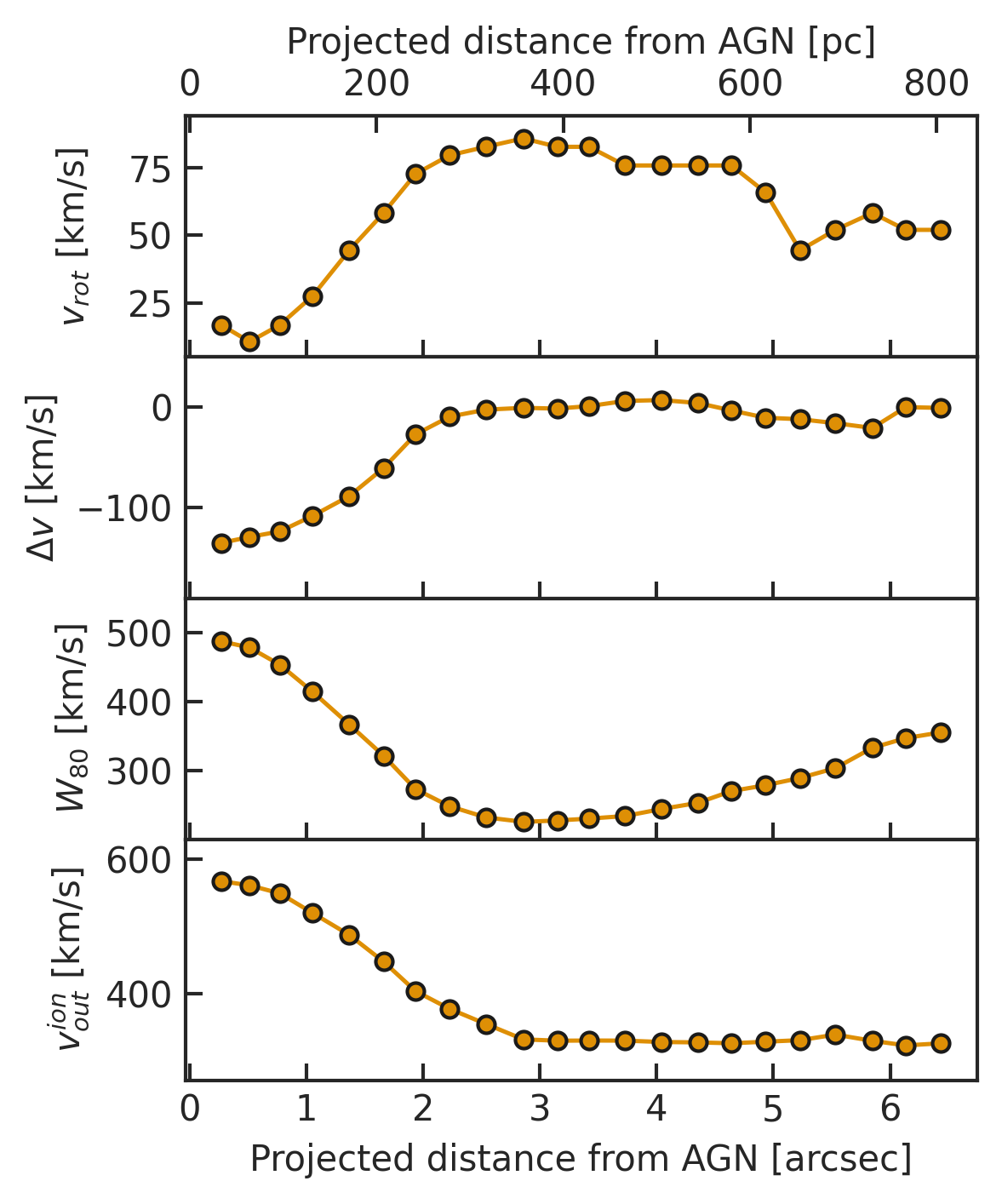}
\caption{Radial profiles for the different 
mean velocities of the [OIII] line,
all as a function
of the projected distance from the AGN.
Panels show, from top to bottom:
the rotational velocity $v_{\rm rot}$
along the kinematic axis,
the broad velocity $\Delta v$,
the $80 \%$ width $W_{80}$, and our estimate
for the mean outflow velocity of the
ionised gas $v_{\rm out}^{\rm ion}$.
}
\label{fig:OIII_radial}
\end{figure}

We isolate the [OIII] rotation velocity ($v_{\rm rot}$)
by taking the median absolute value of $v_p$
along the PA $=265^{\circ}$ line
(the dashed line in the $v_p$ panel
of Fig.~\ref{fig:OIII_velocities})
with a width of 4 pixels
(corresponding to $1.2 \arcsec \sim 150$ pc).
In Fig.~\ref{fig:OIII_radial}, we plot the
mean radial profiles of $v_{\rm rot}$, $\Delta v$,
$W_{80}$, and $v_{\rm out}$.
The [OIII] rotational velocity $v_{\rm rot}$ 
flattens out at 83 km s$^{-1}$ around
$\sim 320$ pc from the centre
(Fig.~\ref{fig:OIII_radial}, top panel),
whereas the CO($3-2$) flattens out at
$193$ km s$^{-1}$ around $r=440$ pc
(Fig.~\ref{fig:barolo_radial}, top panel).
We point out that [OIII] is not the best tracer
for the ionised disc rotation, and 
in fact it is the slowest rotator
among the MEGARA lines: 
H$\alpha$ flattens at 120 km s$^{-1}$,
H$\beta$ at 113 km s$^{-1}$,
[NII] at 120 km s$^{-1}$,
[SII] at 118 km s$^{-1}$,
and [OI] at 110 km s$^{-1}$
(see Appendix~\ref{sec:more_megara_velocities}
for the mean velocity radial profiles
of all the MEGARA lines).

Interestingly,
the ionised gas seems to be rotating
at $60 \%$ the velocity of the molecular gas.
\cite{davis13} found that, in CO-rich
ATLAS\textsuperscript{3D} galaxies
\citep{cappellari11},
the difference between molecular
and ionised rotation velocities
was larger for [OIII]-bright galaxies
(up to a $\Delta v_{\rm rot} \sim 80$ km s$^{-1}$),
due to the different ionisation sources:
a bright [OIII] emission (with respect to
H$\beta$) traces a dynamically hotter 
component of ionised gas than HII
regions embedded in the cold star-forming disc.
Also \cite{levy18} and \cite{su22}
found the ionised gas to rotate slower
than the molecular gas in EDGE-CALIFA 
\citep{bolatto17} and ALMaQUEST
\citep{lin19} galaxies,
but with a smaller difference of
$\sim 25$ km s$^{-1}$.

The radial profiles of $\Delta v$,
$W_{80}$, and $v_{\rm out}$
have similar shapes, with a smooth
decrease of absolute velocities from the centre
up to a radial distance of $\sim 400$ pc.
For these three quantities, the distance
is the projected distance along every direction,
so one has to be careful when comparing them
to $v_{\rm rot}$ or to the molecular radial profiles
of Figs.~\ref{fig:barolo_radial}
and \ref{fig:mol_MOR}.
We used the $v_{\rm out}$ radial
profile of Fig.~~\ref{fig:OIII_radial}
to calculate the other ionised outflow properties,
as the mass outflow rate
(see Section~\ref{sec:ion_outflowrate}).


\begin{figure*}
\centering
\includegraphics[width=0.98\textwidth]{
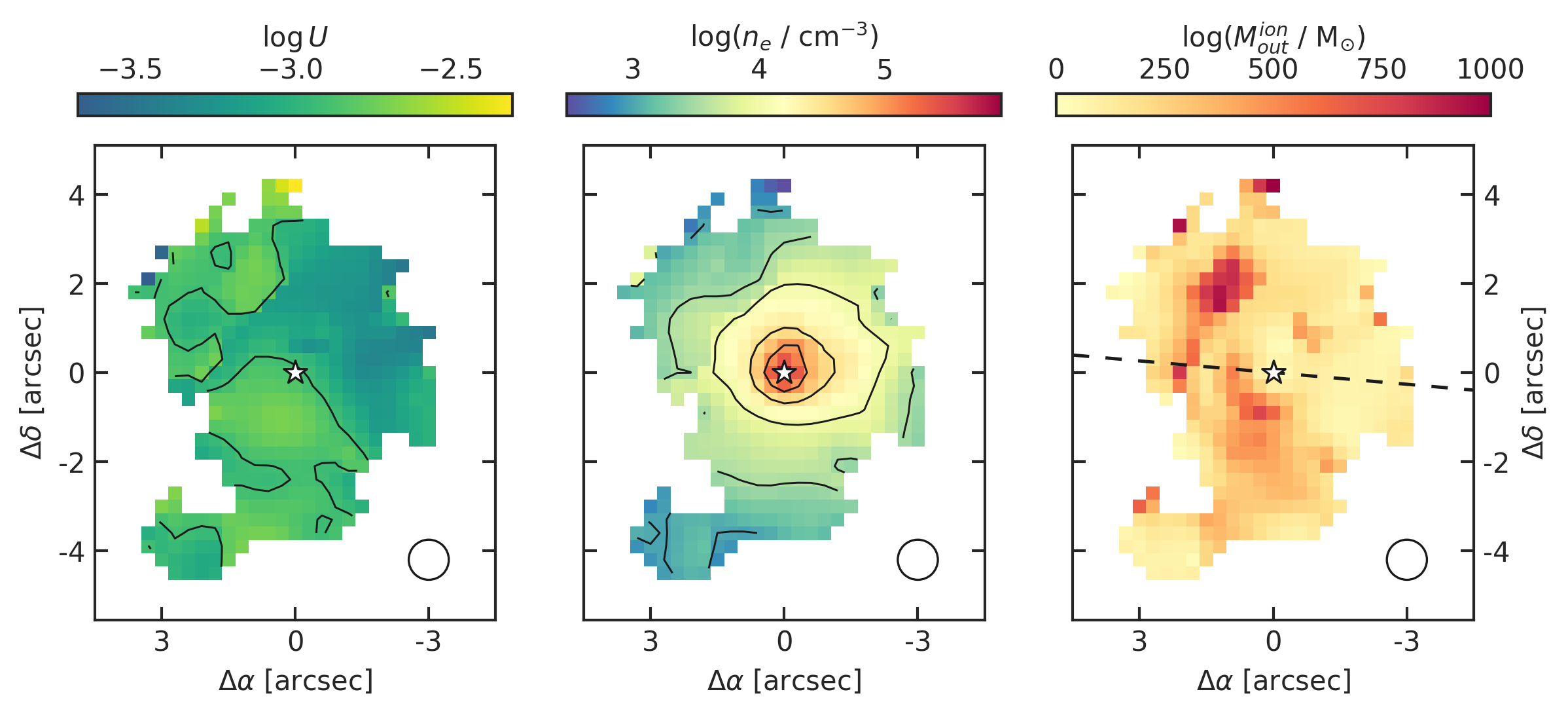}
\caption{From left to right: outflowing [OIII] 
ionisation parameter $\log U$, electron density
$n_e$, and mass $M_{\rm out}^{\rm ion}$.
Contours are at $\log U = -2.9$ (i.e. its median value),
$\log (n_e \ \text{cm}^{-3}) = (3, 3.5, 4, 4.5, 5)$.
The white star symbol marks the AGN position,
and the dashed black line in the top-right panel
is the kinematic major axis (PA $= 265^{\circ}$).
The white circle in the bottom right
of each panel is the MEGARA seeing conditions.
}
\label{fig:ion_3maps}
\end{figure*}

\subsection{The [OIII] outflow mass}
\label{sec:ion_mass}

In this section we calculate the 
electron density
and mass of the ionised outflow.
To do so, we make use of the ALUCINE decomposition
(Section~\ref{sec:alucine}), and we consider
the flux of the broad component as the outflow
(whereas the narrow component is associated to
the ordered gas rotation).
We subsequently explain how we calculated
the outflow properties
using the [OIII] emission line 
(Fig.~\ref{fig:OIII_maps}),
but we also exploited
the modelled broad components
of H$\alpha$, H$\beta$, and [NII]
(Figs.~\ref{fig:halpha}-\ref{fig:nii}).

One of the challenges in the estimation
of the ionised outflow mass is to properly
calculate the gas volume density $n$, 
usually expressed as the electron density $n_e$
(where for ionised gas we expect $n_e \sim n$).
Many studies assume constant fiducial values
for $n_e$ \citep[e.g.][]{harrison14, fiore17}
for all the spaxels 
(or for a whole sample of galaxies).
The most commonly used method to estimate $n_e$
pixel-by-pixel is based on the [SII] doublet ratio
\citep{osterbrock06}.
However, this method has known biases,
one of which is that the doublet ratio 
saturates above $10^4$ cm$^{-3}$.
We refer the interested reader to
\cite{davies20} for a structured discussion
on this topic and for a comparison
between different methods
to estimate $n_e$.

We follow \cite{baron19}, \cite{davies20},
and \cite{peraltadearriba23},
in estimating the ionised gas density
from the ionisation parameter $\log U$,
defined as the number of ionising photons per atom,
$U = Q_{H} / (4 \pi r^2 n_H c)$,
where $Q_{H}$ is the rate of 
hydrogen-ionising photons (in s$^{-1}$ units),
$r$ is the distance from the ionising source,
$n_H \sim n_e$ is the hydrogen density,
and $c$ is the speed of light.
Since $Q_{H}$ can be estimated from the
AGN bolometric luminosity \citep{baron19},
we can find $n_e$ given the ionisation parameter.

Since the [OIII]/H$\beta$ and [NII]/H$\alpha$
line ratios are widely used in AGN studies
\citep{veilleux87, kewley01},
and both depend on $\log U$,
\cite{baron19}, by exploiting a sample of 234
type II AGN with outflow signatures,
empirically determined (with a scatter of 0.1 dex)
the following expression:
\begin{equation} 
\label{eq:logU}
\begin{split}
\log U = & -3.766 + 0.191 \log \left( 
\frac{[OIII]}{H\beta} \right)
+ 0.778 \log^2 \left( \frac{[OIII]}{H\beta} 
\right) \\
& - 0.251 \log \left( \frac{[NII]}{H\alpha} \right)
+ 0.342 \log^2 \left( \frac{[NII]}{H\alpha} \right)
\; \; \; .
\end{split}
\end{equation}

The resulting $\log U$ map for the broad component
of [OIII] is in the left panel of 
Fig.~\ref{fig:ion_3maps}.
There are fewer pixels
than the broad [OIII] map
(Fig.~\ref{fig:OIII_maps}, bottom panels),
since we had to use also the broad
H$\alpha$, H$\beta$, and [NII] maps,
and only the pixels featured in all four maps
are left (with H$\beta$ being the most limiting one).
We recover a median value of
$\log U = -2.9$, 
in agreement with the 
integrated value of $-2.87 \pm 0.12$
found by \cite{davies20} within the X-shooter
FoV ($1.8 \times 1.8$ arcsec$^2$).

From the $\log U$ definition, we follow
\cite{baron19} and calculate the electron density as
\begin{equation}
\label{eq:ne_from_U}
n_e \approx 3.2 \left( 
\frac{L_{\text{bol}}}{10^{45} 
\text{ erg s}^{-1}} \right)
\left( \frac{r}{1 \text{ kpc}} \right)^{-2}
\left( \frac{1}{U} \right) \; \; \text{cm}^{-3}
\end{equation}
where we used the
$\log (L_{\text{bol}} / \text{erg s}^{-1})
= 44.1 \pm 0.09$
obtained by \cite{davies20}
from the X-ray luminosity given by \cite{ricci17}.
For the central pixel we set $r$ equal to
half the pixel size.

We show the spatially resolved $n_e$ map
in the central panel of Fig.~\ref{fig:ion_3maps}.
We find $n_e$ to decrease at increasing distance
from the centre, as found by other works
on local AGN \citep[e.g.][]{
freitas18, shimizu19, davies20, peraltadearriba23}.
The maximum 
$n_{e, \text{max}} = 8.5 \times 10^5$ cm$^{-3}$
is exactly at the AGN position.
To compare our values 
with the results of \cite{davies20} for NGC~5506,
we calculate the median $n_e$ at the edge of 
a $1.8 \times 1.8$ arcsec$^2$ FoV
(the black square in Fig.~\ref{fig:oiii_co_hst}), 
finding $\log (n_e / \text{cm}^{-3}) = 3.95$, 
which is very close to their integrated value of
$\log (n_e / \text{cm}^{-3}) = 4.03 \pm 0.14$.

Before calculating the ionised outflow mass
from the broad [OIII] luminosity, we have to
correct it for the extinction.
To do so, we assume an intrinsic ratio
$H \alpha / H \beta = 3.1$, and we use the
\cite{cardelli89} extinction law
($R_V = 3.1$).
We find the [OIII] outflow area
(i.e. the same of the three panels
in Fig.~\ref{fig:ion_3maps})
to have a median $A_V = 1.9$ mag and
an extinction-corrected total luminosity
$L_{\text{broad [OIII]}} = 10^{41.6}$ erg~s$^{-1}$. 
If we limit the FoV to $1.8 \times 1.8$ arcsec$^2$
we find $10^{41.3}$ erg~s$^{-1}$,
in excellent agreement with the \cite{davies20}
value of $10^{41.2}$ erg~s$^{-1}$.

Finally, the ionised outflow mass $M_{\rm out}^{\rm ion}$
is given by \citep[see][]{rose18, baron19}:
\begin{equation}
\label{eq:Mout}
M_{\rm out}^{\rm ion} = 
\frac{\mu m_H L_{\text{broad [OIII]}}}
{\gamma_{\text{[OIII]}} n_e} \; \; ,
\end{equation}
where $\mu=1.4$ is the mean molecular weight,
$m_H$ is the hydrogen mass,
$L_{\text{broad [OIII]}}$ is the 
extinction-corrected broad [OIII] luminosity,
$n_e$ is the outflowing gas electron density,
and $\gamma_{\text{[OIII]}}$ is the effective
line emissivity, which depends
on the ionisation parameter
\citep[see Equations 5 and 6 in][]{baron19}.
We interpolated the values listed in 
\cite{baron19}, Table 2, to calculate 
$\gamma_{\text{[OIII]}}$ for every spaxel.

The resulting spatial distribution of
the outflowing [OIII] mass is presented in
the right panel of Fig.~\ref{fig:ion_3maps}.
We calculated a total ionised outflowing mass of
$M_{\rm out}^{\rm ion} = 9.8 \times 10^4$ M$_{\odot}$.
In comparison, the mass reported by \cite{davies20} is
$3.2 \times 10^4$ M$_{\odot}$.
The discrepancy arises because the MEGARA
aperture is significantly larger than the X-shooter one 
(as indicated by the white and black squares 
in Fig.~\ref{fig:oiii_co_hst}).
Additionally, \cite{davies20} used
a single value for all the quantities involved
in Equation~\ref{eq:Mout}, whereas we
considered spatial variations, resulting in a more
dispersed distribution of $M_{\rm out}^{\rm ion}$.


\begin{figure}
\centering
\includegraphics[width=0.49\textwidth]{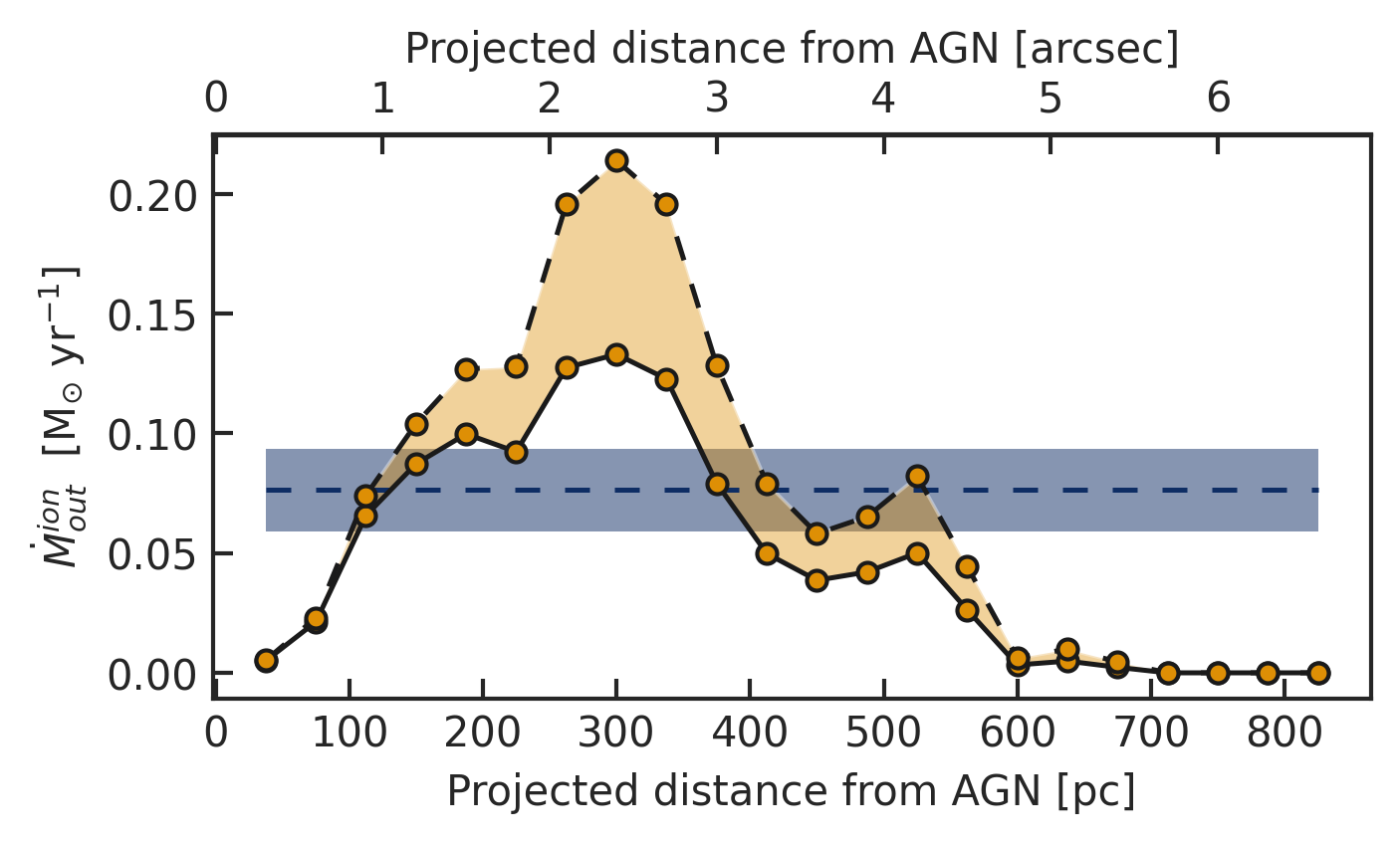}
\caption{Radial profile of the ionised mass outflow rate
$\dot{M}_{\rm out}^{\rm ion}$ as a function of 
the average projected distance
from the AGN. Solid and dashed lines are estimates of
$\dot{M}_{\rm out}^{\rm ion}$ by using the average and the
maximum $v_{\rm out}^{\rm ion}$ at every radius.
The blue dashed line is the integrated mass
outflow rate, $0.076 \pm 0.017$ M$_{\odot}$ yr$^{-1}$,
with the blue shading representing its uncertainty.}
\label{fig:ion_MOR_radial}
\end{figure}

\begin{figure}
\centering
\includegraphics[width=0.49\textwidth]{
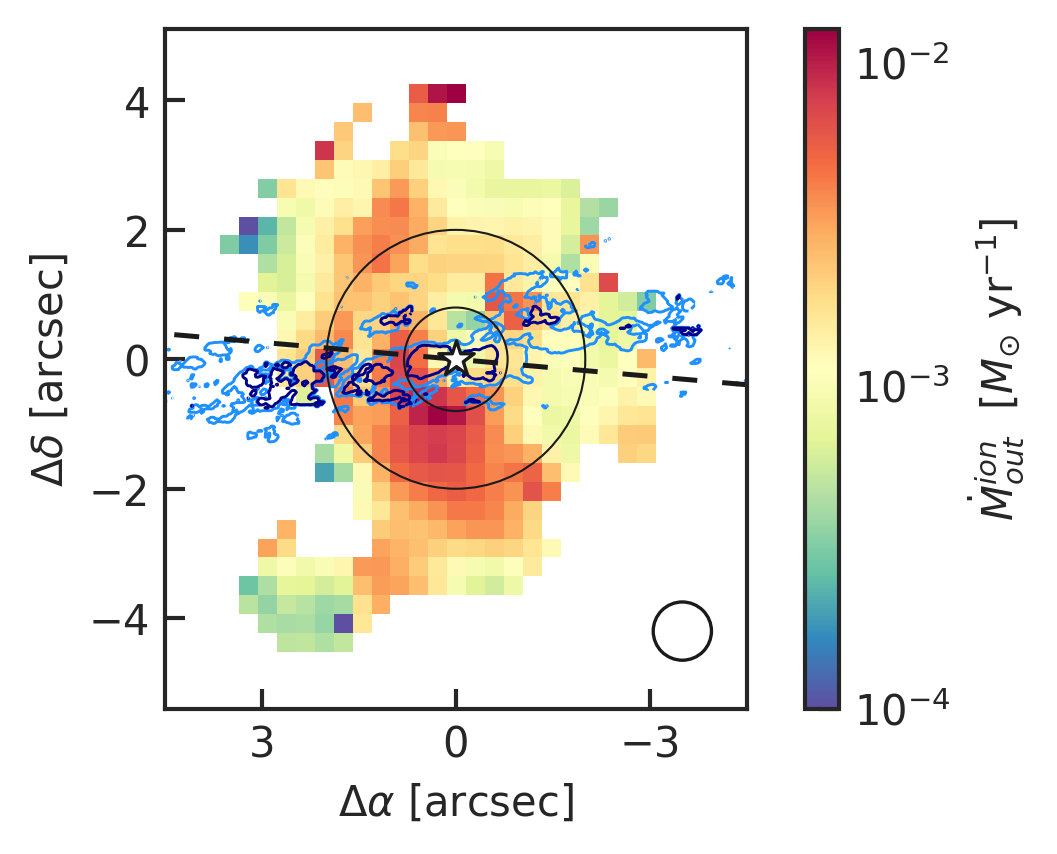}
\caption{Coloured map 
of the ionised mass outflow rate,
with contours of observed 
CO($3-2$) velocity dispersion
(as Fig.~\ref{fig:ngc5506}, bottom panel), at
30 and 50 km s$^{-1}$ in light and dark blue, 
respectively.
The two black circles have a radius of 100 and 250 pc 
(i.e. $\sim 0.8$ and $2$ arcsec, respectively)
from the white star symbol,
which marks the AGN position. 
The dashed line is the kinematic major axis.
The white circle in the bottom right 
is the MEGARA seeing conditions.
}
\label{fig:ion_MOR}
\end{figure}

\subsection{The ionised mass outflow rate}
\label{sec:ion_outflowrate}

We calculate the ionised mass outflow rate
following, as for the molecular gas,
Equation~\ref{eq:Mdot_out}
\citep[as in][]{rose18, baron19, davies20}.
The outflow velocity and mass have been
calculated following
Sections~\ref{sec:nonparam}
and~\ref{sec:ion_mass}.
If we take, as typical outflow radius,
$R_{out,95}^{\rm ion} = 525$ pc (i.e.
the one that contains $95 \%$ of $M_{\rm out}$),
we find $v_{\rm out}^{\rm ion} = 422 \pm 97$ km s$^{-1}$, 
from which we infer a ionised mass outflow rate of
$\dot{M}_{\rm out}^{\rm ion} = 0.076 \pm 0.017$ 
M$_{\odot}$ yr$^{-1}$,
where the uncertainty comes from
the standard deviation of the different
measured radial velocities.

This is significantly lower than the 
$0.21$ M$_{\odot}$ yr$^{-1}$ value
reported in \cite{davies20}.
A factor of $\sim 2$ discrepancy is due to the different
velocity (they measured 792 km s$^{-1}$).
Another difference is the outflow size,
that dilutes the averaged value
(their aperture radius was of 117 pc).
We can recover the \cite{davies20}
value if we plot the radial profile of $\dot{M}_{\rm out}$
(Fig.~\ref{fig:ion_MOR_radial})
using, for each radius, the maximum
outflow velocity available (dashed black line)
rather than the average one (solid black line).

The spatially resolved map of ionised mass 
outflow rate (Fig.~\ref{fig:ion_MOR})
reveals an excess of $\dot{M}_{\rm out}^{\rm ion}$, 
extending from $\sim 0.8 \arcsec$ 
up to $\sim 2.5 \arcsec$ 
south of the AGN.
This is the region where the most extreme
blueshifted velocities of $620$ km s$^{-1}$
reside (see Fig.~\ref{fig:OIII_velocities}).
It is also a region which exhibit some excess of $M_{\rm out}$
(see Fig.~\ref{fig:ion_3maps}, right panel),
hence the local high $\dot{M}_{\rm out}$.
Some minor $\dot{M}_{\rm out}$ clumps are visible 
at $\sim 1.5 \arcsec$ NW 
and $\sim 2.5 \arcsec$ NE from the AGN.
Interestingly, this NE clump
(which is very clear in the $M_{\rm out}$ map)
is located just after the separation
between blueshifted and redshifted velocities
(on the red side) in the bottom-left panel
of Fig.~\ref{fig:OIII_velocities}.
All together these clumps contribute to
the two main bumps in the $\dot{M}_{\rm out}$ radial
profile (Fig.~\ref{fig:ion_MOR_radial}).

The farthest (from the AGN) peak, 
at $\sim 4\arcsec \sim 500$ pc north,
visible in both Fig.~\ref{fig:ion_MOR_radial}
and~\ref{fig:ion_MOR},
is due to the pixels in the northern region
highlighted in the central panels
of Fig.~\ref{fig:OIII_maps},
and whose spectrum is plotted in red
in Fig.~\ref{fig:edgespectra}.
Most of this northern region
has been excluded from our analysis since
it is out of the $\log U$ map
(Fig.~\ref{fig:ion_3maps}, left panel)
and therefore of all the subsequent maps
(this is mainly due to the limited size of
the broad component of the H$\beta$ line,
see Fig.~\ref{fig:hbeta}), but probably
it is part of the ionised outflow.
Interestingly, some molecular clouds
are visible just north of the MEGARA FoV edge
in Fig.~\ref{fig:oiii_co_hst}.

Another region left out by Fig.~\ref{fig:ion_3maps}
is the NW arc with high $W_{80}$
values (Fig.~\ref{fig:OIII_velocities}),
associated with LINER/shock emission
in Figs.~\ref{fig:bpt2} and~\ref{fig:bpt3}.
This arc begins at the western edge 
of the CO($3-2$) emission,
but it may be linked to the high dispersion values
we see going towards NW
(bottom panel of Figs.~\ref{fig:ngc5506}
and \ref{fig:ion_MOR}).
These two regions may indicate that the outflow
(both in the ionised and molecular phases)
has a larger size than the ones we derive
with the present analysis.
However, a more detailed mapping 
of the aforementioned areas
is needed to draw meaningful conclusions.

\begin{figure*}
\centering
\includegraphics[width=0.98\textwidth]{
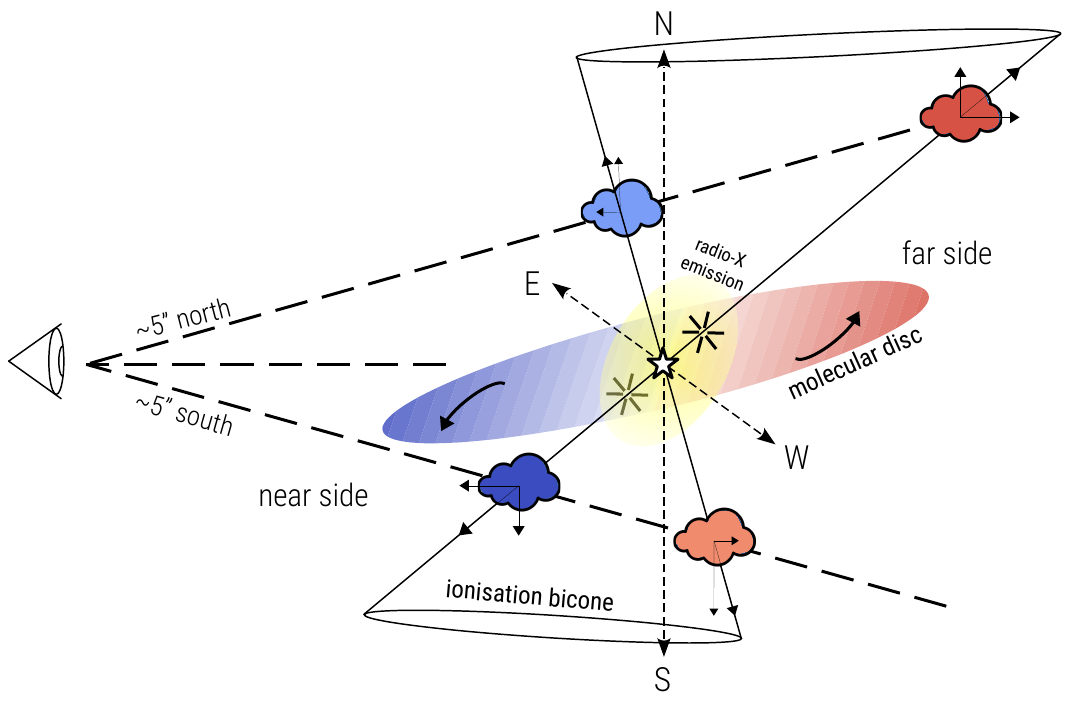}
\caption{Scenario (not to scale)
for the intersection
between the molecular disc
(the red and blue ellipse)
and the ionisation bicone.
In the disc of the galaxy, 
traced by the molecular gas, 
we mark the AGN position (white star) 
and the proposed interactions 
between the two gas phases
(black asterisk symbols).
The 300-pc radio
(at 8.46 GHz) and soft X-ray (below 1 keV)
emission is depicted in yellow.
Along the $\pm 5 \arcsec$ lines of sight,
we draw clouds on the edge of the bicone,
colour-coded depending on whether the gas 
is blueshifted or redshifted.}
\label{fig:cartoon}
\end{figure*}

The immediate vicinity of the AGN is 
relatively devoid of $\dot{M}_{\rm out}^{\rm ion}$,
due to the small amount of $M_{\rm out}^{\rm ion}$ 
(see Fig.~\ref{fig:ion_3maps}) in this region.
This may stem from the observed ionised
wind being a past outflow episode,
now situated $\sim 100$ pc from the centre,
where it encounters resistance from the 
surrounding ISM.
This corresponds to the same distance at which 
we observe a peak in the molecular mass outflow
rate (Fig.~\ref{fig:mol_MOR}),
with the caveat that we are seeing
projected distances for the ionised outflow,
and deprojected distances (on the disc plane)
for the molecular outflow.
We subsequently compare the two phases
in detail in the next section.

\section{Discussion}
\label{sec:discussion}

\subsection{The case for elliptical motions due to a bar}
\label{sec:bar}

Being highly inclined, it is challenging
to prove (or disprove) the presence of
a bar in NGC~5506.
\cite{devaucouleurs91} classified this galaxy
as a peculiar edge-on Sa, while
\cite{baillard11},
by analysing SDSS images, 
signalled the presence of a "barely visible" 
stellar bar (with confidence ranging from "no bar" to 
"bar long about half $D_{25}$").
By inspecting PanSTARRS images
we could in fact recognise an X-shape,
typical of edge-on barred galaxies
\citep{baba22}.

Edge-on barred galaxies typically display
clearly separate components on the gas PV diagrams 
\citep[see e.g. the compilation by][]{bureau99}.
This is not evident on the molecular PV diagrams
of NGC~5506 (Figs.~\ref{fig:barolo_PV_VRAD0}
and~\ref{fig:barolo_PV}), whereas instead
it is noticeable on the major axis of NGC 7172
\citep[][Figs. 8 and 10]{alonsoherrero23},
which is also a highly inclined galaxy.
With the exception of the X-shape
on the minor axis
(explained in Section~\ref{sec:alucine},
see also Fig.~\ref{fig:edgespectra}),
the same applies for the PV
diagrams of the ionised gas
(Fig.~\ref{fig:oiii_pv}).
This, however, could be to an unfavourable
orientation of the bar, being too close to
the minor axis to produce any apparent
perturbation.
It is also worth noting that NLSy1s as NGC~5506
are usually associated with the presence
of a bar \citep{crenshaw03}.

The fact that we see disturbed molecular
clouds on the north-west and south-east
(Fig.~\ref{fig:ngc5506}, bottom panel),
may be an indication of interaction
of the ionised outflow with the 
molecular disc
(Fig.~\ref{fig:ion_MOR}).
In the following section, we aim to provide
a more comprehensive description of the interaction.
However, it is important to note that we 
cannot rule out the potential existence 
of a bar within the central kiloparsec.
Consequently, in our analysis of molecular 
inflow/outflow velocities
(Fig.~\ref{fig:barolo_radial}, bottom panel), 
we treat these results as upper limits.

\subsection{Comparing molecular and ionised outflows}
\label{sec:comparison}

In Section~\ref{sec:mol_model} we modelled
the ALMA CO($3-2$) kinematics, finding
a rotating disc along PA $=265^{\circ}$,
within which the gas is also outflowing.
The most intense region of the molecular outflow
is at $r \sim 100$ pc,
with $v_{out, max}^{\rm mol} = 50$ km s$^{-1}$ and
$\dot{M}_{out, max}^{\rm mol} = 28$ M$_{\odot}$ yr$^{-1}$.
This is also where most of the molecular mass resides.
Another region of interest is at $r \sim 250$ pc,
where we found a second, more modest, peak
of $\dot{M}_{\rm out}^{\rm mol} (250 \text{ pc}) = 11$
M$_{\odot}$ yr$^{-1}$.
We plotted the circles of radii 100 and 250 pc
in Fig.~\ref{fig:ion_MOR}. If we follow the
PA $=265^{\circ}$ dashed line on the eastern side,
we find enhanced values of $\dot{M}_{\rm out}^{\rm ion}$
at such radii.
This could be an evidence of interaction
between the ionised AGN wind and the
molecular disc, where we are seeing perhaps
two different outflow episodes,
in which case, from the 150 pc distance
between the episodes,
we can calculate, given a 500 km s$^{-1}$ velocity,
a $\Delta t_{\rm out} = 0.3$ Myr
\citep[similar to the AGN flickering timescale
derived by][]{schawinski15, king15b}.

We can have a closer look at the interaction
between the ionised and molecular gas by
plotting the CO($3-2$) dispersion contours against
the ionised mass outflow rate map,
as in Fig.~\ref{fig:ion_MOR}:
not only do the $\dot{M}_{\rm out}^{\rm ion}$ regions 
at 100 and 250 pc east from the AGN correlate
with high CO dispersion 
($\sigma_{CO} \geq 50$ km s$^{-1}$), 
but also the region $\sim 1.5 \arcsec$ 
NW has both a local
excess of $\dot{M}_{\rm out}^{\rm ion}$ and high
$\sigma_{CO}$ (up to 61 km s$^{-1}$).
From Fig.~\ref{fig:ion_MOR}
(but also from Fig.~\ref{fig:ngc5506})
it seems NGC~5506 would be 
in the weak coupling scenario
described by \cite{ramosalmeida22}, i.e.
where the biconical ionised outflow intercepts
the molecular disc only partially,
launching a modest molecular outflow
\citep[see also][]{alonsoherrero23}.
This would be in agreement with the bicone model
fitted by \cite{fischer13} for NGC~5506:
they found the inclination between the bicone
and the host galaxy disc to be $32^{\circ}$,
less than the maximum half-opening angle 
of the bicone
\citep[$40^{\circ}$, see Table 6 in][]{fischer13}.

We draw a tentative sketch of the 
relative positions of the molecular disc
and the ionised bicone in
Fig.~\ref{fig:cartoon}.
Every line of sight intercepts both
approaching and receding sides of
the bicone, resulting in a mix of
blueshifted and redshifted velocities
(as in Fig.~\ref{fig:OIII_velocities}).
Once we are far enough from the disc plane
($\sim 5 \arcsec$ north and south),
the edges of the bicone start to appear
distinct on the spectra
(Fig.~\ref{fig:edgespectra}):
this would point out a hollow bicone.
The southern nearest and northern farthest
bicone edges intercept the molecular
disc, hence rising the CO velocity
dispersion and causing the molecular
ring to outflow on the disc plane:
this results in high CO dispersion on the
SE-NW direction (Fig.~\ref{fig:ion_MOR}),
and in an asymmetry in the CO($3-2$) 
PV diagram on the redshifted northern - 
blueshifted southern directions
(Fig.~\ref{fig:barolo_PV}, bottom panel).

An exception to the spatial correlation between
$\sigma_{CO}$ and $\dot{M}_{\rm out}^{\rm ion}$ is
in the immediate vicinity of the AGN:
there the CO line broadening is probably due to
the presence, in a small space,
of multiple components of CO velocities,
even due to ordered rotation alone.
Nevertheless, 
this region also has a deficit of CO emission 
\citep[see Fig.~\ref{fig:ngc5506} and][]{
garciaburillo21},
which may be another indication 
of multiphase feedback.

If we adopt the scenario drawn in 
Fig.~\ref{fig:cartoon}, then the ionised
outflow velocities we measured, 
especially the redshifted
ones in the north and the blueshifted in the south,
are lower limits due to projection effects.
We did not perform a modelling of the 
bicone (so its opening angle in
Fig.~\ref{fig:cartoon} is only qualitative),
but if we adopt an half-opening angle of
$40^{\circ}$ \citep{fischer13},
we can derive a multiplicative factor of
$1 / \sin(40^{\circ}) = 1.56$,
which would result in an average
deprojected $v_{\rm out}^{\rm ion} = 657 \pm 151$ km s$^{-1}$.
Being the deprojected $R_{\rm out}^{\rm ion}$ affected
in the same way, this would not change
the $\dot{M}_{\rm out}^{\rm ion}$.

\begin{table}
\centering
\caption{Results for the molecular and ionised phases
of the AGN outflow. The distance from the AGN $R_{\rm out}$
is in different directions for the two phases.
}
\label{tab:tab_results}
\begin{threeparttable}
\begin{tabular}{lcc}
\toprule
Property & Molecular &  Ionised \\
\midrule
$R_{\rm out}$ [pc]           & 610        &             525 \\
$v_{\rm out}$ [km s$^{-1}$] & $26 \pm 9$ &    $422 \pm 97$ \\
$M_{\rm out}$ [M$_{\odot}$]  & 
$1.7 \times 10^8$ &  $9.4 \times 10^4$ \\
$\dot{M}_{\rm out}$ [M$_{\odot}$ yr$^{-1}$] & 
$8 \pm 3$ &    $0.08 \pm 0.02$ \\
$\dot{E}_{\rm out}$ [$10^{39}$ erg s$^{-1}$] &   
$1.7 \pm 1$ & $4.3 \pm 1.7$ \\
$\dot{P}_{\rm out}$ [$10^{32}$ dyn] & 
$13.1 \pm 6.7$ & $2.0 \pm 0.7$ \\
\bottomrule
\end{tabular}
\begin{tablenotes}
\item \textbf{Notes.}
Molecular $v_{\rm out}$, $\dot{M}_{\rm out}$,
$\dot{E}_{\rm out}$, and $\dot{P}_{\rm out}$
are upper limits, due
to the possible presence of elliptical motions
associated with a nuclear bar.
All the ionised values 
(except $M_{\rm out}$ and $\dot{M}_{\rm out}$)
are lower limits, since $R_{\rm out}$ and $v_{\rm out}$
are projected (on the plane of the sky) measurements.
\end{tablenotes}
\end{threeparttable}
\end{table}

If the AGN wind seen with the [OIII] and the
outflowing CO ring are physically connected, we expect
the kinetic energy rate ($\dot{E}_{\rm out}$)
or the momentum rate ($\dot{P}_{\rm out}$)
to be conserved
\citep[see][and references therein]{king15}.
These two quantities can be 
straightforwardly calculated as
$\dot{E}_{\rm out} = \dot{M}_{\rm out} v_{\rm out}^2 / 2$ and 
$\dot{P}_{\rm out} = \dot{M}_{\rm out} v_{\rm out}$.
The values (listed in Table~\ref{tab:tab_results})
point to a energy-driven 
rather than momentum-driven
outflow \citep{king15, veilleux20}:
in such outflows, the momentum undergoes a boost
\citep[e.g.][]{veilleux20, longinotti23},
which in our case is
$\dot{P}_{\rm mol} / \dot{P}_{\rm ion} = 7$.
However, if we use the values derived
by \cite{davies20} for the ionised outflow,
the ratio of the momentum rates would be
$\sim 1.2$, rather indicating a momentum-driven outflow.
Given the observed $L_{\rm bol}$ and $\lambda_{\rm Edd}$
(see Table~\ref{tab:tab_data}), a radiation pressure-driven
wind would predict an outflow of
$\sim 3$ M$_{\odot}$ yr$^{-1}$ \citep{honig19}, 
not too far from our $\dot{M}_{\rm out}^{\rm mol}$ value:
this also would point to a momentum-driven scenario.

We highlight that the dichotomy between energy 
and momentum conservation refers to single 
or continuous outflow episodes. In the case of NGC~5506,
we may be observing the stratification of multiple 
outflows, a possibility explored also in the next section.
Taking everything into account, if the ionised wind
is pushing and dragging the molecular gas,
it currently seems to impact only the inner part of
the molecular ring. At this stage, the AGN wind
appears to be relatively ineffective in clearing
the entire galaxy
\citep[which is common in local systems, see e.g.][]{
fluetsch19}.

\begin{figure}
\centering
\includegraphics[width=0.49\textwidth]{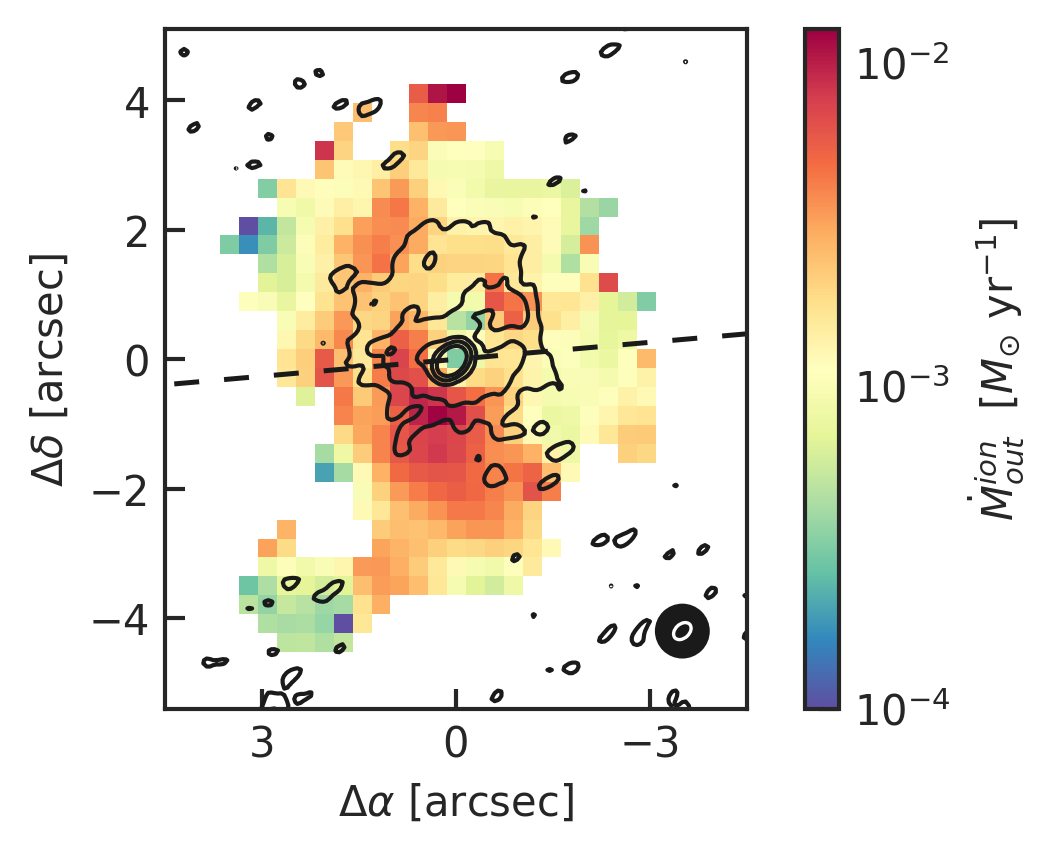}
\caption{Same as Fig.~\ref{fig:ion_MOR} but with
the \cite{schmitt01} 3.6 cm VLA contours, at 
$\log (S_{\nu} / \text{Jy beam}^{-1}) 
= (-4, -3.5, -3, -2.5, -2)$, 
in black.
The black circle in the bottom right is the 
MEGARA seeing conditions,
with the VLA beam ellipse within it in white.}
\label{fig:ion_MOR_VLA}
\end{figure}

\subsection{Extending the spectrum: radio and X-ray literature}

Despite its classification as a radio-quiet galaxy
\citep{terao16},
NGC~5506 has been detected in the radio band
in several studies.
\cite{wehrle87} detected, with the VLA 
at 5 GHz, a radio bubble,
NW from the nucleus, also visible in
the 8.46 GHz VLA A-array continuum image
presented by \cite{schmitt01}.
In Fig.~\ref{fig:ion_MOR_VLA} we plot
the contours of \cite{schmitt01} VLA image
against the ionised mass outflow rate map,
where we can see that the radio bubble
observed by \cite{wehrle87} perfectly
overlaps with the $\sim 1.5 \arcsec$ region
that has both high $\dot{M}_{\rm out}^{\rm ion}$
and high $\sigma_{CO}$.
The extended $\sim 300$ pc VLA emission 
in Fig.~\ref{fig:ion_MOR_VLA} is well
aligned with the galactic disc (PA $=265^{\circ}$),
but extends below and (mostly) over it,
following the [OIII] emission.
\cite{orienti10} measured, for this diffuse radio
emission, a steep spectral index $\alpha = 0.9$,
which combined with the size $< 1$ kpc,
would make it a compact steep spectrum (CSS) 
radio source \citep[e.g.][]{dallacasa13, odea21}.

The VLA contours shown in Fig.~\ref{fig:ion_MOR_VLA}
are also spatially coincident with the soft
X-ray (below 1 keV) emission observed, with the
\textit{Chandra} X-ray Observatory, by \cite{bianchi03}.
Their main explanation is that the photoionised gas
(that we clearly see with MEGARA,
Fig.~\ref{fig:oiii_co_hst}, even if more
extended than 300 pc) is reprocessing
the nuclear X-ray emission.
However, since we detect velocities up
to $\sim 600$ km s$^{-1}$ out to 300 pc from the AGN,
the expected temperature of the shocked emission
is $kT \approx 1.3 (v_{shock}/10^3)^2$ keV 
$\approx 0.5$ keV \citep{fornasini22}, 
which could suggest
a thermal emission for the \textit{Chandra}
soft X-ray observation
\citep[see also][]{paggi12}.

High-resolution radio observations
made with the Very Long Basement
Array (VLBA) at different frequencies 
\citep[1.6 - 15 GHz,][]{roy00},
show a sub-relativistic
($v_{jet} \leq 0.25 c$)
one-sided jet, 
initially oriented $70^{\circ}$
anticlockwise from the north
(so roughly as the CO disc), and then
bending $90^{\circ}$ towards the south
\citep{kinney00},
at 3.4 pc ($\sim 0.03 \arcsec$) 
from the core emission.
In \cite{roy00b} they argue that the
counterjet is not visible because
of free-free absorption
(rather than doppler boosting),
and that the bend might be 
a sign of interaction between 
the jet and the NLR
gas on parsec scales.
\cite{middelberg04} collected
different-epochs EVN, MERLIN and VLBA observations,
and reported a $3\sigma$ upper limit
of $0.50 c$ for the jet motion
with respect to the core.
\cite{gallimore06} argue that the
diffuse emission on the 300-pc scale
\citep[Fig.~\ref{fig:ion_MOR_VLA} and][]{schmitt01},
is attributed to the parsec-scale jet observed
by \cite{roy00}. The misalignment between the
jet trajectory (initially pointing at
$\sim 70^{\circ}$ anticlockwise from north
and later bending $\sim 90^{\circ}$
towards the south) and the elongation of the
diffuse radio emission towards the north
direction can be explained by either jet
precession or jet-ISM interactions
(\citealp{gallimore06}, and also
\citealp{xanthopoulos10} 
come to the same conclusions).

\begin{figure}
\centering
\includegraphics[width=0.49\textwidth]{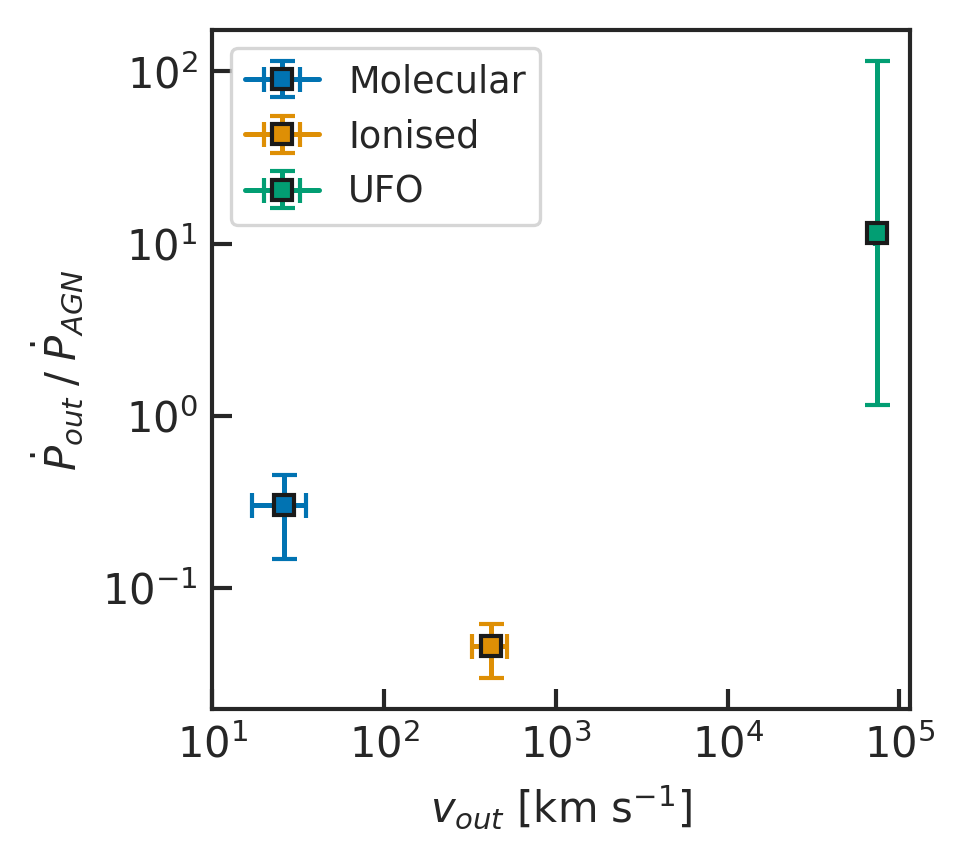}
\caption{Outflow momentum rate 
divided by the AGN radiation
momentum rate $L_{\rm bol}/c$ 
(also called wind momentum load)
as a function of the outflow velocity $v_{\rm out}$
for the molecular gas (in blue), 
the ionised gas (in orange), 
and the UFO (in green).}
\label{fig:Pout_plot}
\end{figure}

Interestingly, such high velocities
are also seen via absorption of
the hard X-ray Fe XXVI Ly$\alpha$ line.
The UFO in NGC~5506 has been observed and studied
by \cite{gofford13} and \cite{gofford15},
where they find $v_{UFO} = 0.246 \pm 0.006$ c.
The momentum rate released by such a UFO
ranges between $5 \times 10^{33}$ and 
$5 \times 10^{35}$ dyn,
where this large uncertainty mostly comes
from the estimation of the distance
between the UFO and the AGN
\citep[see][for a detailed derivation of the UFO parameters]{
tombesi13}.
Even the lower limit of $\dot{P}_{\rm out}^{UFO}$
is $3.8$ times the molecular one
(see Fig.~\ref{fig:Pout_plot}).
If we accept as good all these different measurements,
a plausible explanation for this momentum
decrease (instead of the boost required in the
energy-driven scenario, or the constant $\dot{P}_{\rm out}$
in the momentum-driven scenario)
is, again, that we are seeing different outflow
episodes, among which the UFO is the most recent
\citep[also][suggest multiple activity episodes 
for NGC~5506 from analysing 
polarised radio data]{biny20}.
X-ray observations have shown in fact
continuous rapid variation
among different epochs 
\citep{mchardy87, uttley05, sun18},
even suggesting the presence of a 
supermassive black hole binary system
\citep{manchanda06}.

Both relativistic jets
(e.g. \citealp{mukherjee18, audibert23},
but also low-power jets, 
e.g. \citealp{venturi21, pereirasantaella22})
and UFOs \citep[e.g.][]{
marasco20, longinotti23, salome23}
are thought to be the initial trigger
of galaxy-scale ionised and molecular outflows
\citep[see][for a recent discussion]{singha23}.
Another possibility is that the
VLBI radio structures seen by \cite{roy00}
are shock signatures left by the X-ray UFO
\citep{longinotti18}.


\section{Summary and conclusions}
\label{sec:conclusions}

We presented new GTC/MEGARA 
optical IFU observations 
of NGC~5506, complemented with 
ALMA Band 7 observations 
of the CO($3-2$) transition 
\citep{garciaburillo21}.
NGC~5506 is a nearby ($D=26$ Mpc) luminous 
($L_{\text{bol}} \sim 1.3 \times 10^{44}$ erg s$^{-1}$) 
Seyfert galaxy, part of the GATOS sample
\citep{garciaburillo21, alonsoherrero21}.
The angular resolution of the ALMA observation
($0.21 \arcsec \times 0.13 \arcsec$)
allows us to probe regions on physical 
scales of $\sim 25$ pc for the molecular gas.
The GTC/MEGARA observation, 
with a seeing of $0.9 \arcsec$
(corresponding to $\sim 113$ pc 
at the distance of NGC~5506),
offers a spectral resolution enabling
the analysis of velocities as low as
$\sim 14$ km s$^{-1}$.

The CO($3-2$) map reveals a highly inclined 
($i = 80^{\circ}$) cold molecular gas ring, 
symmetric up to a radius of $3.5 \arcsec \sim 438$ pc,
with an eastern tail extending
up to a $8 \arcsec \sim 1$ kpc radius.
The cold molecular gas mass of the ring
is $\sim 2.3 \times 10^8$ M$_{\odot}$,
calculated assuming a brightness temperature ratio of
$T_{B, CO(3-2)} / T_{B, CO(1-0)} = 0.7$
and a Galactic CO-to-H$_2$ conversion factor.
The CO($3-2$) kinematics reveal a rotating disc,
flattening at $193$ km s$^{-1}$ around $r=440$ pc,
with clear signatures of non-circular motions.
A \textsuperscript{3D}\textsc{Barolo}
model of a rotating disc with a radial velocity
component reproduces reasonably well the 
observed CO kinematics, interpreted as
a rotating and outflowing molecular ring.
Within a $0.4 \arcsec$ radius,
fitted radial velocities are directed
towards the centre, potentially indicating
AGN feeding, though this finding could not
be confirmed since this radius is
very close to the ALMA beam size.
At larger radii, the radial velocity
is directed outwards, decreasing
from a maximum of 50 km s$^{-1}$
to an average of 26 km s$^{-1}$.
The maximum molecular outflow radius
is 610 pc, within which we 
calculate an integrated 
molecular gas mass outflow rate of 
$\sim 8 \pm 3$ M$_{\odot}$ yr$^{-1}$.

We detected several bright emission lines
in the MEGARA spectra, with
[OIII]$\lambda 5007$ standing out as the brightest.
The spatially resolved BPT diagnostic diagrams 
predominantly reveal Sy-like excitation,
ruling out a significant 
contribution from star formation
over a projected region of 
1.5 kpc $\times$ 1.4 kpc.
The [OIII] kinematics appear to be 
dominated by the outflowing gas.
Nevertheless, we separated
disc rotation from non-circular motion spaxel-by-spaxel,
employing both parametric and non-parametric methods.
The ionised gas exhibits a slower rotation
speed than the molecular gas
($\sim 190$ km s$^{-1}$),
with H$\alpha$, [NII] and [SII]
reaching 120 km s$^{-1}$.
Conversely, we detected [OIII] 
radial velocities up to 1000 km s$^{-1}$,
both approaching and receding.
Emplying a non-parametric analysis of the
line wings of [OIII] emission,
we derived an average ionised gas
outflow velocity of 422 km s$^{-1}$ 
within a radius of 525 pc.
To estimate the outflowing mass,
we utilised the broad component intensity maps
from the double Gaussian decomposition.
We calculated the electron density
in every spaxel using the
ionisation parameter method.
This analysis yielded an outflowing mass of
$9.8 \times 10^4$ M$_{\odot}$,
resulting in an ionised mass outflow rate of
$\dot{M}_{\rm out}^{\rm ion} = 0.076 \pm 0.017$ 
M$_{\odot}$ yr$^{-1}$.

We compared the spatially resolved
map of $\dot{M}_{\rm out}^{\rm ion}$
with the CO($3-2$) velocity dispersion map,
identifying spatial correlation between the two.
The ionised outflow does not appear perpendicular
to the plane of the galaxy; instead, it likely lies
at a small angle relative to the disc.
This results in a good geometrical
coupling between the two phases.
We also found diffuse radio and soft
X-ray emission to spatially correlate with
the observed [OIII] emission and
$\dot{M}_{\rm out}^{\rm ion}$.

Various results, both from this study
and the literature, suggest a 
diverse history of outflows for NGC~5506.
These outflows may be associated with the presence
of a parsec-scale radio jet, a $0.25 c$ UFO, 
or a combination of both.
New ALMA and JWST observations, 
offering a higher resolution view
of the nuclear region of NGC~5506, will soon 
become available as part of the GATOS project.
These observations may eventually enhance
our understanding of the complex interactions 
between the sub-parsec radio jet, the UFO, 
the ionised wind, and the molecular torus and disc.


\begin{acknowledgements}
We thank the anonymous referee for suggestions 
and comments that
helped to improve the presentation of this work.
We acknowledge the use of Python \citep{python3}
and the following libraries:
Astropy \citep{astropy1, astropy2},
CMasher \citep{cmasher},
Matplotlib \citep{matplotlib},
NumPy \citep{numpy},
Pandas \citep{pandas},
pvextractor \citep{pvextractor},
Seaborn \citep{seaborn},
SciPy \citep{scipy}, and
SpectralCube \citep{spectralcube}.
This research is based on observations made with the 
NASA/ESA Hubble Space Telescope obtained from the 
Space Telescope Science Institute, which is operated 
by the Association of Universities for Research 
in Astronomy, Inc., under NASA contract NAS 5–26555. 
These observations are associated with program 5479.
This paper makes use of the following ALMA data:
ADS/JAO.ALMA\#2017.1.00082.S. ALMA is a partnership 
of ESO (representing its member states), NSF (USA) 
and NINS (Japan), together with NRC (Canada), MOST 
and ASIAA (Taiwan), and KASI (Republic of Korea), 
in cooperation with the Republic of Chile. 
The Joint ALMA Observatory is operated by ESO, 
AUI/NRAO and NAOJ.
This research has made use of 
"Aladin sky atlas" developed at CDS, 
Strasbourg Observatory, France
\citep{aladin1}.
This research has made use of the 
NASA/IPAC Extragalactic Database (NED), 
which is funded by the National Aeronautics 
and Space Administration 
and operated by the 
California Institute of Technology.
The authors acknowledge the use of computational 
resources from the parallel computing cluster 
of the Open Physics Hub 
(\url{https://site.unibo.it/openphysicshub/en})
at the Physics and Astronomy Department in Bologna.
FE acknowledges support from grant 
PRIN MIUR 2017-20173ML3WW$\_$001,
and funding from 
the INAF Mini Grant 2022 program “Face-to-Face
with the Local Universe: ISM’s Empowerment (LOCAL)".
AAH and LHM acknowledge support from research grant 
PID2021-124665NB-I00 funded by 
MCIN/AEI/10.13039/501100011033 
and by ERDF "A way of making Europe".
SGB acknowledges support from the Spanish grant
PID2022-138560NB-I00, funded by
MCIN/AEI/10.13039/501100011033/FEDER, EU.
IGB acknowledges support from STFC 
through grant ST/S000488/1 and ST/W000903/1.
LPdA acknowledges financial support from grant
PGC2018-094671-B-I00 funded by 
MCIN/AEI/10.13039/501100011033
and by ERDF "A way of making Europe".
MPS acknowledges funding support from the 
Ram\'on y Cajal program of the Spanish Ministerio de Ciencia 
e Innovaci\'on (RYC2021-033094-I).
CRA acknowledges support from
EU H2020-MSCA-ITN-2019 Project 860744 
"BiD4BESt: Big Data applications for black hole 
Evolution STudies", 
from project "Feeding and feedback in active galaxies", 
with reference PID2019-106027GB-C42, 
funded by MICINN-AEI/10.13039/501100011033, 
and from project "Tracking active galactic nuclei 
feedback from parsec to kiloparsec scales", 
with reference PID2022-141105NB-I00.
EB acknowledges the Mar\'ia Zambrano program 
of the Spanish Ministerio de Universidades 
funded by the Next Generation European Union 
and is also partly supported by grant 
RTI2018-096188-B-I00 funded by the Spanish Ministry 
of Science and Innovation/State Agency of Research 
MCIN/AEI/10.13039/501100011033.
OGM acknowledges finantial support from UNAM 
through the project PAPIIT IN109123 
and from CONAHCyT through the project 
"Ciencia de Frontera 2023" CF-2023-G-100.
CR acknowledges support from Fondecyt Regular grant 
1230345 and ANID BASAL project FB210003.
\end{acknowledgements}

\bibliographystyle{aa} 
\bibliography{bibespo}

\begin{appendix}


\section{Goodness of ALUCINE fit}

\begin{figure}
\centering
\includegraphics[width=0.49\textwidth]{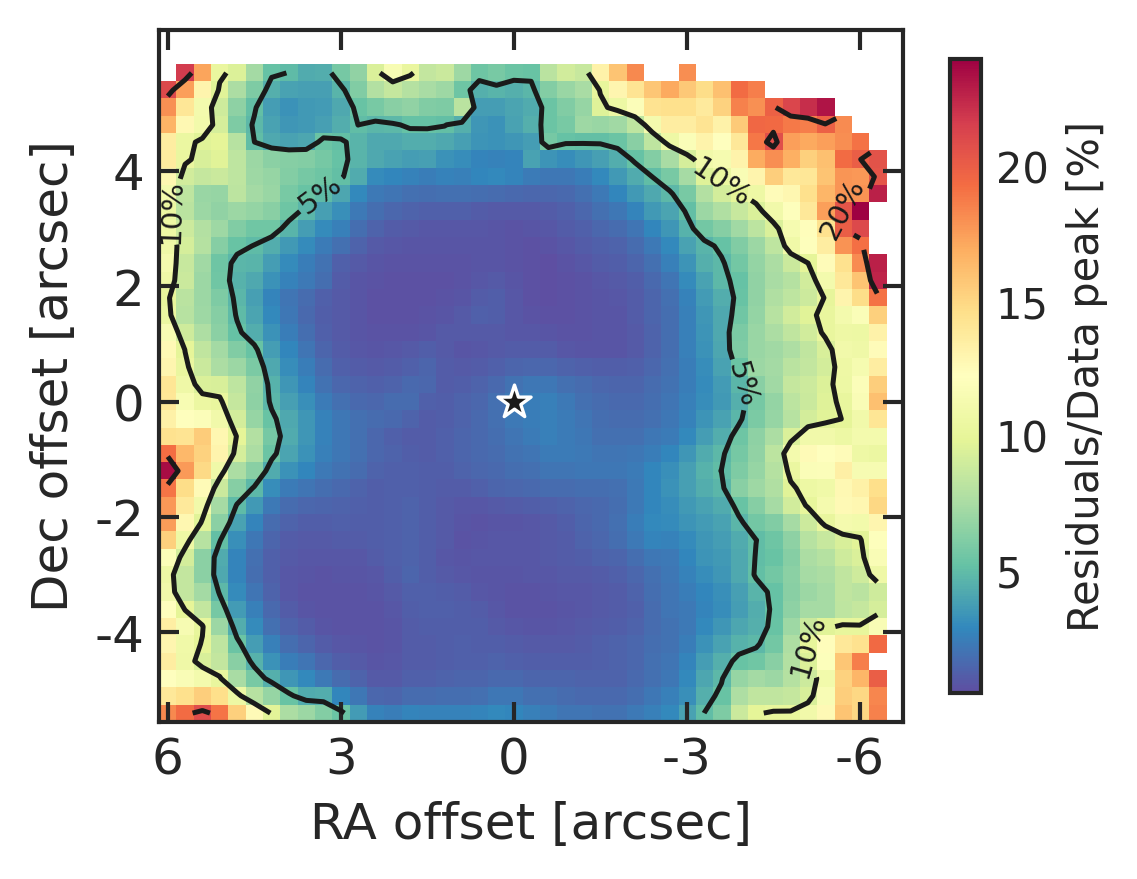}
\caption{Residuals of the [OIII] ALUCINE fit
(defined as |observations - model|) divided by the
observed peak flux, in every spaxel, 
in percentage.}
\label{fig:residuals}
\end{figure}

See Fig.~\ref{fig:residuals}.


\section{Gaussian decomposition of MEGARA lines}
\label{sec:more_megara}

\begin{figure*}
\centering
\includegraphics[width=0.99\textwidth]{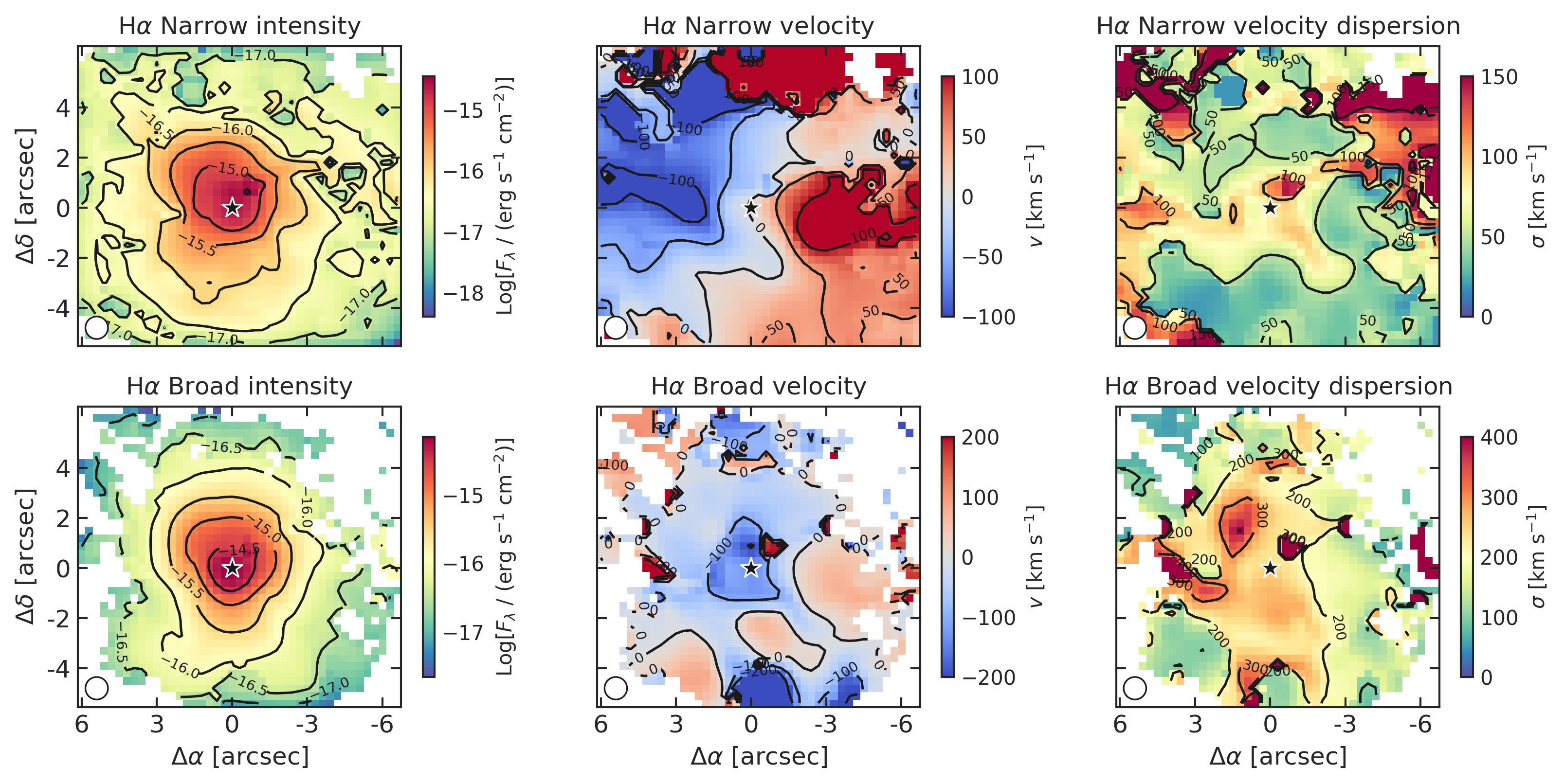}
\caption{Gaussian decomposition made by ALUCINE for
the H$\alpha$ line. Top and bottom panels
are for narrow and broad component, respectively,
while in the three columns, from left to right, 
are the intensity, velocity, 
and velocity dispersion maps. 
The AGN position is marked with a 
black star symbol, and distances are measured from it.
The white circle in the bottom left
is the MEGARA seeing conditions.}
\label{fig:halpha}
\end{figure*}

\begin{figure*}
\centering
\includegraphics[width=0.99\textwidth]{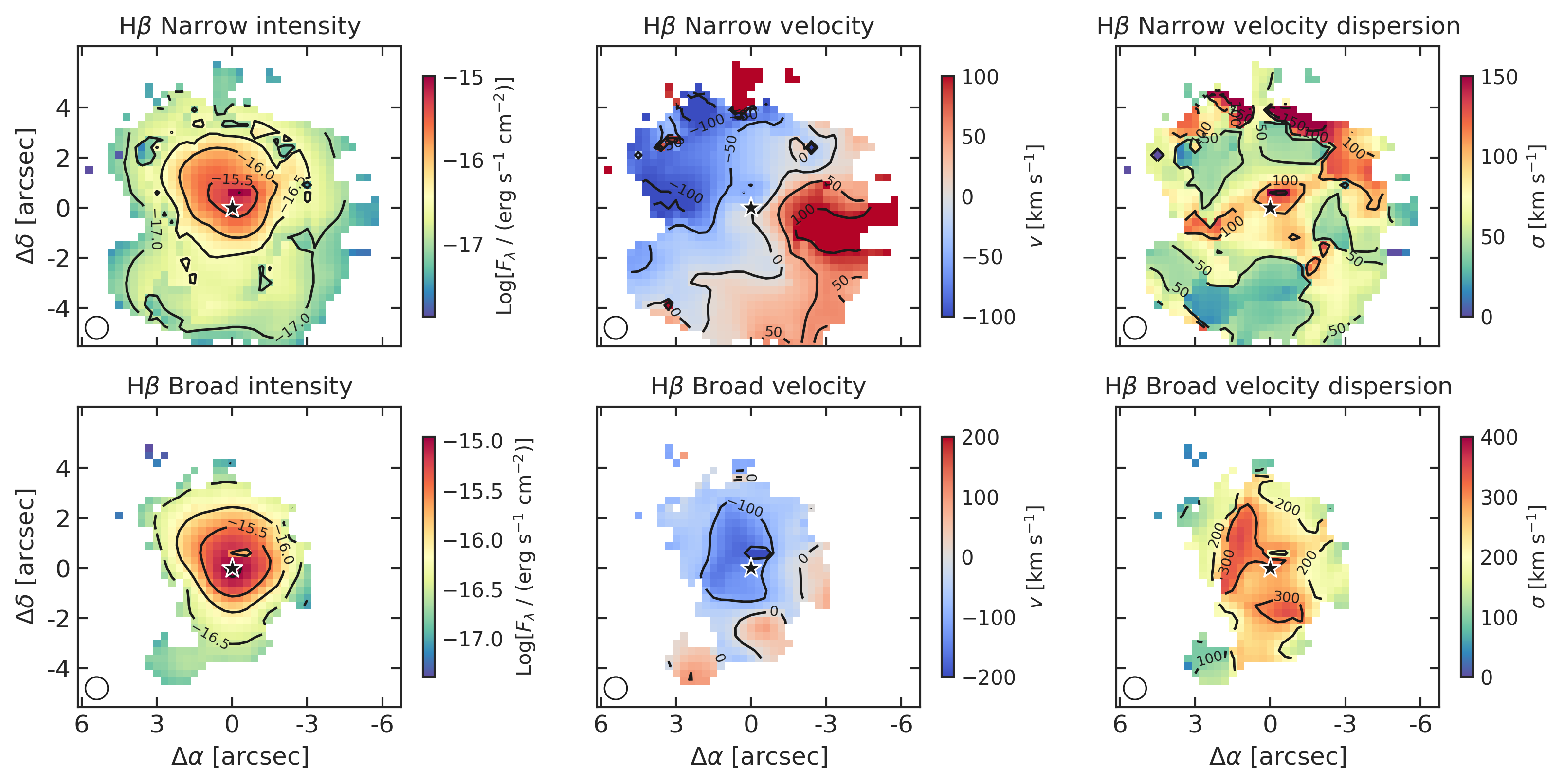}
\caption{Same as Fig.~\ref{fig:halpha}, but for
the H$\beta$ line.}
\label{fig:hbeta}
\end{figure*}

\begin{figure*}
\centering
\includegraphics[width=0.99\textwidth]{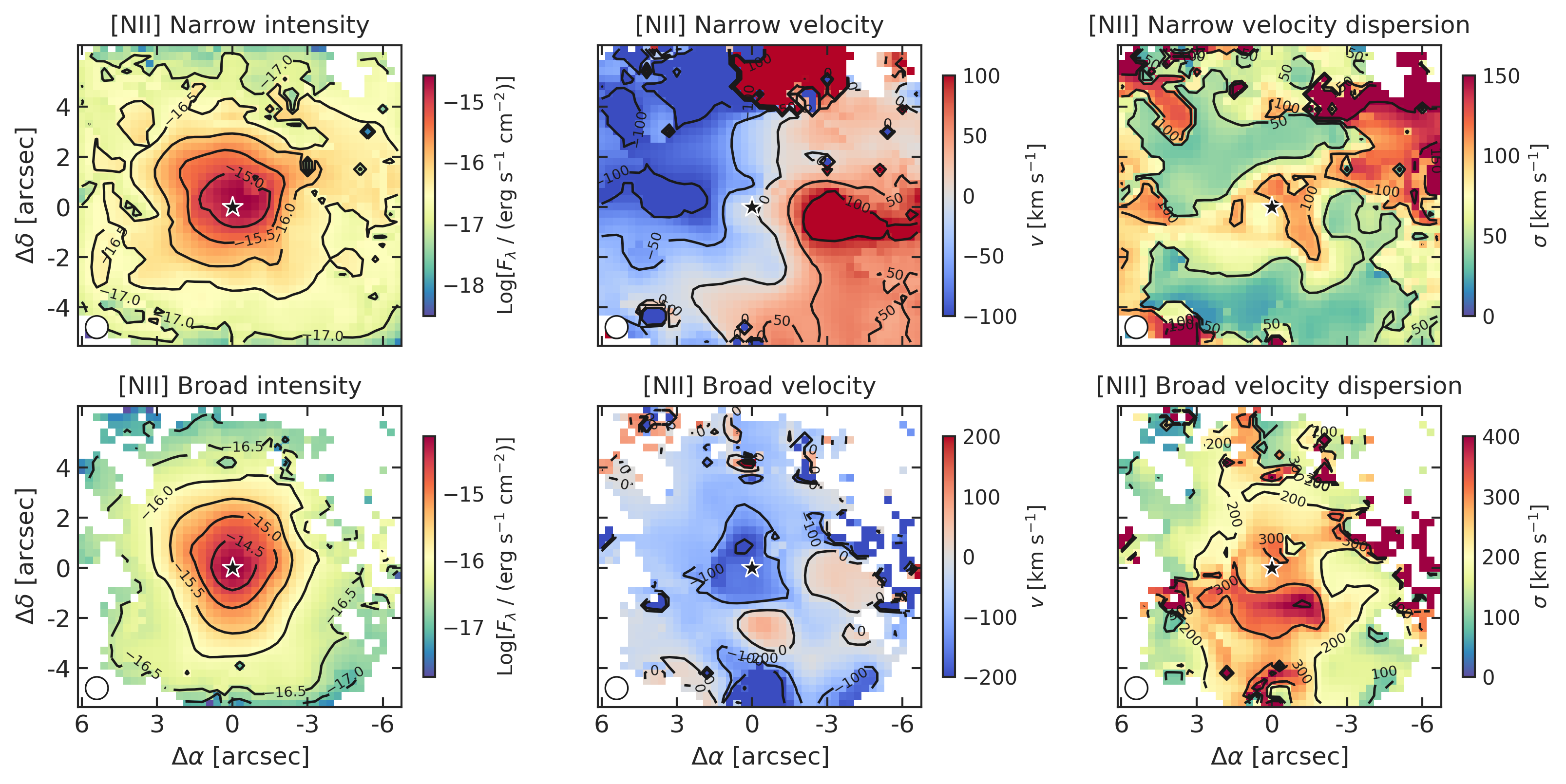}
\caption{Same as Fig.~\ref{fig:halpha}, but for
the [NII]$\lambda\lambda6548, 6583$ doublet. 
The intensity maps show the sum 
of the two lines of the doublet, while the
kinematics are assumed to be the same
between the two lines.
The [NII] emission is fitted together
with the blended H$\alpha$ line (Fig.~\ref{fig:halpha})
to properly separate their kinematics.}
\label{fig:nii}
\end{figure*}

\begin{figure*}
\centering
\includegraphics[width=0.99\textwidth]{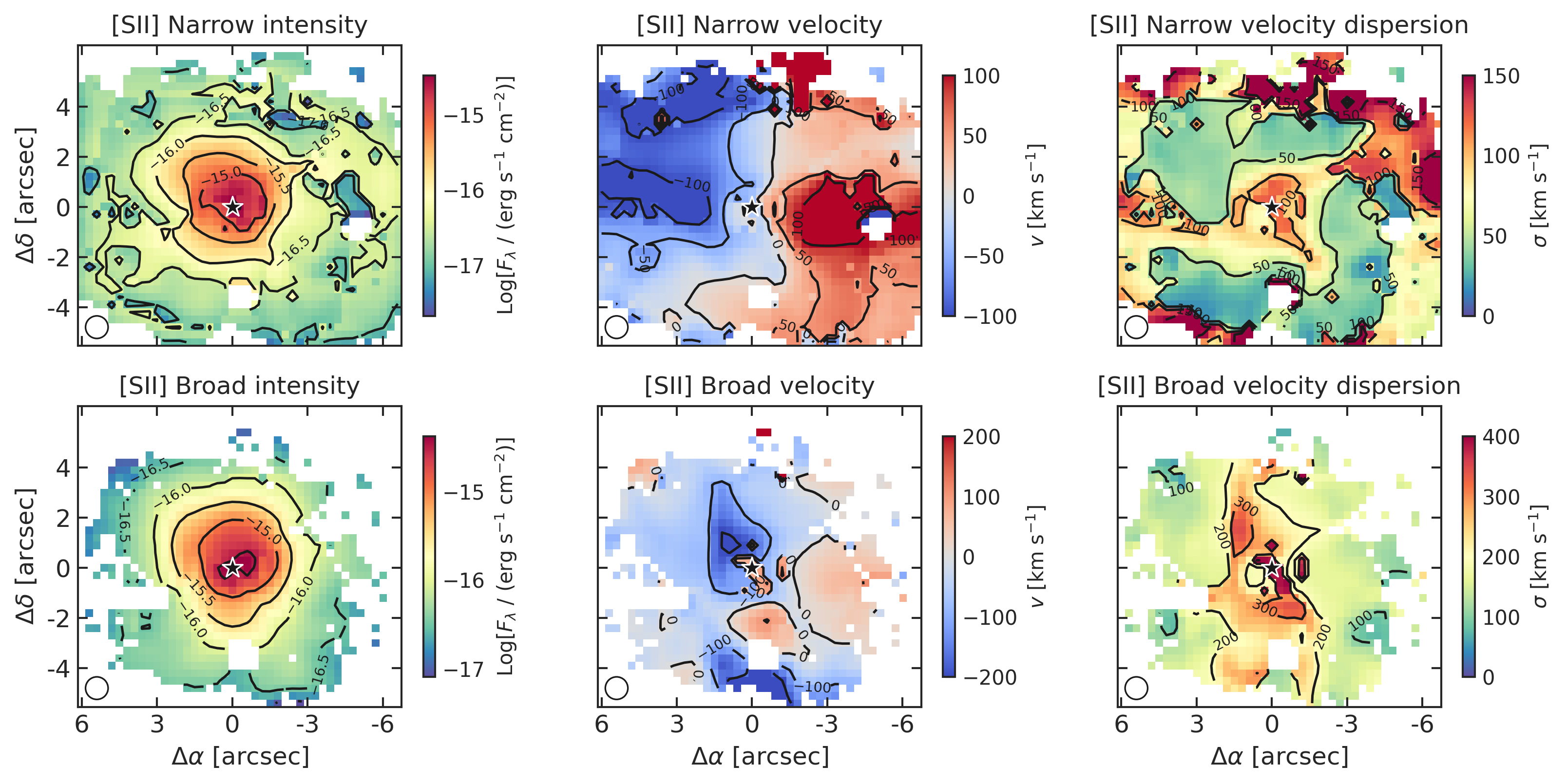}
\caption{Same as Fig.~\ref{fig:halpha}, but for
the [SII]$\lambda\lambda6716, 6731$ doublet. 
The intensity maps show the sum 
of the two lines of the doublet, while the
kinematics are assumed to be the same
between the two lines.}
\label{fig:sii}
\end{figure*}

\begin{figure*}
\centering
\includegraphics[width=0.99\textwidth]{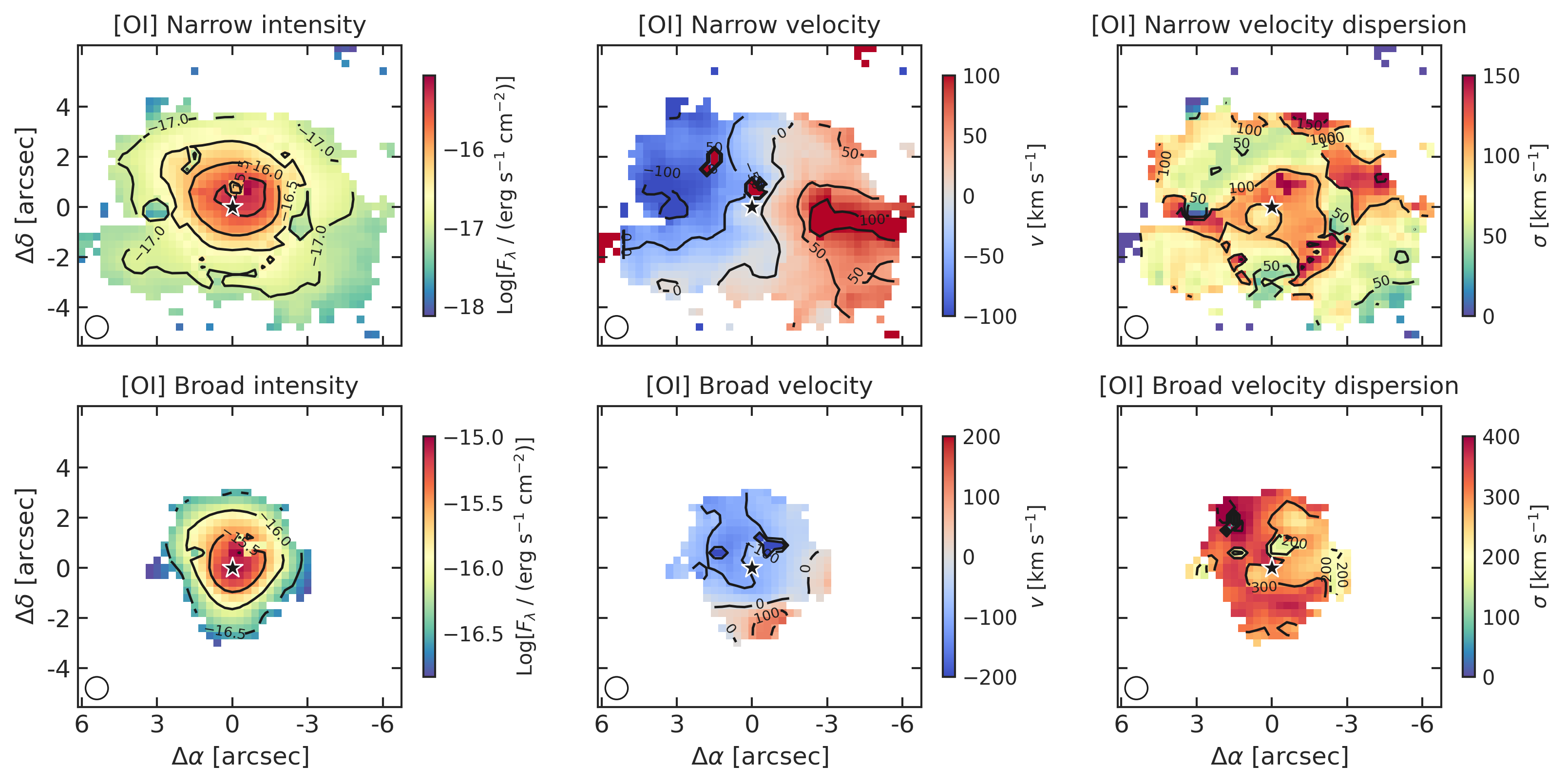}
\caption{Same as Fig.~\ref{fig:halpha}, but for
the [OI]$\lambda6300$ line.}
\label{fig:oi}
\end{figure*}

See Figs.~\ref{fig:halpha}, 
\ref{fig:hbeta}, \ref{fig:nii},
\ref{fig:sii}, and \ref{fig:oi}.


\section{BPT diagrams of MEGARA lines}
\label{sec:bpt}

\begin{figure}
\centering
\includegraphics[width=0.49\textwidth]{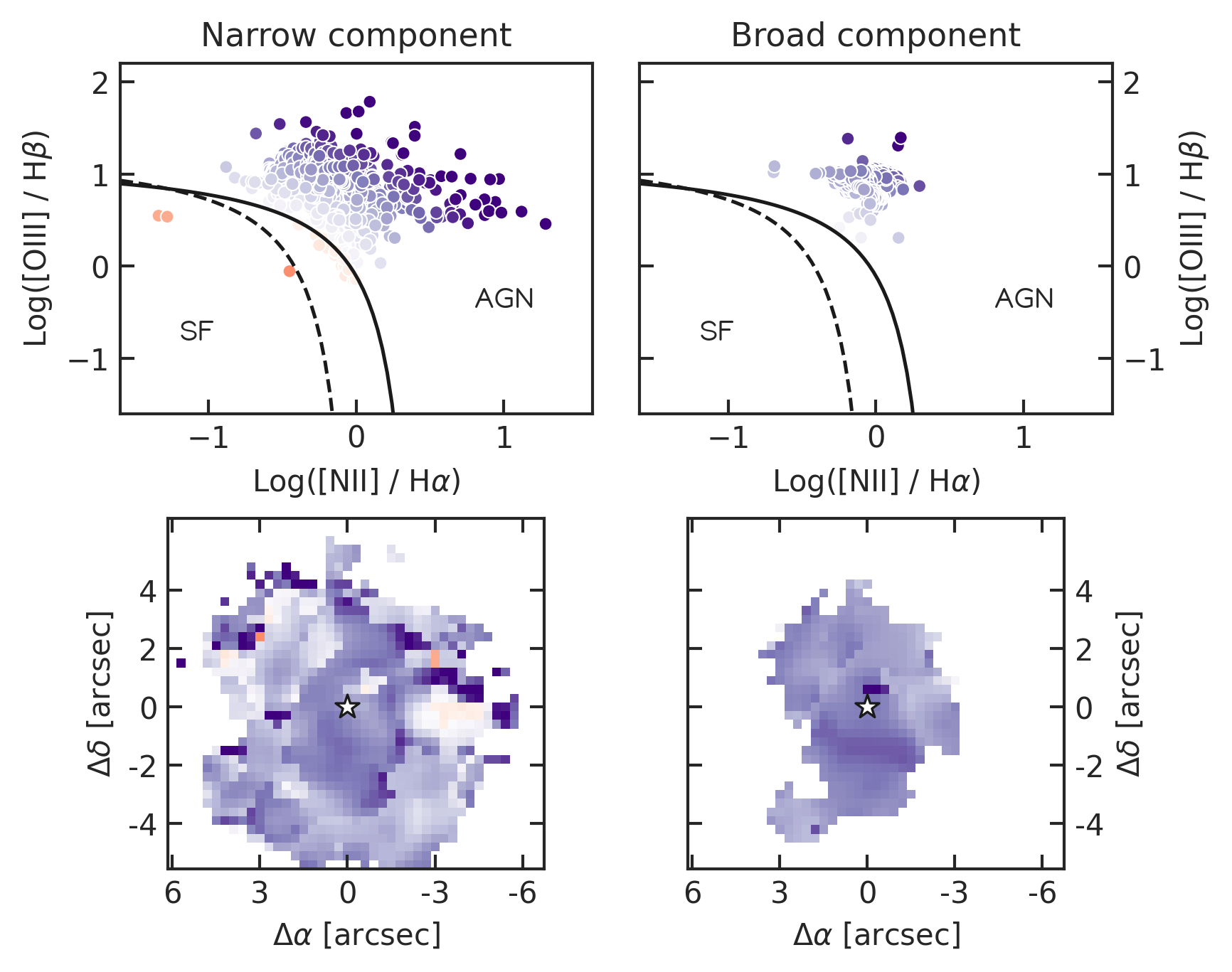}
\caption{Diagnostic Baldwin, Phillips, Telervich
(BPT) diagram \citep{baldwin81, veilleux87}
of [OIII]/H$\beta$ vs [NII]/H$\alpha$ line ratios,
for the narrow (\textit{left panels}) and
broad (\textit{right panels}) components of the
ionised gas. 
Solid and dashed black lines
are empirical curves derived by
\cite{kewley01} and \cite{kauffmann03},
that separate different excitation mechanisms, 
marked on the plots as SF (star formation) and AGN.
The spaxels are coloured depending on their distance
from the separation lines, and are plotted with the 
same colours in the spatially resolved maps
(bottom panels).
The black star marks the AGN position.}
\label{fig:bpt1}
\end{figure}

\begin{figure}
\centering
\includegraphics[width=0.49\textwidth]{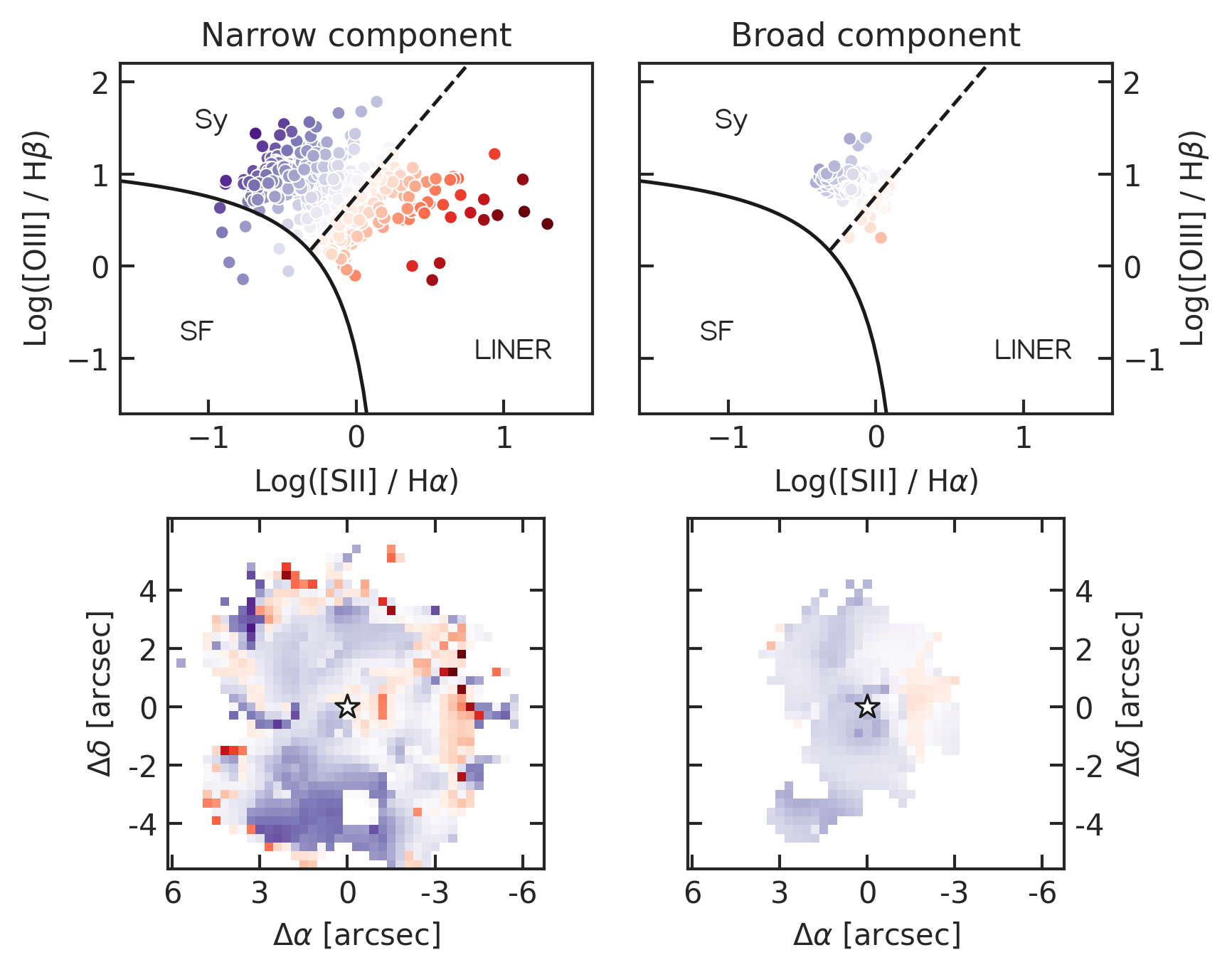}
\caption{Same as Fig.~\ref{fig:bpt1}, but for the
[OIII]/H$\beta$ vs [SII]/H$\alpha$.
The different excitation mechanisms are marked
on the plots as SF (star formation), 
Sy (Seyfert), and LINER 
(low-ionisation nuclear emission-line region),
with separation lines from \cite{veilleux87}.
}
\label{fig:bpt2}
\end{figure}

\begin{figure}
\centering
\includegraphics[width=0.49\textwidth]{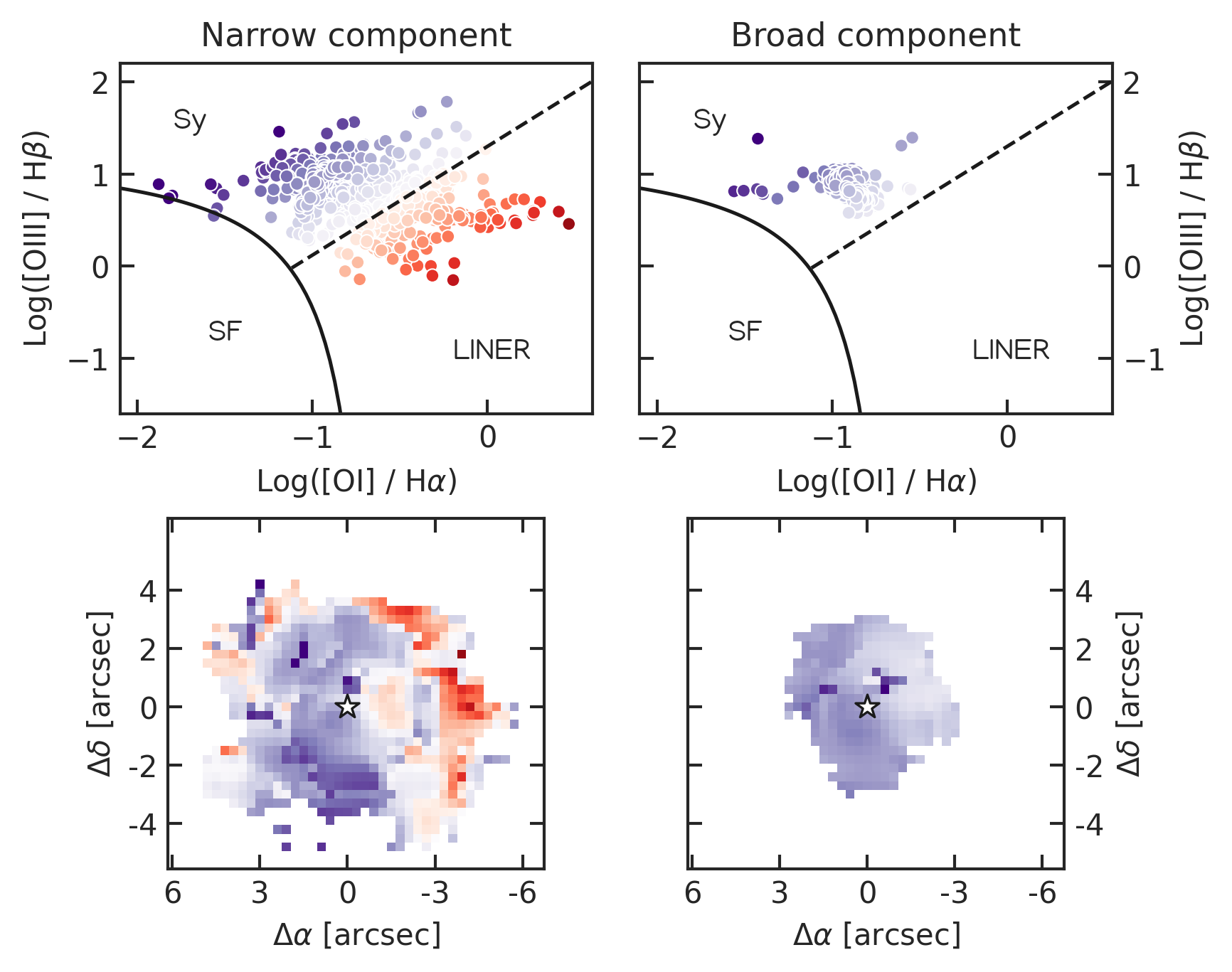}
\caption{Same as Fig.~\ref{fig:bpt1}, but for the
[OIII]/H$\beta$ vs [OI]/H$\alpha$.
The different excitation mechanisms are marked
on the plots as SF (star formation), 
Sy (Seyfert), and LINER 
(low-ionisation nuclear emission-line region),
with separation lines from \cite{veilleux87}.
}
\label{fig:bpt3}
\end{figure}

See Figs.~\ref{fig:bpt1}, 
\ref{fig:bpt2} and \ref{fig:bpt3}.


\section{Mean velocities radial profiles of all MEGARA lines}
\label{sec:more_megara_velocities}

\begin{figure}
\centering
\includegraphics[width=0.49\textwidth]{
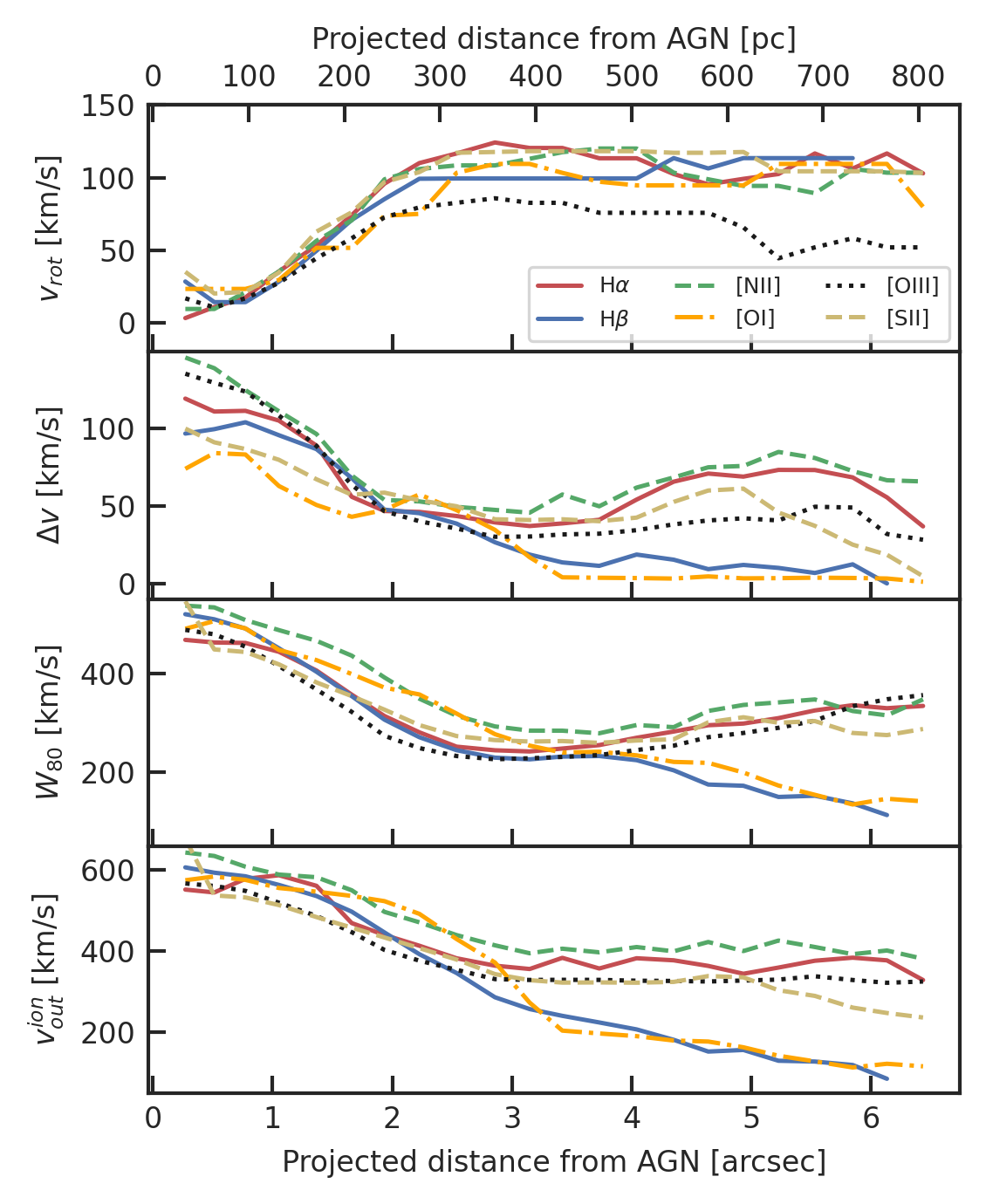}
\caption{Same as Fig.~\ref{fig:OIII_radial},
but for all the ionised gas emission lines.}
\label{fig:megara_velocities}
\end{figure}

See Fig.~\ref{fig:megara_velocities}.

\end{appendix}

\end{document}